\documentclass[ final, reprint,  amsmath,amssymb, nofootinbib, superscriptaddress, showkeys, pra, aps]{revtex4-2}

\usepackage{url}
\usepackage{graphicx}%
\usepackage{amsmath,amssymb,amsfonts}%
\usepackage{float}
\usepackage{mathtools}%
\usepackage{makecell}
\usepackage[gen]{eurosym}
\usepackage{xcolor}
\usepackage{soul}
\usepackage{placeins}

\usepackage{lettrine}
\usepackage{booktabs} 

\usepackage{color} 

\begin{document}

\title{Combined climate stress testing of supply-chain networks and the financial system with nation-wide firm-level emission estimates}

\author{Zlata Tabachová}
\email{tabachova@csh.ac.at}
\affiliation{Complexity Science Hub, Metternichgasse 8, 1030 Vienna, Austria}

\author{Christian Diem}
\email{christian.diem@smithschool.ox.ac.uk}
\affiliation{Smith School of Enterprise and the Environment, University of Oxford, South Parks Road, Oxford, OX1 3QY, United Kingdom}
\affiliation{Institute of New Economic Thinking, University of Oxford, University of Oxford, Manor Road, Oxford, OX1 3UQ, United Kingdom}
\affiliation{Complexity Science Hub, Metternichgasse 8, 1030 Vienna, Austria}

\author{Johannes Stangl}
\affiliation{Complexity Science Hub, Metternichgasse 8, 1030 Vienna, Austria}

\author{András Borsos}%
\affiliation{Complexity Science Hub, Metternichgasse 8, 1030 Vienna, Austria}
\affiliation{National Bank of Hungary, Szabadság tér 9, 1054 Budapest}

\author{Stefan Thurner}
\email{corresponding author e-mail: stefan.thurner@meduniwien.ac.at}
\affiliation{Complexity Science Hub, Metternichgasse 8, 1030 Vienna, Austria}
\affiliation{Medical University of Vienna, Spitalgasse 23, A-1090 Vienna, Austria}
\affiliation{Santa Fe Institute, Santa Fe, 1399 Hyde Park Rd, NM 75791, USA}
\affiliation{Supply Chain Intelligence Institute Austria, ASCII, }

\date{\today}





\begin{abstract}
\noindent 
On the way towards carbon neutrality, climate stress testing provides estimates for the physical and transition risks that climate change poses to the economy and the financial system. Missing firm-level CO\textsubscript{2} emissions data severely impedes the assessment of transition risks originating from carbon pricing. Based on the individual emissions of all Hungarian firms (410,523), as estimated from their fossil fuel purchases, we conduct a stress test of both actual and hypothetical carbon pricing policies. Using a simple 1:1 economic ABM and introducing the new carbon-to-profit ratio, we identify firms that become unprofitable and default, and estimate the respective loan write-offs. We find that 45\% of all companies are directly exposed to carbon pricing. At a price of 45 EUR/t, direct economic losses of 1.3\% of total sales and bank equity losses of 1.2\% are expected. Secondary default cascades in supply chain networks could increase these losses by 300\% to 4000\%, depending on firms' ability to substitute essential inputs. To reduce transition risks, firms should reduce their dependence on essential inputs from supply chains with high CO\textsubscript{2} exposure. We discuss the implications of different policy implementations on these transition risks.
\end{abstract}

\maketitle


\lettrine[lines=2, lhang=0.0, loversize=0.0]{T}{he}  climate crisis imposes large economic costs  \citep{protocol1997kyoto, agreement2015paris, mercure2018macroeconomic, lamperti2019public, IPCC_2022, bressan2024asset} and creates additional risks for financial stability \citep{battiston2017climate, campiglio2018climate, lamperti2019public, battiston2021accounting}. To alleviate the consequences of climate change, numerous policies for reducing CO$_2$ emissions have been implemented \citep{dincer1999energy, directive2003ETSI, directive2023ETSII,regulation2023CBAM, regulation2023SCF}.
While policies like carbon pricing aim to reduce CO$_2$ emissions \cite{stechemesser2024climate}, they will impose significant costs on economic actors, leading to so-called \emph{transition risks} \cite{campiglio2018climate, mercure2018macroeconomic, fierro2024macro}. 
\textit{Climate stress testing} (CST) is the main framework for estimating the potential economic impacts of physical- and transition risks for the economy and the financial system  \citep{battiston2017climate, acharya2023climate}. Many central banks and international organizations use CST for assessing transition- \citep{allen2020climate, vermeulen2021heat, guth2021oenb, roncoroni2021climate, sever2021climate, loschenbrand2024credit} and physical risks \cite{alogoskoufis2021ecb, hallegatte2022bank, LeporeFernando2023, EEA2024, WorldBankClimatePortal}, while climate related risks are included into bank regulation  \citep{baselClimateMeth, baselClimateRiskTransmission, baselClimatePrinciples}.  


The most important policy for reducing emissions to date is carbon pricing \citep{IPCC_2022, stechemesser2024climate}.
For example, the new pricing mechanism of the European Union is the \textit{Emissions Trading System II (EU ETS II)} \citep{directive2023ETSII}. It covers emissions from fuel combustion in buildings, road transport, and other industrial sectors, starting in 2027. Initially, the price per ton of CO$_2$ equivalent emissions will be capped at $45$ EUR/t and will float freely from 2030 onward. 
To limit global warming to 2 degrees, integrated assessment models estimate that carbon prices will need to range between 5–220 USD/t of CO$_2$-equivalent in 2030, and 45–1050 USD/t in 2050 \citep[Section 2.5.2]{IPCC_2022}. 
In such a framework, many firms will face a potentially large cost shock depending on their emissions and future carbon prices. In the short-term, firms either have to pass on increased costs along their supply chains or reduce profit margins \cite{duprez2018price}; if neither is feasible, they face bankruptcy, affecting additional firms through supply-chain dependencies.   

To assess the transition risks of carbon pricing, reliable CO$_2$ emission estimates for individual firms are essential, however, barely available \citep{loschenbrand2024credit}. Reliable estimates only exist for a tiny fraction of firms --- covered by the existing emission trading schemes like, ETS I, \citep{NGFS_emission_data, loschenbrand2024credit} or future reporting requirements for the largest firms 
\citep{EU2022CSRD}. This lack of data makes it {\em de facto} impossible for CST models to comprehensively assess the direct effects of carbon prices on firms, and the corresponding consequences for financial risks faced by banks and the financial system. 
Currently, the state-of-the-art for estimating (financial) transition risks is the taxonomy of Climate Policy Relevant Sectors (CPRS)  \citep{battiston2017climate, battiston2020austrian,kosztowniak2023climate,battiston2023climate} that categorizes NACE 4-digit industry sectors, based on their fossil fuel use, but lacks the actual emissions of firms. Using sector-level instead of firm-level data for assessing transition risks can lead to substantial underestimations of economic losses due to supply network effects \citep{diem2024estimating}. To capture the \textit{indirect} economic impacts of carbon pricing, firm-level supply chain network (SCN) data and appropriate shock propagation models are crucial \citep{barrot2016input, inoue2019firm, carvalho2021supply, diem2022quantifying, pichler2023building, diem2024estimating, tabachova2024estimating}. Yet, CST models today predominantly rely on industry-level data \cite{tabachova2024estimating}.

Here, we conduct the first data driven firm-level-based climate stress test for an entire country. First, we estimate the   CO$_2$ emissions of all 410,523 firms in Hungary that pay value added tax (VAT), based on their oil and gas purchases. We use unique transaction-level VAT records that cover all supply chain links (10 million) between Hungarian firms. Second, we quantify how different CO$_2$ price levels affect firms' profitability and calculate whether firms are likely to default due to  carbon price shocks, should they fail to adapt. Third, comprehensive country-wide bank-firm loan records allow us to assess the impact on banks' financial stability. Fourth, we estimate the effect of carbon-price-induced firm defaults spreading along supply chains which potentially amplifies the \textit{direct} carbon price shock. This is done with the  supply chain contagion model presented in \cite{diem2022quantifying, reisch2022monitoring, diem2024estimating, stangl2024firm, diem2024supply}. Finally, we simulate one pessimistic scenario, where firms can't substitute essential inputs sourced from defaulted suppliers and one optimistic scenario, where full substitution is allowed for. We estimate the resulting losses to the banking system with the financial stress testing framework developed in \cite{tabachova2024estimating}. 


\begin{figure*}[]
	\centering
	\includegraphics[width= .425\textwidth]{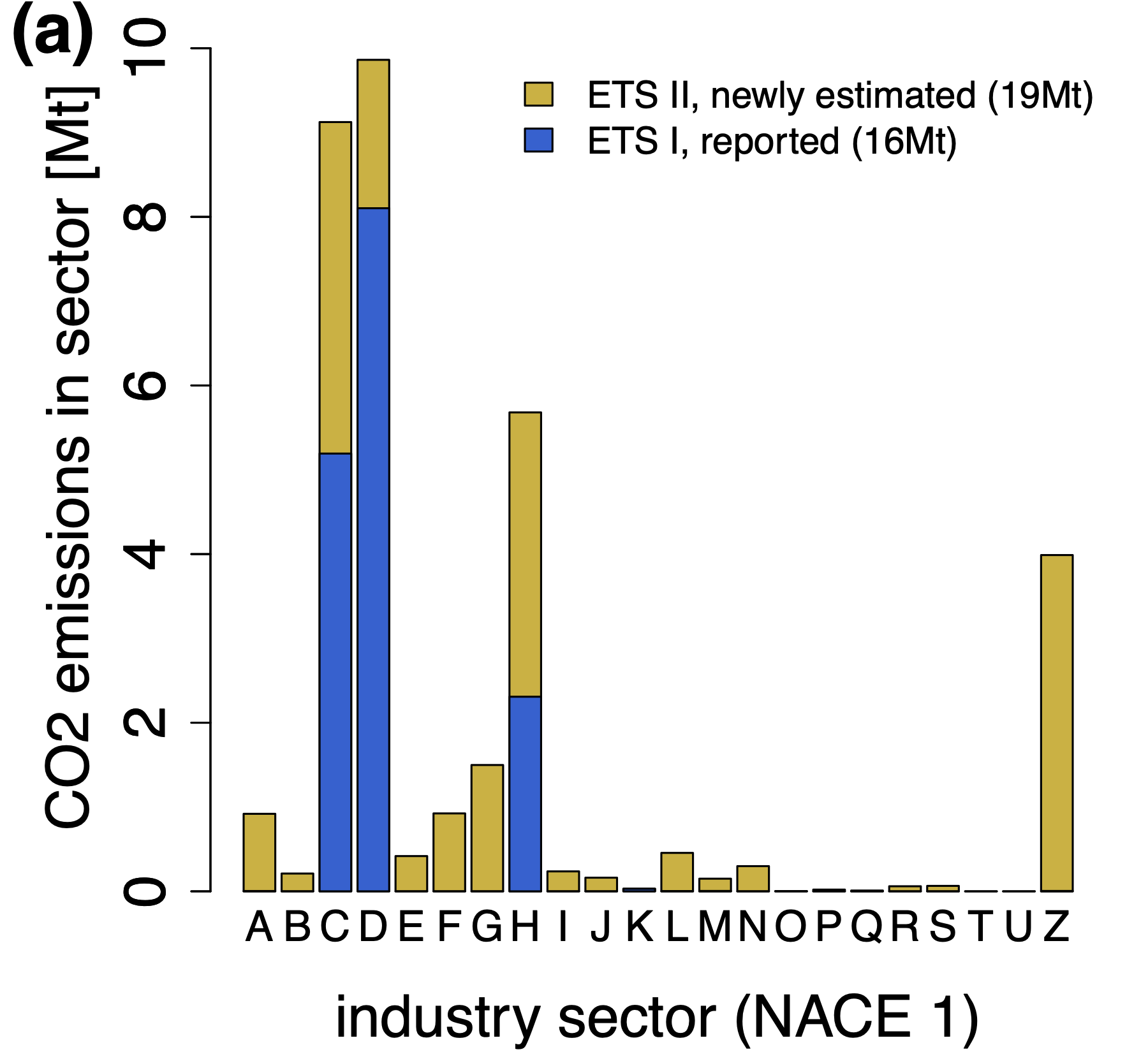}
	\includegraphics[width= .425\textwidth]{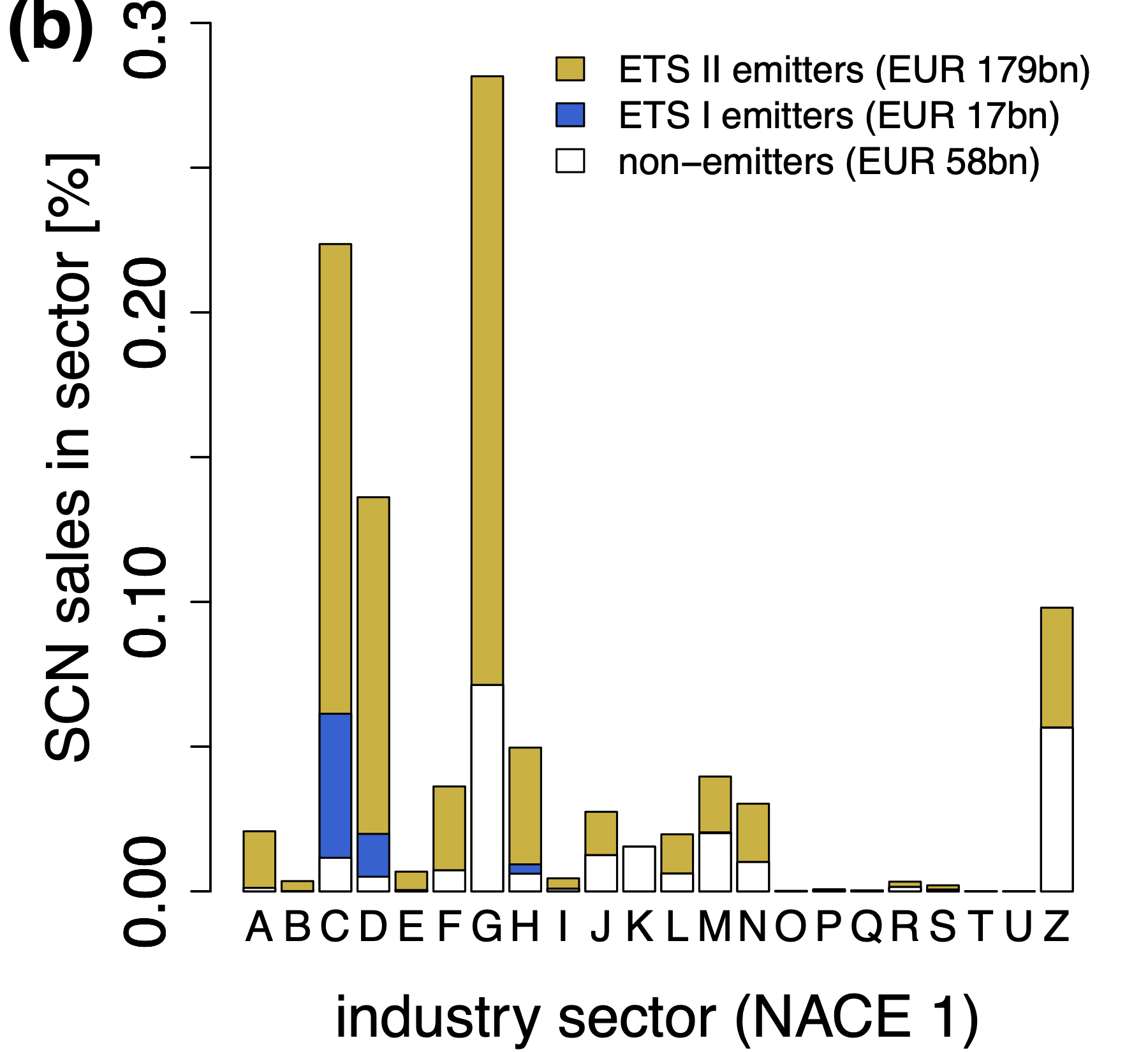}
 	\includegraphics[width= .425\textwidth]{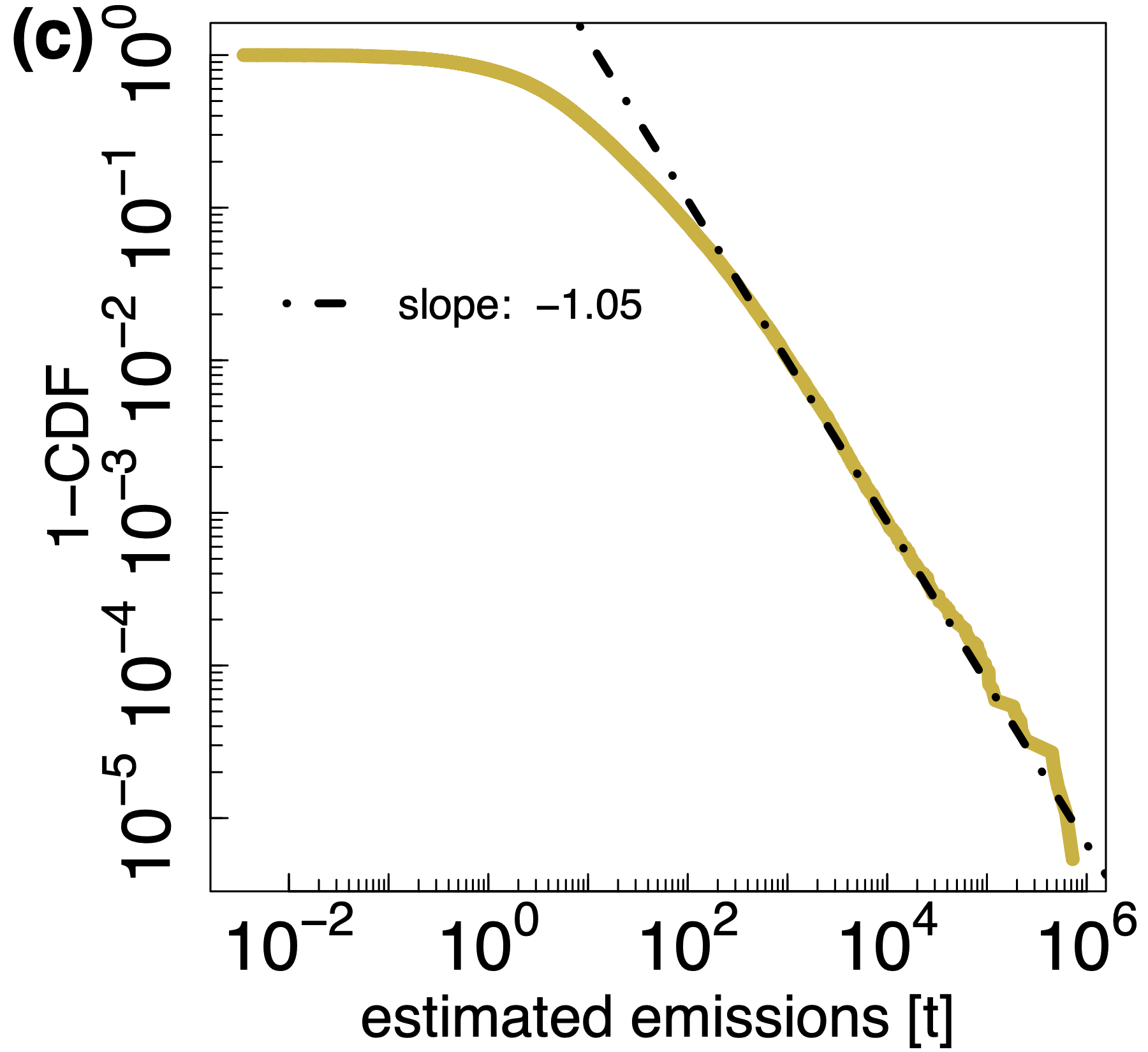}
   	\includegraphics[width= .425\textwidth]{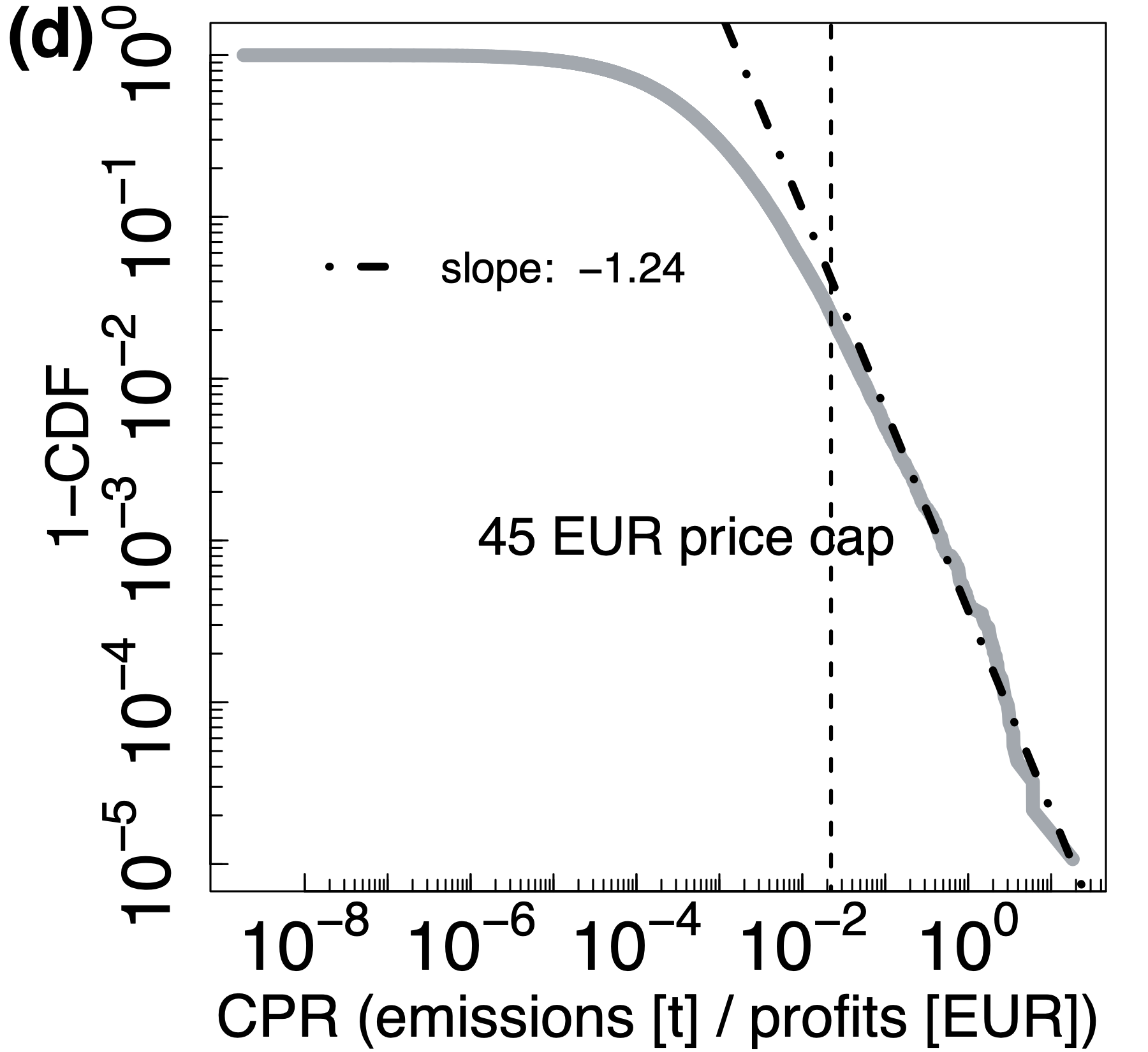}
  	\caption{\textbf{Estimated and reported emissions of firms in Hungary.}  \textbf{(a)} CO$_2$ emissions ($y$-axis) of firms aggregated to 21 NACE 1-digit industry sectors ($x$-axis). Blue bars show emissions of 106 companies reported in EU ETS I, responsible for $16$ Mt. Dark yellow bars show the emissions estimates of $185,783$ (45\%) companies accounting for $19$ Mt. We find missing emissions of $3.9$ Mt, for firms in sector C-Manufacturing, followed by H-Transportation $\&$ Storage ($3.3$ Mt) and G-Wholesale $\&$ Retail ($1.5$ Mt). Sector Z (firms with unknown classification), accounts for $3.9$ Mt. Firms in sector K-Financial $\&$ Insurance Activities are excluded, see Section Methods.  \textbf{(b)}  Fraction of sales of firms (y-axis) within each NACE 1-digit sector (x-axis), (bars add to 1). White bars show sales of firms without emissions; these account for total sales of EUR $58$ bn ($23\%$), ETS I firms (blue bars) account for EUR $17$ bn ($7\%$), while EUR $179$ bn ($70\%$) are from firms with positive emission estimates (yellow), i.e., which will be subject to the new EU ETS II policy. 
   The largest outputs are in sectors G, C and D, in which ETS II firms are responsible for $21\%$, $16\%$ and $11\%$ of total sales respectively. \textbf{(c)} and \textbf{(d)}  Counter cumulative distributions (1-CDF) of estimated emissions and Carbon-to-Profit ratios (CPR) of firms (log-log scale), respectively.  \textbf{Legend:} \textbf{A} Agriculture, Forestry $\&$ Fishing,
  \textbf{B} Mining $\&$ Quarrying,
  \textbf{C} Manufacturing,
  \textbf{D} Electricity, Gas, Steam, Air Conditioning,
  \textbf{E} Water Supply, Sewerage, Waste,
  \textbf{F} Construction,
  \textbf{G} Wholesale $\&$ Retail Trade,
  \textbf{H} Transportation $\&$ Storage,
  \textbf{I} Accommodation $\&$ Food Service,
  \textbf{J} Information $\&$ Communication.,
  \textbf{K} Financial $\&$ Insurance Activities,
  \textbf{L} Real Estate Activities,
  \textbf{M-U} Other Service Activities,
  \textbf{Z} Undefined 
   } 
	\label{fig1_emissions}       
\end{figure*}

\section*{Results}\label{sec_results}

To estimate the impacts of future carbon pricing policies,\footnote{In particular  we are interested in the introduction of the ETS II in 2027 and the removal of the ETS II price cap in 2030.} we use exceptionally comprehensive data sets representing firms (supply chains, loans, income statements) and banks (loans, equity) in 2022\footnote{Note the delay in the arrival process of data.}. We calculate the additional costs of firms from carbon prices  ---  from 10 to 1,000 EUR/t --- based on their 2022 data. We implicitly assume that firms do not change their business model and do not rewire their supply network links before carbon prices are introduced (in 2027 and 2030), i.e., a disorderly transition scenario. In practice firms will  adapt in anticipation of future carbon prices and actual losses are likely to be lower than our stress test suggests.   

\subsection*{Estimating firm-level CO$_2$ emissions with supply chain data}

Currently, only 119 Hungarian firms report CO$_2$ emissions for ETS I\footnote{106 are contained in the VAT based SCN data. Firms active in ETS I vary over time, with 119 emitters in 2022. The total number of firms in the ETS I list does not exceed 300 \citep{ClimateActionEU}.}. Here we estimate the carbon emissions of all $410,523$ individual firms. Comprehensive firm-level supply chain network data allows us to accurately compute the value of firms' purchases from national suppliers in the natural gas and mineral oil sectors. The CO$_2$ emissions for the oil and gas sector in Hungary (household shares are subtracted) are 13.7 Mt and 12.5 Mt, respectively \citep{budget2023global}. We distribute these emissions to firms proportional to their purchases of oil and gas divided by all sales of oil and gas in Hungary, respectively. For details on the estimation procedure see Methods Section Step, 1 and for a validation of the emission estimates based on firms where exact emissions are known, see SI Section \ref{appendixB_add_info_onestimation} Fig. \ref{figSI_scatterplot_emissions}. 

Figure \ref{fig1_emissions}(a) shows the massive improvement of CO$_2$ emissions coverage of firms in the Hungarian economy across 21 NACE 1-digit industry sectors. The $y$-axis shows emissions in Mt of 106 firms covered by the ETS I and SCN data (blue bars) and of $185,783$ firms for which we estimate positive emissions (dark yellow bars), for each industry sector ($x$-axis), respectively\footnote{Note that the blue bars show reported emissions of the 106 firms appearing both in the ETS I data and the supply chain network data. The yellow bars show our emission estimates for all firms not in ETS I, of which $185,783$ (45\%) do have a positive estimate. The others do neither buy oil nor gas.}. While existing ETS I emissions cover 16 Mt that for firms predominantly in sectors C (manufacturing), D (energy), and H (transport), our emission estimate covers an additional 19 Mt carbon emission for firms across all sectors; for further details, see  SI Section \ref{appendixB_add_info_onestimation} and Fig. \ref{figSI_scatterplot_emissions}. 

It is apparent that new carbon prices will likely affect $185,783$ firms that are responsible for 70\% of gross output within Hungary. Figure \ref{fig1_emissions}(b) shows the aggregate sales of firms within Hungary ($y$-axis) for each NACE 1-digit sector ($x$-axis). The sales of firms with known emissions (ETS I) account for only 17 bn (7\%) of sales (blue bars), whereas the $185,783$ firms with newly estimated positive emission (dark yellow) cover 179 bn (70\%) of sales --- this amount of sales will be directly affected by carbon prices (if firms do not adapt). Firms for which our estimate yields zero emissions from oil and gas purchases (white bars) are responsible for 58 bn (23\%) of sales. 

For assessing the transition risks of carbon pricing we need to relate the costs of CO$_2$ emissions of individual firms to their profits. Figure \ref{fig1_emissions}(c) shows the cumulative probability distribution of CO$_2$ emissions in tons; the $y$-axis is the probability of a firm having more emissions than the value indicated on the $x$-axis. Note the log-log scale; firms with 0 emissions are not shown. The distribution is heavy tailed --- few firms have very large emissions, most firms have small emissions. For a reference, the dash-dotted line shows a power-law with exponent $-1.05$. Only 0.09\% of the firms have more than $10^4$t of emissions, and 0.009\% of firms have emissions larger than $10^{5}$t. The distribution of firms' emissions across NACE 1 sectors is shown in SI Section \ref{appendixB_add_info_onestimation} Fig.\ref{figSI_boxplots_emissions}.  

To analyze the sensitivity of individual firms to  carbon pricing, we calculate their \textit{Carbon-to-Profit Ratio (CPR)}, as a firm's emissions in tons ($\mathcal{E}_i$) divided by its net profits ($\mathcal{P}_i$), i.e., $\textrm{CPR}_i = \frac{\mathcal{E}_i}{\mathcal{P}_i} $. A CPR of 0.1 (0.02) means that a firm becomes unprofitable at a carbon price, $\pi$, of 10 EUR/t (50 EUR/t), and hence, goes out of business. Knowing the CPR of firms before introducing carbon pricing highlights, which firms need to adapt the most to avoid bankruptcy. Figure \ref{fig1_emissions}(d) plots the CPR of firms ($x$-axis) against the probability ($y$-axis) of firms having a CPR higher than the $x$-axis value. 2,328 firms (2\% of all firms with non-negative profits and estimated emissions) will become unprofitable at the ETS II price cap of 45 EUR/t (dashed vertical line) ---  given their current business model and assuming they can not pass on the costs\footnote{Note, that the profit variable is not available for all the firms in the supply chain network, therefore, the CPR ratio can be quantified for a subset of firms. The number of firms with positive profits and estimated emissions is 93,215.}. 


\subsection*{Estimating firms' direct output losses from carbon pricing}

Assessing transition risks at specific carbon price levels (EUR/t) requires estimates of their effects on the real economy (measured in gross output) and financial system losses (measured in bank loan write-offs). To estimate \textit{direct} real economy losses for a given carbon price, $\pi$, in EUR/t, we estimate which firms become unprofitable due to  additional costs from carbon pricing, and, hence should shut down.\footnote{If for a given carbon price, $\pi$, a firm becomes unprofitable, continuing its operations would result in losses for its owners and they should decide to close the firm and take out the remaining equity. In accounting, the contribution margin becoming negative is called \textit{shut down point}.} Given the emission estimate, $\mathcal{E}_i$, of firm, $i$, (see Fig.\ref{fig1_emissions}(c)) and the carbon price, $\pi$, firm $i$ faces costs of $\pi \cdot \mathcal{E}_i$. Hence, $i$ becomes unprofitable if its current profits, $\mathcal{P}_i$, are smaller than the additional carbon costs $\pi \cdot \mathcal{E}_i$. To arrive at the economy-wide \textit{direct} production (gross output) losses for a given price, $\pi$, we sum the sales, $s_i^{out}$, of all firms $i$ becoming unprofitable and divide by the total sales of all firms; for details, see Section Methods, Step 2.  

\begin{figure*}
	\centering
	\includegraphics[width= .75\textwidth]{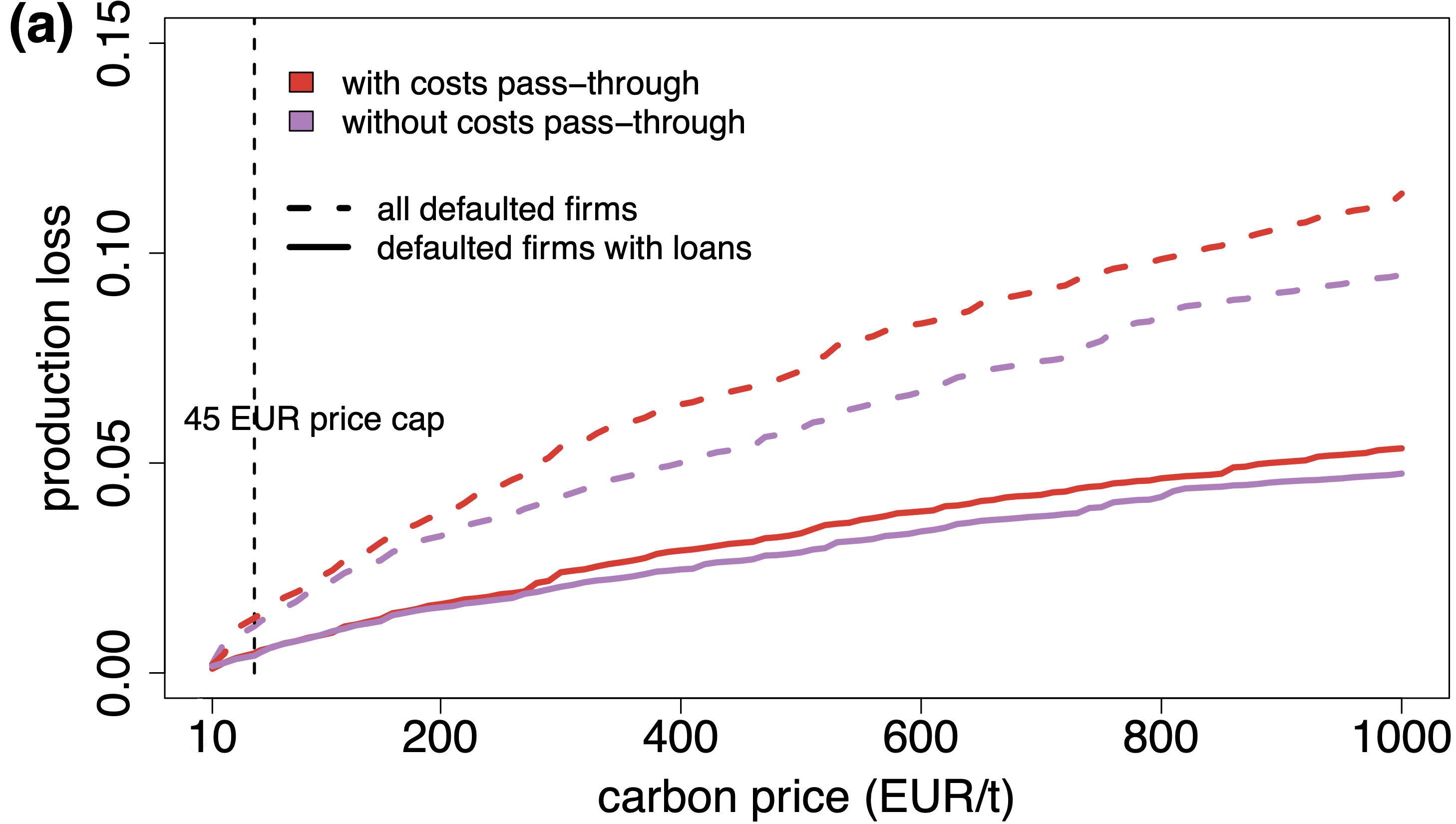}
 	\includegraphics[width= .75\textwidth]{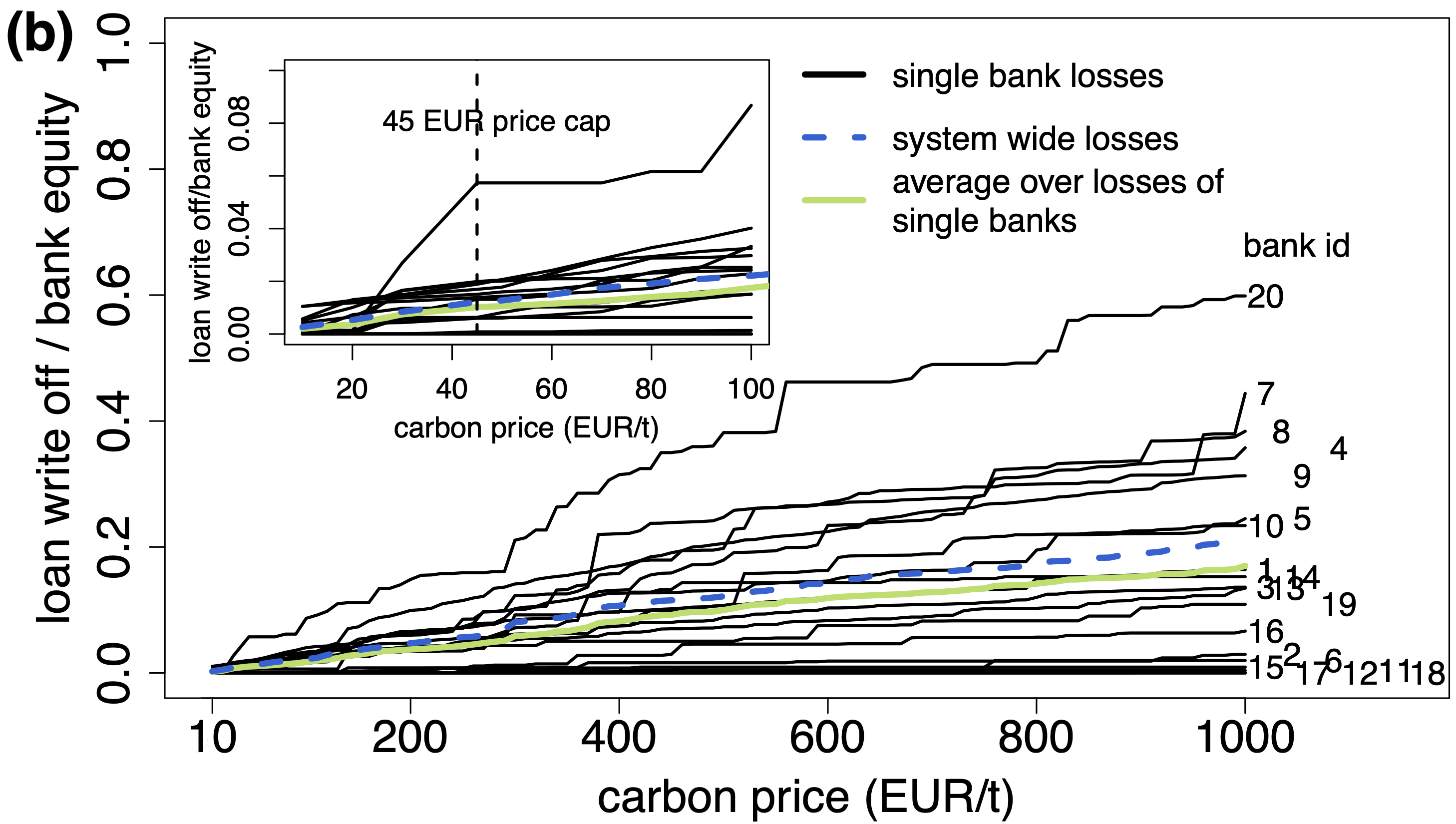}
  	\caption{\textbf{Direct losses to the real economy and the financial system from carbon pricing.} Losses are calculated from firms becoming unprofitable at a given carbon price ($x$-axis). \textbf{(a)} Fraction of total sales lost from firms defaulting due to carbon pricing ( $x$-axis). Dotted lines show losses considering all firms in the SCN, solid lines consider only those with bank loans. Red indicates losses computed with a carbon cost pass-through; purple means no pass-trough.  
    The total sales lost without pass-through start at 0.2\% for 10 EUR/t, and reach 1\%, 2\% and 3.3\% at 45, 100 and 200 EUR/t, respectively. Production losses are higher with cost pass-through, (0.2\%, 1.3\%, 2.3\%, and 3.8\% for 10, 45, 100 and 200 EUR/t, respectively). When considering only firms with loans, losses are roughly a factor two lower. 
    \textbf{(b)} Direct total bank equity losses ($y$-axis) from firms defaulting at a given carbon price scenario ($x$-axis). Black lines show losses of individual banks, IDs are shown on the right; compare Fig.\ref{fig2_two_risk_approaches}.    
    The inset shows that for a price of 10 EUR/t loan write offs range from 0 to 1.5\% of equity across banks, increasing only slightly with price, and range from 0 to 8.5\% for a price of 100 EUR/t. Banking system-wide equity losses (blue dashed line) range from 0.3\%, 1.2\%, 2.2\%, 4.7\% and 21\% for carbon prices of 10, 45, 100, 200 and 1000 EUR/t, respectively. Average losses across banks (green line) are slightly lower.
    }
	\label{fig4_banks_dir}       
\end{figure*}

Figure \ref{fig4_banks_dir}(a) shows the economy-wide \textit{direct} production losses ($y$-axis) for carbon price scenarios of 10, 20, $\dots,$ to 1,000 EUR/t ($x$-axis) for all firms as a purple, dashed line.\footnote{Production here is measured in terms of value of products sold to other firms, i.e., sales.} The total  sales losses are 0.2\%, 
 2.0\%, 
 3.3\%, 
 and 9.5\% 
at 10, 100,  200 and 1,000 EUR/t, respectively. At the ETS II price cap of 45 EUR/t, losses are estimated to affect around 1\% of economy-wide sales.
Within the carbon price range compatible with 2 degree warming (estimated at 5–220 USD/t in 2030, \citep{IPCC_2022}) losses could increase up to 3.3\% at 200 EUR/t. The purple solid line shows corresponding production losses when only firms with bank loans (relevant for loan write offs) are considered. 

Firms would likely attempt to pass on additional  carbon costs to their customers. To capture this, we use a simple \emph{cost pass-through} mechanism that yields updated costs for each firm. The model assumes that firms can pass on a fraction of their additional carbon costs proportional to their market share (within their NACE 4-digit sector), for details, see Section Methods, Step 2. Figure \ref{fig4_banks_dir}(a) shows the economy-wide \emph{direct} production losses, given the \emph{cost pass-through}-based carbon costs of firms as red dashed line. Interestingly, the production losses are slightly higher when allowing for carbon costs being passed on downstream along supply chains,  0.2\%, 1.3\%, 2.3\%, and 3.8\% for 10, 45, 100 and 200 EUR/t, respectively. 
The difference in losses increases to about 1.5 percentage points for prices above 300 EUR/t. When considering the output losses from only firms with loans (solid lines), the difference is negligible for prices below 300 EUR/t and small for prices above. The subsequent results are obtained using carbon costs including pass-through. Note here we consider sales losses due to firms shutting down, but ignore that firms might sell less due to higher prices. 
Direct and in-direct output and bank equity losses and the corresponding amplification factors are summarized in SI Section \ref{SI_loss_comparison_table} Table \ref{SI_tab_loss_comparison_per_co2price}.


\subsection*{Estimating direct financial system losses from carbon pricing}

Next, we assess the exposure of the banking system to transition risks from carbon pricing. We analyze the largest $m=20$ banks reporting CET1 capital data (totaling EUR 13 bn) and all their commercial loans to $56,595$ Hungarian firms (totaling EUR 24 bn); for details on the data processing steps, see Section Data.\footnote{We exclude firms with negative profits from the default calculations. This yields 36,255 firms with a loan volume of 17 bn.} 
SI Section \ref{appendixD_bank_exposures_and_cprs} Fig. \ref{fig2_two_risk_approaches} shows for every bank the amount of loans outstanding to firms affected by carbon pricing. We find that exposures based on the CPRS taxonomy \citep{battiston2022nace} can substantially under- or overestimate banks' exposures to carbon pricing, see Fig. \ref{figSI_loans_NACE4}, while firms' actual costs from carbon pricing are not quantifiable with CPRS. 

To estimate banks' losses from carbon pricing, we calculate the amount of loans that must be written off when firms become unprofitable due to additional carbon costs, see Methods Section Step 3. Figure \ref{fig4_banks_dir}(b) shows the \textit{direct} losses from loan write-offs for every bank (solid black lines) as a fraction of their equity ($y$-axis) for carbon price scenarios ranging from 10 to 1,000 EUR/t ($x$-axis); bank IDs are shown on the right. It is immediately visible that the fraction of lost equity due to carbon prices strongly varies between banks. The inset shows that for prices at 10 EUR/t and 100 EUR/t loan write offs range from 0 to 1.5\%, and, 0 to 8.5\% of equity across banks, respectively.  
At the ETS II price cap of 45 EUR/t, 12 banks suffer less than 1\% of losses, with the highest individual bank's loss amounting to 6\%. Banking system-wide equity losses range from 0.3\%, 1.2\%, 2.2\%, 4.7\%, and 21\% for carbon prices of 10, 45, 100, 200, and 1,000 EUR/t, respectively. This indicates that with a carbon price cap of 45 EUR/t \textit{direct} bank equity losses will be small. Within the carbon price range compatible with 2 degree warming at approx. 200 EUR/t losses could increase up to 4.7\%. 
This means that banks should definitely also monitor the expected carbon costs of their debtors. 


\begin{figure*}[t]
	\centering
	\includegraphics[width= 0.9\textwidth]{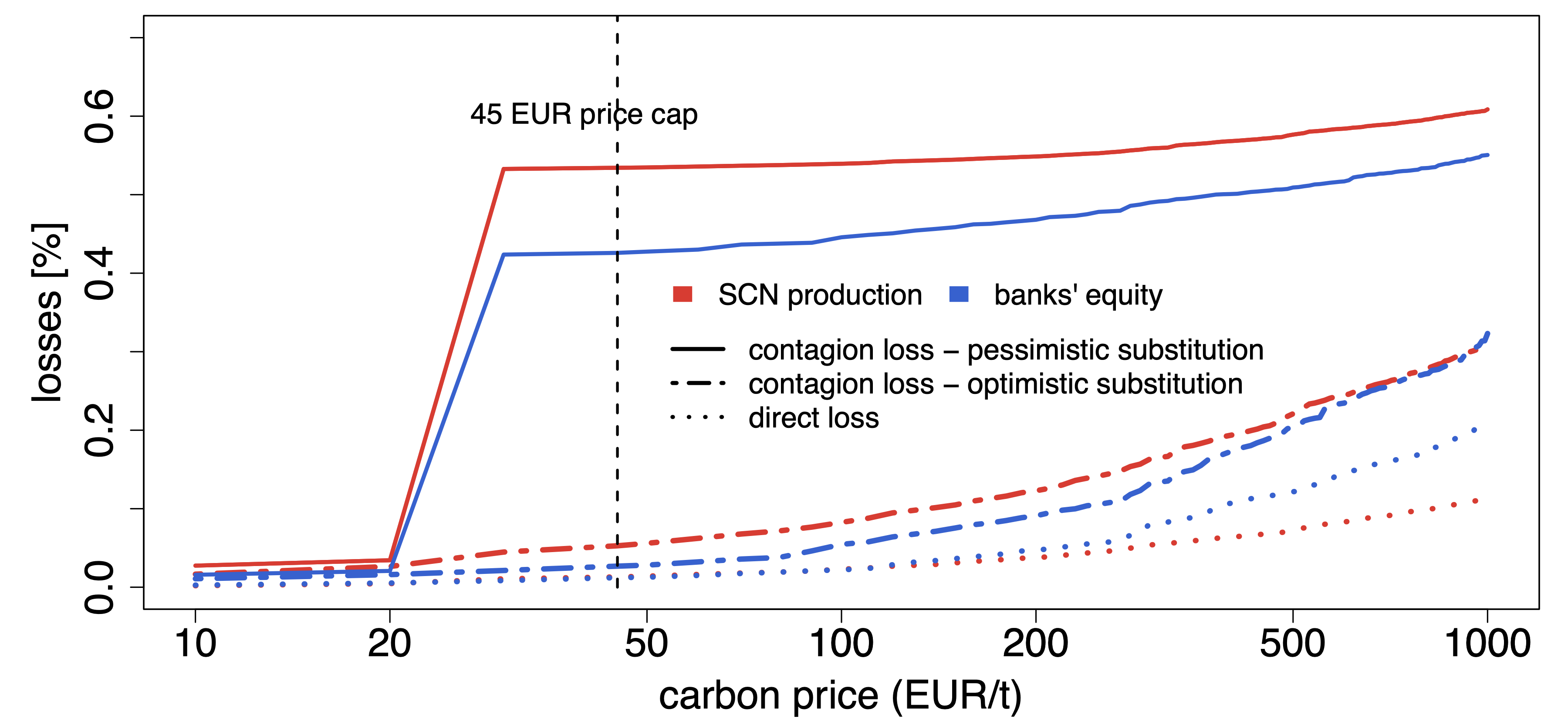}
  	\caption{\textbf{Amplification of economic and financial  losses due to supply chain contagion.}  
    The fraction of output and  total equity losses ($y$-axis) are shown for various CO$_2$ price scenarios ($x$-axis; log-scale). Losses are based on carbon costs with SCN cost pass-through. The EU ETS II price cap of $45$ EUR/t is shown as a dashed vertical line.  The \textit{direct} gross output losses from Fig.\ref{fig4_banks_dir}(a) (dotted red line) and  
    \textit{direct} bank equity losses Fig.\ref{fig4_banks_dir}(b) (dotted blue line) are shown as a reference. 
    In the \textit{optimistic} scenario (red dash-dotted line) the total gross output losses with SCN contagion range between 1.7 and 12.3\% across for prices from 10 to 200 EUR/t. Total bank equity losses (blue dash-dotted line), range between 1.1 and 9.1\%.
    The \textit{pessimistic} scenario shows a jump at 30 EUR/t from about 3\% to more than 50\% of gross output losses, while bank equity losses jump to 43\% (blue solid line). 
    } 
	\label{fig4_fin_prod_loss}       
\end{figure*}

\subsection*{Amplification of transition risks due to supply chain contagion}

The COVID-19 pandemic and natural disasters demonstrated that \textit{direct} shocks to firms propagate along SCNs which can lead to a substantial amplification of economic losses (\textit{indirect shocks}) \citep{hallegatte2008adaptive, barrot2016input, inoue2019firm, boehm2019input, carvalho2021supply}. 
In our context, firms that default at a given carbon price  stop purchasing from their suppliers and supplying to their customers, causing demand and supply shocks that propagate across the SCN. The size of shock propagation cascades is largely determined by how easily firms can replace defaulted customers and suppliers and substitute their missing essential inputs \citep{pichler2020production, diem2022quantifying, diem2024estimating}. We model this uncertainty by simulating one \emph{optimistic} scenario using linear production functions allowing for full substitution between firms' inputs and one \emph{pessimistic} scenario based on Generalized Leontief production functions, where firms cannot substitute essential inputs. For details on these scenarios and shock propagation model calibration, see Methods Section, Step 4 and \cite{diem2022quantifying, diem2024estimating}\footnote{In the simulations it is implicitly assumed that firms do not adapt prior to the introduction of the carbon price.}.

\subsubsection*{Indirect supply network losses}

Figure \ref{fig4_fin_prod_loss} shows the fractions of economy-wide output loss and banking system-wide equity loss (y-axis) as a function of the CO$_2$ price, including cost pass-through. For results without cost pass-through, see SI Section \ref{AppendixF_ad_info_systemic_risky_firms} Fig. \ref{figSI_fsri_esri_costs_pass_through}.  Direct gross output losses from Fig. \ref{fig4_banks_dir}(a) (dotted red line) and direct bank equity losses Fig. \ref{fig4_banks_dir}(b) (dotted blue line) are shown as reference. 

In the \textit{optimistic} scenario, total gross output losses (red dash-dotted line) are initially small with almost 1.7\% at 10 EUR/t, increase noticeably to 5.3\% at 45 EUR/t, reach 12.3\% at 200 EUR/t, and increase further up to 31\% for a price of 1,000 EUR/t.\footnote{Indirect losses correspond to the difference between dotted and dashed-dot lines.}\footnote{\citep{diem2024estimating} show that for COVID-19 like shocks, propagation effects in a standard Cobb-Douglas general equilibrium model are on average approximately a third smaller than when using linear production functions in our model. Here we expect this relation to be similar.} 
For low prices the supply chain contagion amplification of  transition risks is relatively small in absolute terms,  for the ETS II cap of 45 EUR/t gross output losses are amplified by a factor of 4. In absolute terms the amplification at 200 EUR/t is substantial, gross output losses increase from 3.8\% to 12.3\%. This implies that if carbon prices were to start at high levels and firms would not adapt beforehand, transition risks would cause substantial economic losses even when firms can easily substitute inputs. 

In the \textit{pessimistic} scenario (essential inputs can not be substituted), the gross output losses including shock propagation (red solid line) start at 2.8\% for 10 EUR/t and jump at 30 EUR/t to above 50\%. This remarkable jump occurs because a firm that belongs to the {\em systemic risk core} of the SCN \cite{diem2022quantifying} defaults due to its additional carbon costs. Firms in the systemic risk core are inter-linked through essential supply relations that lack alternative suppliers (the corresponding inputs can't be substituted in the pessimistic scenario). Hence, the failure of one core firm cascades to other core firms that will suffer large production losses that then propagate to a large fraction of the SCN. For details on high systemic risk firms, see SI Section \ref{AppendixF_ad_info_systemic_risky_firms}, for details on the core \citep{diem2022quantifying}. This implies that if a systemic core firm was to become unprofitable due to carbon costs and firms were not able to substitute essential inputs, even a relatively low carbon price could bear potentially significant transition risks. Consequently, policy makers should monitor the timely transition of firms that are systemically critical in their country's SCN.

\subsubsection*{Indirect losses in the banking system} 
Supply chain contagion causes temporary declines of firms' production levels, reducing their sales and purchases. To assess the SCN-amplification effects on financial stability, we translate production losses to financial losses by updating the firms' income statements and balance sheets accordingly. Then, we assess which firms become insolvent or illiquid and default in this counterfactual scenario. Finally, we calculate additional bank equity losses from writing off loans given to firms that default due to supply chain contagion. For details, see Section Methods, Step 5 and \cite{tabachova2024estimating}.

For the \textit{optimistic} scenario Fig. \ref{fig4_fin_prod_loss} shows, that total bank equity losses (blue dash-dotted line), start at 1.1\% and gradually increase to 32\% at 1,000 EUR/t. At the ETS II cap of 45 EUR/t the supply chain contagion amplifies the losses from 1.2\% to 2.7\%, for 200 EUR/t losses are amplified substantially from 4.7\% to 9.1\%. 
This means that transition risks from prices below the ETS II cap could be approx. doubled by supply chain contagion, but at relatively low levels, hence, are unlikely to affect financial stability. 
In contrast, the 2030 upper bound of 2-degree compatible CO$_2$ prices (5–220 USD/t) could involve more substantial risks to the banking system, \textit{if} firms and banks do not adapt in advance.  
In the \textit{pessimistic} scenario, due to the failure of a systemic risk core firm, bank equity losses (blue solid line) jump to  43\%  (EUR 5.6 bn) at 30 EUR/t. This implies that regulators and central banks should monitor whether  high systemic risk firms adapt in time to reduce the potential threats for financial stability.  
Overall, the contagion results show that banks are not only exposed to transition risks through their own debtors' CO$_2$ emissions, but also the emissions in their debtors' supply chains; for details, see SI Section \ref{appendix_bank_losses_45_pessimistic}. 

SI Section \ref{appendix_45_N1_losses} shows that banks are exposed to climate transition risk not only from firms in high emission sectors (C, H, D or G), but also from firms in sectors with relatively small emission shares like (L and F) due to network effects. 
For details on how individual banks are affected in the 45 EUR/t scenario, see SI Section \ref{appendix_bank_losses_45_pessimistic}.
For robustness checks with respect to supply network topology, see SI Section \ref{AppendixF_ad_info_systemic_risky_firms}. 


\section*{Discussion}\label{sec_discussion}

By estimating CO$_2$ emissions for all Hungarian firms paying VAT, we show that future carbon pricing policies such as the EU ETS II scheme will affect 45\% of all companies (which account for 70\% of total sales), across all industrial sectors. Existing emissions data (ETS I) only accounts for 119 Hungarian firms, concentrated in manufacturing, utilities, and transport industries, with combined sales of 7\%.  The loans granted by banks to firms affected by carbon pricing exceed their CET1 capital in nearly half of the banks.
CPRS-based exposure estimates \cite{battiston2017climate} can substantially misestimate the CO$_2$-estimates based exposures of banks to carbon pricing. 

At a carbon price of 45 EUR/t (EU ETS II 2027-2030 price cap) we find that \emph{direct} economic losses are relatively small. Those firms that become unprofitable account for about 1.3\% of economy-wide gross output and 1.2\% of the total bank equity, which are at risk if firms do not adapt in time. A rapid carbon price introduction from 0 to 200 EUR/t (the upper bound of the CO$_2$ price range in 2030 compatible with 2 degree warming \citep{IPCC_2022}) yields 3.8\% of \emph{direct} gross output and 4.7\% of bank equity losses, respectively. 

Analyzing how firm defaults spread along supply chain networks, we find that supply chain contagion is a  substantial amplification factor for transition risks caused by carbon pricing. When assuming that firms can fully substitute their inputs, gross output loss estimates are at a noticeable 5.3\% at 45 EUR/t, and reach 12.3\% at 200 EUR/t, while bank equity losses amount to 2.7\% and 9.1\% for 45 EUR/t and  200 EUR/t, respectively. 
Assuming that firms cannot substitute essential inputs, direct losses are amplified by factors of up to 40, as high systemic risk firms \citep{diem2022quantifying} default due to carbon pricing and their output can not be substituted. This pessimistic scenario particularly highlights the importance of systemically relevant firms transitioning early, in line with \cite{stangl2024firm}. Simulating unlikely pessimistic scenarios in stress testing gives a more complete picture of potential risks to decision makers \citep{battiston2024enhanced}. Importantly, our findings show that banks are not only exposed to transition risks from their debtors CO$_2$ emissions, but also to emissions in debtors' supply chains. This matters for banks' transition risk management strategies.

The estimated economic losses presented here are \emph{short-term} losses before recovery from the initial carbon pricing shock sets in and are \textit{not} long-term reductions of output. 
These estimates suggest non-negligible transition risks in response to a sudden carbon price increase from 0 to 200 EUR/t (upper price bound compliant with 2 degrees warming) when firms are not preparing for high CO$_2$ prices. This is consistent with recent assessments of {long-term macro-economic} implications (employment, GDP, etc.) of CO$_2$ price-paths based on \textit{aggregate} data \citep{fierro2024macro}. Given that the decarbonization of the economy must happen fast \cite{fankhauser2022meaning} with a minimum of economic cost, implementation matters. Our results suggest that starting carbon pricing with a low, but fast enough increasing price cap, reduces transition risks when compared to the current ETS II plan, where a 45 EUR/t price cap over 3 years is followed by a rapid transition to free floating prices in 2030 with the risk of a large and unpredictable price jump, see also SI Section \ref{SI_limitations_and_policy}. The presented CST methodology can aid policy makers in assessing transition risks of carbon pricing implementation strategies. 

We neglected positive economic effects from growth in green energy technologies \citep{bddotnucker2023employment} and the fact that technological progress in renewable energy will  decrease energy costs \citep{way2022empirically}. Both effects will offset costs and economic losses from carbon pricing. As mentioned, our model neglects the anticipation and adaptation of firms and banks before CO$_2$ price implementation. Consequently, our loss estimates should be interpreted as upper bound estimates for the losses when no adaptation takes place and should not be interpreted as forecasts of actual future losses. Note also that economic losses  could be considerably higher in a scenario where no transition policies are implemented due to global warming \citep{IPCC_2022}.  
Another limitation is that the SCNs considered here are static, new models would be needed to capture the re-linking dynamics of firms and the evolution of supply chains as a whole. That would also provide a more realistic handling of growth and price reductions in green sectors. For more model limitations, see SI Section \ref{SI_limitations_and_policy}.

Transition risks in terms of direct economic losses of currently planned carbon pricing schemes appear manageable. However, contagion effects along supply chains could become a significant amplifier of transition risk that has been overlooked so-far. Indirect supply chain effects can become decisive when comparing different  implementation schemes of carbon pricing. Predictable implementations with a continuously increasing price likely lead to less transition risks, than price trajectories with large unpredictable discontinuities. 



%

\section*{Declarations}

\textbf{Acknowledgments.} This work was supported by Jubilaeumsfonds of the Austrian central bank project under P18696 and UK Research and Innovation MSCA Postdoctoral Fellow guarantee under EP/Z003199/1. 

\textbf{Competing Interests.} The authors declare that there are no conflicts of interest. 

\textbf{Data availability.} Data is confidential and cannot be shared, but is accessible through the central bank of Hungary. 
Replication codes are available at \url{https://github.com/zlatataa/Supply_chain_adjusted_financial_climate_stress_testing}. 
 
\textbf{Author contribution.} 
C.D., Z.T., and S.T. conceived the work and wrote the paper. A.B. prepared SCN data. Z.T. wrote the code. J.S. calculated the emissions. 


\newpage
\section*{Methods}\label{sec_methods}

\subsection*{Data}\label{sec_data}

The data used in this study is available at the Central Bank of Hungary. It includes information about Hungarian firms in 2022 gathered from several sources: the Hungarian firms registry, firms' VAT reports, bank-to-firm-level loan data from the credit registry and CET1 capital of banks collected by the regulator. The VAT reports include firm-to-firm transactions recorded throughout the year, which are used to reconstruct the supply chain network, $W$. As in earlier works \citep{diem2022quantifying, diem2024estimating, tabachova2024estimating} we filter out one-time purchases, to retain only links that appear at least twice in two different quarters, for the shock propagation application, but not for emissions estimation. This ensures that only stable links representing supply chain links are included for shock propagation calculation. The final network, $W$, reflects annual transactions, with each element, $W_{ij}$, representing the total purchase volume of firm $j$ from firm $i$. Further, company registry data, including balance sheets and income statements, is used to extract financial variables such as equity or profits of firms in the supply chain network. Finally, we merge the credit registry data containing all loans between banks and firms, with the supply network and company registry data. We don’t differentiate between loan types, and if a firm has multiple loans with the same bank, we combine them into a single amount. The bank-firm loans are represented in the matrix $B$, where each element, $B_{ik}$, reflects the total exposure of bank $k$ to firm $i$.

Overall, the data includes 410 523 firms in the supply chain network, out of which 56 595 have loans with 20 banks that report CET1 capital. The total output of the network (sum over all links) equals to EUR 254 bn (see Tab.\ref{tab:thresholded_W}), GDP in 2022 was approx. EUR 169 bn, see \url{https://www.ksh.hu/stadat_files/gdp/en/gdp0004.html}. The loan volume (sum over outstanding principals of firms) is EUR 24 bn, and CET1 capital of 20 banks is equal to EUR 13 bn. The exchange rate used in this study is 1 EUR = 400 HUF. 

\begin{table*}[]
\centering
\begin{tabular}{ c c c c c c}
\makecell{link weight \\ threshold [EUR]}&\# firms & \# links & \makecell{total sales \\ \newline [bn EUR]} & \makecell{loan volume \\ \newline [bn EUR]} & \makecell{estimated \\ emissions [Mt]} \\
\hline
none& 410 523 (100\%) & 10 698 769 (100\%) & 254 (100\%) & 17.2 (100\%) & 19.8 (100\%) \\
25 000& 203 592 (50\%) & 733 251 (7\%) & 232 (91\%)  & 16.9 (98\%)&19.5 (98\%) \\
250 000& 51 618 (13\%) & 105 949 (1\%) & 187 (74\%) & 15 (87\%) & 17.5 (88\%) \\
\hline
\end{tabular}
\caption{\textbf{Summary characteristics of the full and subsetted networks.} The first row provides information about the full network, $W$. The subsetted networks, $W^s$, (line 2 and 3) are derived by removing all links below the specified link weight thresholds. Percentage shares in parentheses are with respect to the full network. Loan volume is equal to loans of firms, that satisfy conditions of the default definitions given by eq.(\ref{chi_dir}) and (\ref{chi_indir}). Note that emission estimation is performed only on the full network and emissions of the sub-networks presented in the table are those estimated emissions of firms that remain in the sub-network after thresholding.}
\label{tab:thresholded_W}
\end{table*}

In Section Methods we introduce \textit{direct} and \textit{indirect defaults} (see eq.(\ref{chi_dir},\ref{chi_indir})), which are well defined only if revenue, material costs, equity, liquidity, operating and net profits of a firm are non-negative. Firms reporting negative values for these variables  are excluded from steps 3 and 5 of our model that involve defaults, see below. By doing so we reduce number of firms, that can initiate the contagion and cause direct and indirect losses to banks. The non-negativity of the mentioned financial variables is satisfied by 299 830 (73\%) out of 410 523 firms (see SI Section \ref{appendixC_data} Tab.\ref{tab:fin_var_410K} and Tab.\ref{tab:fin_var_299K}). The loan volume then reduces from EUR 24 bn (of 56 595 firms) to 17 bn (of 36 255 firms), which decreases upper possible losses that we obtain in our stress testing. Nevertheless, excluded firms are retained in the remaining steps that do not include default checks. They can still propagate shocks within the supply chain network or pass carbon costs on their buyers.

\begin{figure*}[]
	\centering
	\includegraphics[scale=.3, keepaspectratio]{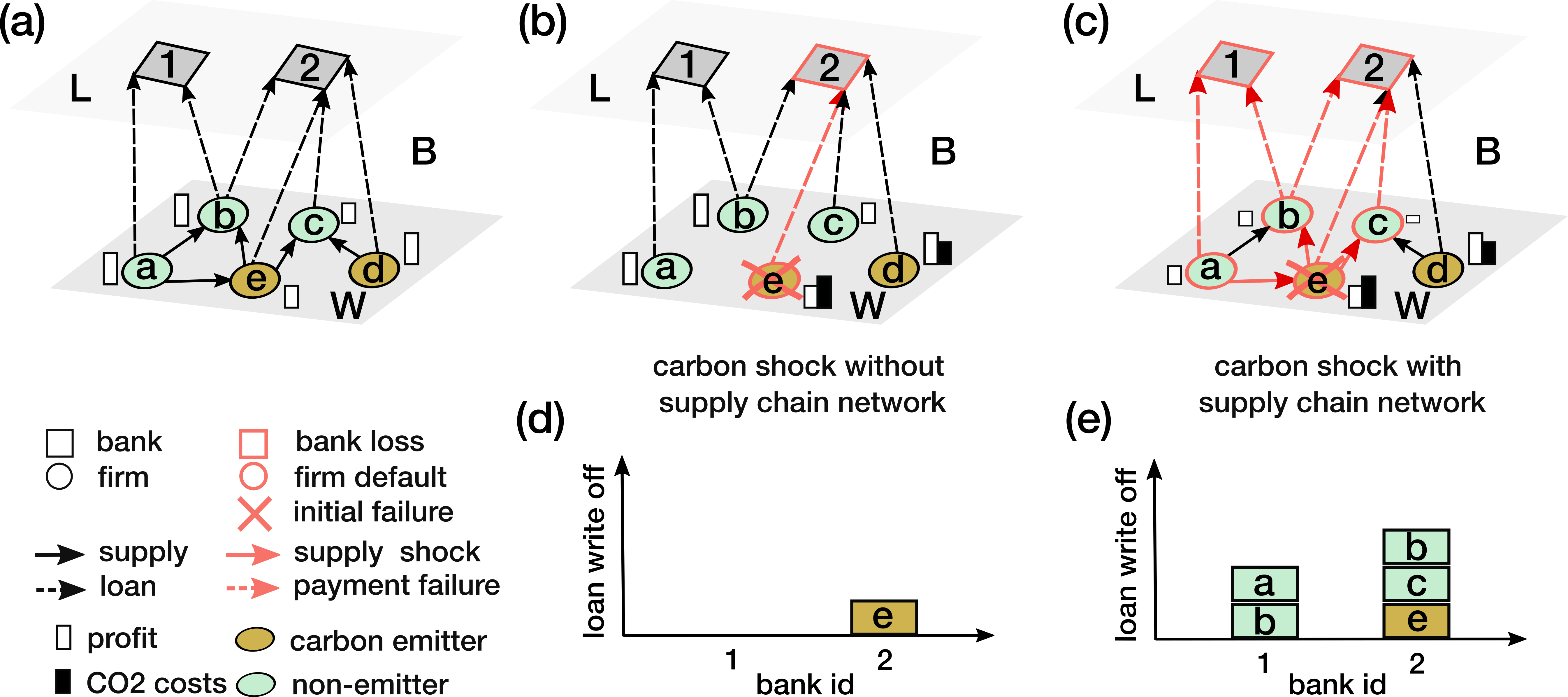}

  	\caption{\textbf{Schematic view of firm-to-bank contagion on the multi-layer network with and without supply chain links.} Panel \textbf{(a)} shows the two-layer network representation of the interconnected economic and financial system. The real economy is represented by the supply chain network, $W$. Every node is a firm (circle) and a link, $W_{ij}$, (solid arrow) indicates a supplier-buyer relation, from supplier $i$ to buyer $j$, with the weight equal to the transaction volume between the two firms. Additionally, yellow color of circles indicates that firms $e$ and $d$ are carbon intensive, while the light green color marks non-carbon intensive firm $a$,$b$ and $c$. White bars next to the nodes indicate profit levels of firms proportional to the heights of these bars. The top layer represents the inter-bank network, $L$, where nodes are banks (square), but links are omitted as we focus on the transmission of real economy shocks to the banking system, but not shock propagation within the banking system. The interbank layer, $L$, is connected (dashed arrows) with the supply chain network, $W$, via bank-firm loans represented by matrix, $B$. $B_{ik}$ is equal to the outstanding principal that firm $i$ has towards bank $k$. Panels \textbf{(b)} and \textbf{(c)} schematically depict how real economy shocks can spread to the interbank network  without and with supply chain network, respectively. Black bars next to the two carbon intensive firms $e$ and $f$ indicate the amount of additional CO2 costs these firms need to pay for their emissions. Since profits of  firm $e$ are smaller than the new costs, it defaults and goes out of business. Consequently, its lender bank number $2$ writes off its loans and suffers losses as depicted in panel \textbf{(d)}. However, when including the supply chain network, production of supplier $a$ and buyers $b$ and $c$ of the failed firm $e$ are reduced and their profits drop. As a result of supply chain contagion they also default and banks need to write off their loans. Panel \textbf{(e)} shows supply chain contagion adjusted losses of banks caused by carbon pricing of firm $e$. Bank number $1$ suffers losses from $2$ carbon non-intensive clients $a$ and $b$. Bank number $2$ faces losses from $3$ of its clients $e$, $b$ and $c$, with the latter two being initially climate non-risky. This example demonstrates that climate risk of banks can arise from debtors directly and indirectly from their supply chains.
    }
	\label{fig0_toy}       

\end{figure*}

\subsection*{Economic network representation} 

The real economy and the financial system are represented as two-layer network of economic transactions, see Fig.\ref{fig0_toy}(a). The bottom layer, $W\in \mathbb{R}^{n,n}_{0,+}$, is the \textit{supply chain network (SCN)} with nodes (circles) representing $n\in\mathbb{N}$ individual firms. A directed link (solid arrow), $W_{ij}$, represents the value of goods and services firm $i$ sells to firm $j$. Every node (firm) has \textit{out-strength} $s_i^{\text{out}}\equiv \sum_{j=1}^n W_{ij}$ and \textit{in-strength} $s_i^{\text{in}}\equiv \sum_{j=1}^n W_{ji}$. $s_i^{\text{out}}$ and $s_i^{\text{in}}$ are the revenue and material costs, respectively, of firms' transactions within Hungary. The upper layer, $L$, represents the financial system consisting of $m\in\mathbb{N}$ banks (squares). The two layers are connected by the bank-firm loan matrix, $B\in\mathbb{R}^{n,m}$, where $B_{jk}$ represents a loan from bank $k$ to firm $j$, or a liability from firm $j$ to bank $k$ (dashed arrows).

The multi-layer network given by $W$, $L$ and $B$ is reconstructed from 2022 empirical data consisting of firms’ transactions, income statements, balance sheets and outstanding principles of banks’ commercial loan portfolios in the respective year (see Section Data for details). 
In this manner it represents an initial state of the economy in the year 2022, not affected by carbon pricing. To perform a climate stress test on it, we apply a carbon price shock on firms for which we estimate positive CO$_2$ emissions and use the newly developed financial stress testing model of \cite{tabachova2024estimating} to propagate this shock first to other firms in the SCN and then to banks. This allows us to calculate projected balance sheets of firms in a counterfactual scenario assuming carbon pricing and, eventually, estimate losses of firms and banks adjusted for supply chain contagion. We describe the model in 5 steps: 

\begin{enumerate}
    \item Designing the stress scenario (initial shock): Additional carbon costs of firms based on their CO$_2$ emission estimates and carbon prices.
    \item Identifying firms' shutdown points for carbon pricing and cost pass through: Comparing additional costs of carbon with firm profits. Shutdown occurs when the new cost structure renders profits structurally negative, making continued operations non-viable, leading to default.
    \item Calculating losses of banks from firms' direct defaults due to carbon pricing.
    \item Calculating indirect production losses of firms from supply chain contagion: Upstream and downstream shock propagation due to firms defaulting in Step 3, (drops of input supply and output demand to suppliers and customers). 
    \item Calculating indirect defaults of firms and SCN-adjusted losses of banks from indirect production losses causing illiquidity and insolvency of firms
\end{enumerate}

\subsection*{Step 1. Estimating firms' CO$_2$ emission to assess additional costs from carbon pricing}

Firm-level emissions in Hungary are known only for less than 119 firms \citep{stangl2024firm, ClimateActionEU} which report them as the EU ETS I participants (106 can be matched to the supply network data). Emissions of firms which will fall under the scope of the EU ETS II are unknown, and we need to estimate them. For this purpose we utilize the supply chain network, $W$,  to identify oil and gas purchases of firms. Fig.\ref{figSI_emissions_method}  illustrates the emission estimation procedure. First, based on the \textit{Statistical Classification of Economic Activities in the European Community on the four digit level (NACE)}, we use  NACE 4-digit industry affiliation of firms, to identify suppliers of gas, oil and petroleum products to other firms within the Hungarian economy. Gas selling sectors are \textit{D35.2.1 - Manufacture of gas}, \textit{D35.2.2 - Distribution of gaseous fuels through mains} and \textit{D35.2.3 - Trade of gas through mains}. 
Oil and petroleum distributors are picked from NACE 4-digit sectors \textit{C19.2.0 - Manufacture of refined petroleum products}, \textit{G46.7.1 - Wholesale of solid, liquid, and gaseous fuels and related products} and \textit{G47.3.0 - Retail sale of automotive fuel in specialized stores}. Overall, we identify 47 distributors of gas and 729 distributors of oil in our dataset (see Tab.\ref{tab:gas_oil_summary} for more details). We denote  firms belonging to these gas- and oil-supplying sectors with index sets $I_g$ and $I_o$, respectively. Total output of these sectors, --- excluding trade within or between those sectors to avoid double counting, --- are given as $s^{\text{out},gas}=\sum_{i\in I_g} s_i^\text{out}$ and $s^{\text{out},oil}=\sum_{i\in I_o} s_i^\text{out}$.

We then quantify the gas- and oil-related  in-strength each firm receives from gas- or oil-supplying firms as

\begin{equation}
    s^{\text{in},gas}_i=\sum_{j \in I_g} W_{ij} \quad,\quad s^{\text{in},oil}_i=\sum_{j \in I_o} W_{ij} \quad.
\end{equation}

Finally, to estimate emissions of these firms we distribute the total emissions  related to oil and gas consumption --- $\mathcal{E}^{gas} = 12.5$ Mt and $\mathcal{E}^{oil} = 13.7$ Mt ---  based on the relative in-strength each firm receives. Therefore emissions of firm, $i$, stemming from the consumption of gas and oil products can simply be estimated as
\begin{equation}
    \mathcal{E}_i= \frac{s^{\text{in},gas}_i}{s^{\text{out},gas}} \mathcal{E}^{gas} + \frac{s^{\text{in},oil}_i}{s^{\text{out},oil}} \mathcal{E}^{oil} \quad.
\end{equation}

Aggregated emissions  related to oil $\mathcal{E}^{oil}$ and gas $\mathcal{E}^{gas}$ are based on publicly available data on CO$_2$ emissions from fossil fuels for the entire Hungarian economy for the year 2022, obtained from the Global Carbon Project \citep{budget2023global}. Household consumption of gas and oil is excluded. External sources indicate that in Hungary approximately one-third of gas is used for residential heating; hence, this portion is subtracted from the total gas emissions before distributing the remainder among firms \citep{IEA_report}. The use of oil products in private cars is also estimated and deducted from the total oil-related emissions before distributing the remainder among firms. Oil has a negligible role in the residential heating sector in Hungary \citep{IEA_report}. $73.6\%$ of the total oil consumption is estimated to be used in the commercial sector, and the respective emissions are distributed among firms accordingly.

Note that we manually exclude financial firms—classified under the NACE 1-digit code K, as some firms labeled under this sector code are actually energy trading or insurance companies. Including these firms would distort the analysis, as their attributed emissions are not linked to production processes. Our analysis focuses on production-related emissions tied to firms' operations, as pricing these emissions leads to production cost increases, which may pose financial risks to banks. Based on this procedure, we arrive at a good estimate of firms' CO$_2$ emissions as the comparison of our estimates with the reported emissions of firms listed in the EU ETS I the correlation of log ETS I emissions and our log emission estimates is 0.61 (see Fig. \ref{figSI_scatterplot_emissions}). Additionally, firms in the NACE4-level sector \textit{G46.1.2 - Agents involved in the sale of fuels, ores, metals, and industrial chemicals} are excluded from the stress-testing analysis. These firms, whose aggregated estimated emissions amount to $1.2$ Mt, are likely not direct consumers of fossil fuels in their production processes but rather operate as distributors of fossil fuels or derived products. But due to limitations in the VAT data used for the emissions estimation, it is not possible to determine what proportion of their out-links represent sales of gas or oil products. As a result, we are unable to meaningfully allocate emissions to these firms or to further distribute their estimated emissions to the firms purchasing from them. Consequently, they are excluded from our analysis. In comparison to the overall emissions the resulting mis-estimations are likely to be relatively small. This assumption can lead to underestimation of emissions of some firms that purchase oil or gas from agents in sector G46.1.2. Further limitations of the emissions estimates arising from data availability are discussed  in Supplementary Information Section  \ref{SI_limitations_and_policy}.

Based on the firm-level emission estimates, we can assess exposure of individual firms and banks to  climate transition risk related to carbon pricing,  taxation, etc. .

\begin{table}[h]
\centering
\begin{tabular}{ r r r r }
\makecell{NACE 4-digit\\sector code} &\# firms  & \# out-links & $s^{out}$ [M EUR] \\
\hline
\textbf{D 35.2.1}& 5 & 11 & 2.4 \\
\textbf{D 35.2.2}& 11 & 1 933 & 34 \\
\textbf{D 35.2.3}& 31 & 102 459 & 12 255 \\
\hline
\textbf{C 19.2.0}& 5 & 105 817 & 5 247 \\
\textbf{G 46.7.1}& 159 & 24 889 & 1 664 \\
\textbf{G 47.3.0}& 565 & 261 019 & 5 247
\end{tabular}
\caption{\textbf{Summary of gas (D) and oil (C,G) distributing sectors.}}
\label{tab:gas_oil_summary}
\end{table}

\subsection*{Step 2. Carbon price shock with and without costs pass-through} \label{step_2}

Firms with positive CO$_2$ emissions, will fall under the scope of the EU ETS II policy and incur additional carbon costs, $\mathcal{E}_i\pi$, where $\pi$  is the price of carbon emissions in EUR per tonne. Typically, an increase in production costs results in price changes of the produced goods and services, leading to adjustments in the buyers demand for a good (price elasticities of demand). Hence, additional costs are at least partially  passed on to buyers \citep{duprez2018price}. The ability of firms to increase prices (pass costs on) depends on factors such as market concentration and demand elasticity (how demand reacts to price changes). As we lack product-level price and quantity information, here, we cannot model demand adjustments through price elasticities. Instead, we develop a heuristic cost pass through mechanism that assumes short-term inelastic demand for all companies and that bargaining power between supplier and buyer is proportional to the suppliers market share. This means that whenever a supplier can pass its carbon costs onto a buyer, the buyer will accept the price increase and still buys the same amount as previously. The fraction of the additional carbon costs a firm can pass on to its customers is based on its market share within its NACE 4-digit industry sector. 

First, we calculate the market shares of firms belonging to the 600 distinct NACE 4-digit industry sectors in our data.
We denote the NACE 4-digit industry affiliation of firms with the industry affiliation vector $\eta$, where $\eta_i = k $ means firm  $i$ belongs to NACE 4-digit sector $k$. The market share of firm $i$ within its industry, $k$, is calculated as $\mu_i = 
\frac{s_i^\text{out}}{\sum_{j=1}^{n} s_j^\text{out} \delta_{\eta_i,\eta_j}}$, where the Kronecker delta, $\delta_{\eta_i,\eta_j} $, equals one if $\eta_i  = \eta_j = k$ and zero otherwise. 
Then, a firm with additional costs from its emissions $\mathcal{E}_i\pi$ will keep only the fraction, $(1-\mu_i)\mathcal{E}_i\pi$, of the costs and passes the remaining fraction, $\mu_i\mathcal{E}_i\pi$, on to its buyers. In turn, the buyers will experience additional costs, which they can pass on downstream again based on their market shares and so on, until the initial carbon costs, $c(0)=\mathcal{E}\pi$, of emitters are distributed among the firms in the network. Note, that this set up doesn't allow for upstream bargaining power of firms.
For other network based approaches on attributing carbon emissions and hence costs along supply chains see \cite{van2023network}.

The cost pass-through heuristic is formalized by a simple algorithm.  $\bar{W}$ denotes the row normalized supply network, $W$, i.e. $\bar{W}_{ij} = W_{ij} / s_i^\text{out}$ denotes the value $i$ sells to $j$ as fraction of $i$'s overall sales, $s_i^\text{out}$. The initial carbon cost of firm $i$ are  $c_i(0) = \mathcal{E}_i \pi$ at iteration step $t=0$, according to the bargaining assumption $(1-\mu_i)c_i(0)$ is the fraction of costs $i$ must keep, while it can pass on $\mu_i c_i(0)$. At $t=0$, we set the retained cost vector, $\gamma_i(0)$, to $\gamma_i(0) = (1-\mu_i)c_i(0)$.
Then, for iteration $t+1$ the costs firm $i$ receives from other firms, $c_i(t+1)$, and the costs firm $i$, keeps, $\gamma_i(t+1)$, are updated iteratively as

\begin{eqnarray}
    && c_i(t+1) = \sum_{j=1}^{n} \mu_j \bar{W}_{ji} c_j(t) \quad,\\
    &&\gamma(t+1) = (1-\mu_i) c_i(t+1) + \gamma_i(t) \quad.
    \label{eq:cpt}
\end{eqnarray}
 
We iterate Eq. (3-4) until until 99.9999\% of the initial carbon costs are distributed among firms at iteration $T$ and the vector $\gamma(T)$ is the new carbon cost vector of firms after the initial carbon costs have been passed along the supply chain. In the main text all results referring to carbon costs with pass-through mechansims are based on $\gamma(T)$.
Note that the vector $c(t)$ converges to zero as $t$ grows as $\mu_i \leq 1$ and $\bar{W}_{ji} \leq 1$, thus, $\gamma(t)$ also converges, see SI Section \ref{SI_cost_pass_through}.
Above we also assume that a firm produces only one type of product given by its NACE 4 sector, and buyers don't change their supplier and the quantity demanded. Additionally, the demand for oil and gas is inelastic, meaning that firms' consumption of these resources remains unchanged despite rising emission prices.

\subsection*{Step 3. Direct defaults of firms and loan write offs by banks}\label{step_3}

The additional carbon costs, $\gamma(\pi)$, faced by emitters or their buyers (from the cost pass through mechanisms) are fully absorbed by their \textit{net profits}, $\mathcal{P} \in \mathbb{R}_{0,+}^n$\footnote{Note that here we assume non-negativity of net profits which should be true for any well functioning business. Nevertheless, some firms in the dataset have negative profits. It is impossible for us to assess financial viability of these firms, and, thus, we exclude them from this part of the simulation (for more details see Section \ref{sec_data} Data section). }.  This can happen only if  the 2022 profits are larger than the carbon price costs, $
    \mathcal{P}_i \geq \gamma(\pi)$,
where $\pi$ is the price in euro per tonne of CO$_2$. We assume, that firms with negative profits when a carbon price is in place have a non-sustainable business model and should exit the market. To identify all firms that default from the additional carbon costs at price $\pi$, we define the direct default vector, $\chi^{\text{dir}}$, with binary elements 

\begin{equation}
    \chi^{\text{dir}}_i\equiv 
    \begin{cases} 
  1 \qquad  \text{if} \; \mathcal{P}_i \leq \gamma(\pi)  \quad, \\ 
  0 \qquad \text{if} \; \mathcal{P}_i > \gamma(\pi)  \quad . \\ 
  \end{cases}
  \label{chi_dir}
\end{equation} 
The \textit{direct default indicator} keeps track of firms failing as they become unprofitable due to a carbon price $\pi$, $\chi^{\text{dir}}=\chi^{\text{dir}}(\pi)$. Note that for the calculations without cost pass trough we use the initial carbon costs $c(0) =  \pi \mathcal{E} $ instead of $\gamma(\pi)$.

The resulting economy wide direct production losses, $\Lambda^\textrm{dir}$, resulting from these direct defaults are quantified as the total annual sales of the failed firms expressed as a proportion of the entire transaction volume within the network, i.e.

\begin{equation}
    \Lambda^\textrm{dir}(\pi,cpt)\equiv\sum_{i=1}^n \frac{s_i^\textrm{out}}{\sum_{j=1}^n s_j^\textrm{out}} \chi^{\textrm{dir}}_i(\pi,cpt) \quad ,
    \label{eq_dir_prod_losses_main}
\end{equation}
where $s_i^{\text{out}}\equiv \sum_{j=1}^n W_{ij}$ is out-strength (all sales within the supply network) of firm $i$, and $\chi_i^{\textrm{dir}}\in \{0,1\}$ is the indicator of \textit{directly defaulted firms} at carbon price $\pi$, and $cpt\in\{0,1\}$ indicates presence or absence of carbon costs pass-through mechanism in simulations.

The resulting banking system loss occurs, as banks are required to allocate capital reserves to cover potential losses from loans written off due to client defaults. When a loan is deemed unrecoverable and written off, the bank must account for the loss by adjusting its loan loss provisions and ensuring compliance with regulatory capital requirements. Hence, equity is reduced by the loan write-off. We refer to these as \textit{direct loan losses}, $\mathcal{L}_k^\textrm{dir}(\pi)$, of the carbon price. The direct loan losses are calculated as the sum of the outstanding principals of directly defaulted clients divided by the bank's Tier 1 equity, i.e.,

\begin{equation}
    \mathcal{L}_k^\text{dir}\equiv \sum_{j=1}^n \chi^{\text{dir}}_j  \frac{\kappa_{jk}B_{jk}}{e_k} \quad ,
\end{equation}

for every bank $k \in \{1,2,...,m\}$. $\chi^{\text{dir}}_j \in \{0,1\} $ indicates whether firm, $j$, directly defaulted due to the carbon price, $B_{jk}$ is the outstanding principal from loans firm $j$ received from bank $k$, $e_k$ is the equity of bank $k$ and $\kappa_{jk}\in [0,1]$ is the \textit{loss given default} parameter. For simplicity we assume $\kappa_{jk}=1$ for all loans, implying that our climate stress test gives the upper bound of potential losses. Note, bank equity losses  scale with the LGD, making it intuitive to infer outcomes for lower values. Additionally, we do not account for possible measures such as installment postponements, firm restructuring in case of financial difficulties, the use of collateral by banks, or other forbearance measures.

To asses direct losses of the entire banking system, $\mathcal{L}^\text{dir}$, we need to re-weight the fraction of lost equity of individual banks by their relative equity sizes and sum them up, i.e.

\begin{equation}
    \mathcal{L}^{\text{dir}}=\sum_{k=1}^m \frac{e_k}{\sum_{\ell=1}^m e_\ell} \mathcal{L}_k^{\text{dir}} \quad.
\end{equation}

In Fig.\ref{fig0_toy}(b) and (d) we demonstrate this process on a toy-model network. Firms $e$ and $d$ are emitters with additional carbon costs (black bars), that need to be absorbed by their net profits (white bars). In the case of firm $e$ profits are higher than carbon costs, but in the case of firm $e$ carbon costs exceed profits and the firm becomes unprofitable and defaults. As a result, bank number 2 writes off loan $B_{e2}$, and suffers losses $\mathcal{L}_2^\text{dir}=\frac{B_{e2}}{e_2}$, where $e_2$ is the equity of the bank 2. Losses of both banks are schematically indicated in panel (d). Bank number 1 doesn't experience any direct losses from carbon pricing.

\subsection*{Step 4. Supply chain network contagion and production losses}\label{step_4}

Firms becoming unprofitable due to the carbon price, go out of business (when $\chi_i^{\text{dir}} = 1$) and consequently stop their production, which in turn affects their suppliers and buyers upstream and downstream in the production network. Here we employ the supply chain contagion model introduced by \cite{diem2022quantifying, diem2024estimating}. 
The model simulates short-term production losses caused by supply chain network contagion before the initially failed firms are replaced and missing inputs substituted. 
In the model, every firm in the SC network, $W$, uses a \textit{Generalized Leontief production function} (GL) to produce its output, $x_i$ given the supply network, $W$,
\begin{align} \label{GL}
       x_i(t+1)  = & \min \left[ \min_{k\in \mathcal{I}_{i}^{\textrm{es}}} \left[ \frac{1}{\alpha_{ik}} \sum_{j | \eta_j = k} W_{ji}(t) \right], \right. \notag \\
       & \left. \quad \qquad \beta_i + \frac{1}{\alpha_i} \sum_{k \in \mathcal{I}_{i}^{\textrm{ne}}} \sum_{j | \eta_j = k} W_{ji}(t) \right]
\end{align}
where the parameters $\alpha_{ik}$, $\alpha_i$ and $\beta_i$ are calibrated from the data, such that $x_i=s_i^\textrm{out}$ for every $i\in \{1,2,...,n\}$. Note that $j | \eta_j = k$ refers to all firms $j$ that belong to sector $k$, i.e., where $\eta_j = k$. $\mathcal{I}_i^\textrm{es}$ and $\mathcal{I}_i^\textrm{ne}$ are sets that contain all \textit{essential} and \textit{non-essential} inputs of firm $i$,  $\eta_j$ gives the industry, $k$,  of firm $j$. Essential inputs are treated in a Leontief, non-linear, manner (the first term in the outer brackets), meaning that the absence of one essential input completely halts production. Non-essential inputs are treated in a linear manner (the second term in the outer brackets), meaning that their absence will reduce the final output only by the relative cost share of that input out of all others.
Before a stress test given by the initial carbon cost, the SC network is in a steady state in which every firm produces $100\%$ of its capacity. This \textit{production level of a firm} at time step $t_0$ is denoted by $h_i(t_0)$ and defined as $h_i(t_0)\equiv\frac{x_i(t_0)}{s_i^\textrm{out}}=1$. When the initial shock is applied at time step $t_1$ and then propagated through the network, production levels of firms can drop and $h_i$ changes. Thus, $h_i(t)\in [0,1]^n$ is fraction of the original production level  of firm $i$ at time step $t$.

At time $t_1$, carbon costs, $\gamma(\pi)$, (or $c(0)$ for no cost pass through) are imposed on firms, and, as described before, firms $i$ where $\chi_i^\text{dir}=1$ become unprofitable, and go out of businesses. A failure of a firm, $i$, is leads to the complete production termination, given by $h_i(t_1)=0$. The updated vector $h(t_1) \in \{ 0,1\}^n$ is used as the initial shock to the production network, $W(t_1)=W(t_0)\circ h(t_1)$ and $x_i(t_1)=f(W(t_1))$ given by Eq.(\ref{GL}). The shock propagates along links of the SC network and affects production levels of firms based on availability of essential or non-essential inputs. The propagation is iterated until every firm stops to adjust its production levels, i.e. $h_i(t_\tau) - h_i(t_{\tau +1}) < \epsilon$, and the system converges to a new stationary state at time $T := t_{\tau+1}$. This state is represented by the final remaining production levels, $h(T) \in \left[0,1\right]^n$. Losses incurred by the entire production network in a scenario with carbon price $\pi$ are equal to

\begin{equation}
    \Lambda(\pi)=\sum_{i=1}^n \frac{s_i^\textrm{out}}{\sum_{j=1}^n s_j^\textrm{out}} \left(1-h_i(T,\pi)\right)\quad.
    \label{eq_prod_losses}
\end{equation}
This can be rewritten as the sum of direct production losses before the contagion
\begin{equation}
    \Lambda^\textrm{dir}(\pi)=\sum_{i=1}^n \frac{s_i^\textrm{out}}{\sum_{j=1}^n s_j^\textrm{out}} \left(1-h_i(t_1,\pi)\right)
    \label{eq_dir_prod_losses}
\end{equation}
and indirect production losses from the contagion, $\Lambda^\textrm{indir}=\Lambda-\Lambda^\textrm{dir}$.

As outlined in the main text we use an optimistic and a pessimistic scenario for how shocks propagate in the production network based on the substitutability of essential inputs. We utilize the GL production function, as given by eq.(\ref{GL}) for the pessimistic scenario, and a Linear production function, by assuming all inputs for all firms are in $\mathcal{I}^\text{ne}$, i.e., the second term in eq.(\ref{GL}). The final contagion losses suffered by the production network with GL and Linear production functions will be denoted by $\Lambda^\textrm{GL}$ and $\Lambda^\textrm{L}$ respectively, or alternatively as production losses with \textit{pessimistic} and \textit{optimistic} substitution. More details on the differences in contagion effects between the two functions can be found in \cite{diem2022quantifying}.  \cite{diem2022quantifying} use the model to calculate the \textit{Economic Systemic Risk Index (ESRI)} of each firm, that ranks firms based on the size of the shock propagation cascade they cause upon their hypothetical failure.  Firms with the highest rankings are considered \textit{systemically risky} or \textit{systemically important}. Significant contagion triggered by the failure of single firms occurs in simulations involving the GL production function, where the absence of essential inputs can create bottlenecks in production processes. To distinguish essential and non-essential inputs of NACE industries, we follow \cite{diem2024estimating} and utilize the results of   a survey on how critical inputs are for different industry sectors  \citep{pichler2020production}.

\subsection*{Step 5. Translating supply chain contagion into financial losses and loan write offs by banks}\label{step_5}

We translate the production levels, $h(T)$, at $T$ (when supply chain contagion subsided) to financial losses of firms by updating their 2022 income statements and balance sheets such that they reflect the losses from  supply chain contagion and costs of carbon prices. For this purpose we utilize the stress testing model  developed in \cite{tabachova2024estimating}. The decreased production levels, $h(T)$, reduce firms' revenues, $r\in \mathbb{R}^n_{0,+}$, and material costs, $c\in \mathbb{R}^n_{0,+}$. This in turn affects firms' \textit{operating profits}, $p \in \mathbb{R}_{0,+}^n$, which changes the equity and liquidity positions in firms' balance sheets, and eventually their ability to repay loans to banks \footnote {Note that here we assume non-negativity of operating profits, revenues and material costs which should be true for any well functioning business. Nevertheless, for some firms in the dataset some or all of these values are negative. It is impossible for us to assess financial viability of these firms, and, thus, we exclude them from this part of the simulation mechanism (for more details see the Data section)}. We assume that a firm $i$'s revenue, $r_i$ and material costs, $c_i$, decrease proportionally with the fraction of lost production, $1-h_i(T)$, while other income statement variables remain unchanged. Thus, the \textit{profit reduction}, $\Delta p_i$, of a firm $i$ is calculated as the difference between the operating profit, $p_i$, without shock and profit, $\tilde{p}_i$, in the counterfactual scenario where the contagion occurs, and we can write
\begin{align}
        \Delta p_i &=p_i - \tilde{p}_i\\
               &= (r_i - c_i + o_i) - (h_i(T)(r_i-c_i)+o_i)\\
               &=(1-h_i(T))(r_i-c_i) \quad,
               \label{eq:profit_loss}
\end{align}
where $o_i$ represents the sum of income statement variables other than operating profits and are not affected by supply chain contagion. The income statement variables, $r_i, c_i$, represent the business year 2022, as we rescale them by $h_i(T)$, we implicitly assume that the shock on sales and costs lasts for one year. This assumption can be easily changed by rescaling the losses.   

Next, the production losses lead to a change of \textit{equity}, $z\in \mathbb{R}^n_{0,+}$, and \textit{liquidity} represented by the \textit{short term assets} $a\in \mathbb{R}^n_{0,+}$, in firms' updated balance sheets. We update a firms' balance sheet by adjusting equity ($z\rightarrow \tilde{z}$) and liquidity ($a\rightarrow \tilde{a}$) based on the  changes of profits $\Delta p_i$.
The updated equity, $\tilde{z}_i$, of firm, $i$, is calculated from the equity, $z_i$, from the beginning of the year before the shock, plus \textit{retained earnings}, $\zeta_i$, and reduced by the profit loss, $\Delta p_i$, from the production contagion, and less the carbon costs $\gamma_i(\pi)$, i.e., 
\begin{equation}
    \tilde{z}_i=z_i+\zeta_i-\Delta p_i-\gamma_i \quad.
    \label{eq:proj_eq}
\end{equation}
We proxy the retained earnings with the retained earnings balance sheet item of the year 2022.
We do not model changes (positive or negative) of other balance sheet variables. Derived from the cash-flow statement, the liquid asset position, $a_i$, is reduced by the profit reductions and the additional carbon costs, i.e.,
\begin{equation}
    \tilde{a}_i=a_i-\Delta p_i-\gamma_i \quad.
    \label{eq:proj_liq}
\end{equation}
The updated equity and liquidity positions determine the default of firms. If $\tilde{z}_i$ or $\tilde{a}_i$ turn negative, firm, $i$, will become insolvent or illiquid, respectively, see Insolvency Code  \cite{Act} (Section 27) and  \cite{tabachova2024estimating}. Insolvency or illiquidity of a company indicate unlikeliness to repay loans and hence, firms need to be considered defaulted by banks and their loans written off. Thus, in analogy with direct default indicator we define \textit{indirect default indicator}
\begin{equation}
  \chi_i^\text{indir}  = \begin{cases} 
  1 \qquad  \text{if} \quad (\tilde{z}_i\leq 0  \quad \text{or} \;\;  \quad  \tilde{a}_i \leq 0) \quad \textrm{and} \quad \chi^{\textrm{dir}}_i=0 ,\\ 
  0 \qquad \text{else} 
  \quad . \end{cases}
  \label{chi_indir}
\end{equation}


In analogy with the direct loan write offs (or losses), $\mathcal{L}^{\text{dir}}_k$, of banks we define the \textit{indirect losses} of bank $k$ as
\begin{equation}
    \mathcal{L}_k^\text{indir}\equiv \sum_{j=1}^n \chi^{\text{indir}}_j  \frac{\kappa_{jk}B_{jk}}{e_k} \quad.
    \label{eq:fin_loss_bank_dir}
\end{equation}
These losses arise entirely from the SCN contagion and can be underestimated if network data is not available in stress testing procedures. Hence, the \textit{SCN contagion-adjusted financial losses of banks} are 
\begin{equation}
    \mathcal{L}_k = \mathcal{L}_k^{\text{dir}}+\mathcal{L}_k^{\text{indir}}\quad.
    \label{eq_bank_losses}
\end{equation}
To asses the SCN contagion-adjusted financial losses of the entire banking system, $\mathcal{L}$, we need to re-weight the losses of individual banks by their relative equity sizes and sum them up

\begin{equation}
    \mathcal{L}=\sum_{k=1}^m \frac{e_k}{\sum_{\ell=1}^m e_\ell} \mathcal{L}_k\quad .
    \label{eq_total_fin_losses}
\end{equation}

The example in Fig.\ref{fig0_toy}(c) shows  how the  failure of one firm triggers supply chain contagion that causes indirect defaults of firms and equity losses of banks. As in  panel (b), firm $e$ shuts down due to the high carbon costs (black bar) that can not be covered by its profit (white bar) and defaults. Thus, $h(t_0)=(1,1,1,1,1)$, $\chi^\text{dir}=(0,0,0,0,1)$ and $h(t_1)=(1,1,1,1,0)$. Its failure causes supply chain contagion downstream to its buyers $b$ and $c$, and upstream to its supplier $a$, yielding a new production vector $h(T)$. As result of the production losses from contagion, firms $a,b,c$ default indirectly, thus $\chi^\text{indir}=(1,1,1,0,0)$. Banks exposed to the directly and indirectly defaulted firms write off loans from their balance sheets.  Banks 1 and 2 have equity of $e_1=e_2=1$, and  loans of $B_{a1}=B_{b1}=B_{b2}=B_{c2}=B_{d2}=B_{e2}=0.1$ (dashed links). Using Eq. (\ref{eq:fin_loss_bank_dir},\ref{eq_bank_losses}) and assuming that the loss given default is $100\%$ ($\kappa=1$) for all the loans, we obtain that $\mathcal{L}_1^\text{dir}=0$, $\mathcal{L}_2^\text{dir}=B_{e2}=0.1$, $\mathcal{L}_1^\text{indir}=B_{a1}+B_{b1}=0.2$, $\mathcal{L}_2^\text{indir}=B_{b2}+B_{c2}=0.2$. Panel (e) shows the total losses,  $\mathcal{L}_k=\mathcal{L}_k^\text{dir}+\mathcal{L}_k^\text{indir}$,  for the two banks, of$\mathcal{L}_1=0.1$ and $\mathcal{L}_2=0.3$, respectively. The total banking system losses are $\mathcal{L}=\frac{1}{2}0.1+\frac{1}{2}0.3=0.2$. Hence,  after the supply chain contagion caused by the initial failure of firm $e$ (due to carbon pricing), banks number $1$ and $2$ write off $10\%$ and $30\%$ of their equities respectively, and the total losses of the financial system are equal to $20\%$ of the system's equity. Additionally, the example demonstrates how bank number $1$ can underestimate its climate risk exposure. Without the supply chain links in panel (b) it has no exposure to climate-risky debtors, but when their debtors' production depends on suppliers or buyers with emissions, the climate risk (here from carbon pricing) can be inherited.

\clearpage

\newpage


\begin{widetext}


\renewcommand{\thesection}{S\arabic{section}}  
\setcounter{section}{0}

\renewcommand{\thefigure}{S\arabic{figure}}
\setcounter{figure}{0}

\renewcommand{\thetable}{S\arabic{table}}
\setcounter{table}{0}

\renewcommand{\theequation}{S.\arabic{equation}}
\setcounter{equation}{0}

\section*{Supplementary Information}

\vspace{2cm}
\tableofcontents

\newpage

\FloatBarrier

\clearpage
\newpage

\section{Additional information on emission estimates for NACE 1 sectors}\label{appendixB_add_info_onestimation}

The newly estimated firm-level emission that account for 19 Mt of country's emissions which is $19\%$ higher than the reported $16$ Mt of emissions from ETS I firms. Fig.\ref{fig1_emissions}(a) shows, that the new emissions are measured in sectors A, B, E, F, G that are not carbon intensive based on reported emissions, but now account for $3.9$ Mt. Additional $9$ Mt of emissions is estimated in carbon intensive sectors C, D and H (yellow bars). The service sectors I-U account for $1.4$ Mt and the remaining $3.9$ Mt is attributed to firms with unknown sector affiliation. Note that our method (Methods Step 1) excludes firms from the sector K, Financial $\&$ Insurance Activities, hence it has zero emissions. Estimated emissions of each sector individually are in Tab.\ref{tab:emissions_nace1} and of single firms in Fig.\ref{figSI_boxplots_emissions}. 
These firms are distributed across all sectors, which shows that the scope of the ETS II will be much wider than the one of the ETS I. Especially, the most affected sectors  will be G-Wholesale $\&$ Retail Trade, and H-Transportation. Average emissions per firm in a sector are the biggest in sector D with 2,497 tonnes, followed by 1,078 tonnes per firm in sector B and 482 tonnes per firms in sector E (see Tab.\ref{tab:emissions_nace1}). Sectors C and H with the highest estimated total emissions have average emissions of 221 and 411 tonnes per firm respectively. 

Additionally, we show comparison of estimated and reported emissions of ETS I companies in Fig.\ref{figSI_scatterplot_emissions}, see caption for more details. 
Since our estimates are based solely on domestic oil and gas consumption, they do not account for emissions related to electricity use, certain chemical reactions, waste handling, or oil and gas purchases from abroad, yet for the majority of firms in the supply network this mis-estimation should be small as very few firms have import relations \citep{dhyne2021trade}. For two firms our method yields zero emissions. It means that these firms do not buy oil or gas in Hungary. In 46 cases, reported emissions are overestimated (dots below the diagonal). We assume that all out-links of oil and gas distributors represent sales of oil and gas, however, these transactions may also involve sales of other products unrelated to fossil fuel combustion. This limitation of our method is due to the unavailability of product-level information.

\begin{figure}[hb]
	\centering
	\includegraphics[scale=.1, keepaspectratio]{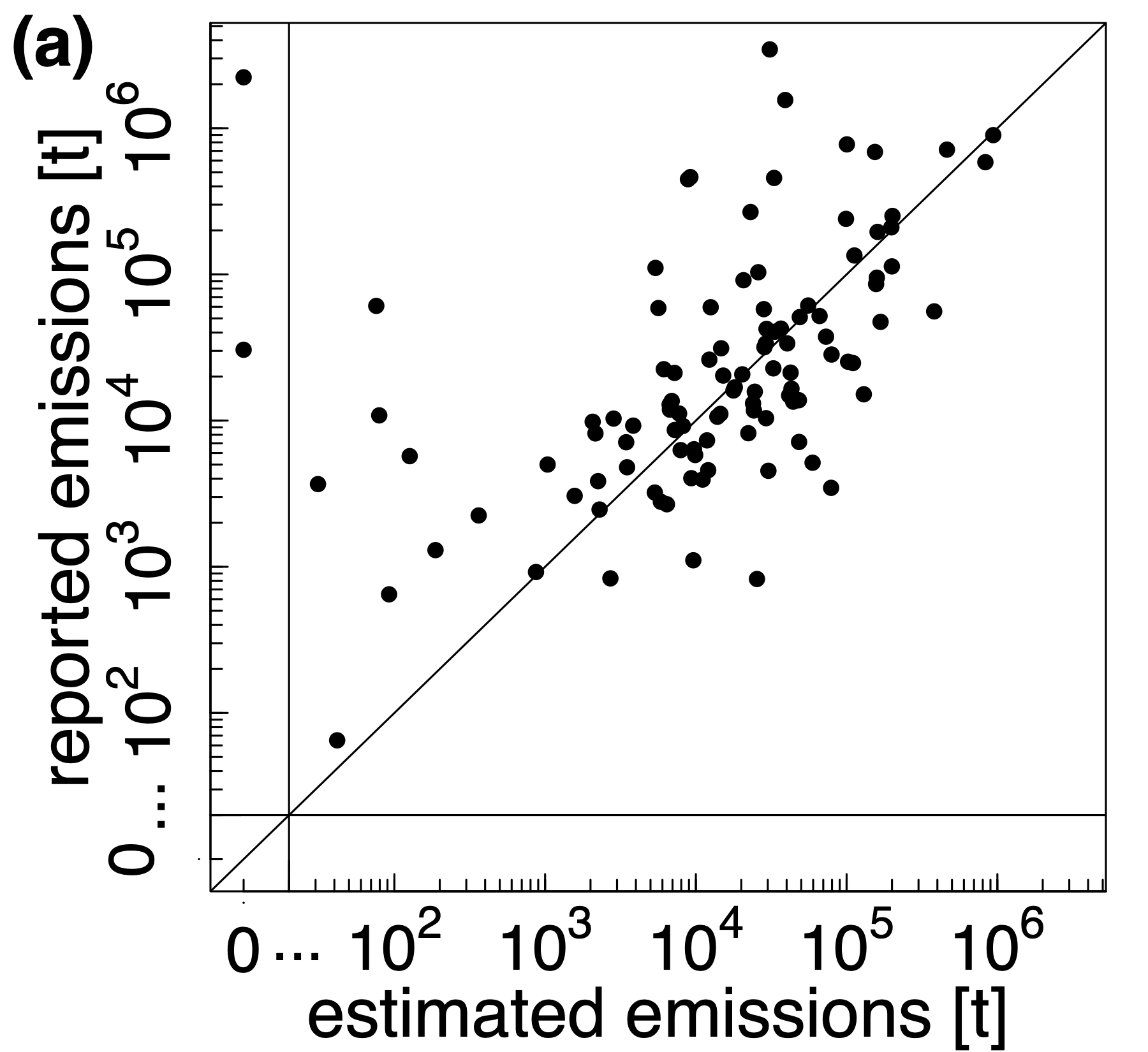}
	\includegraphics[scale=.1, keepaspectratio]{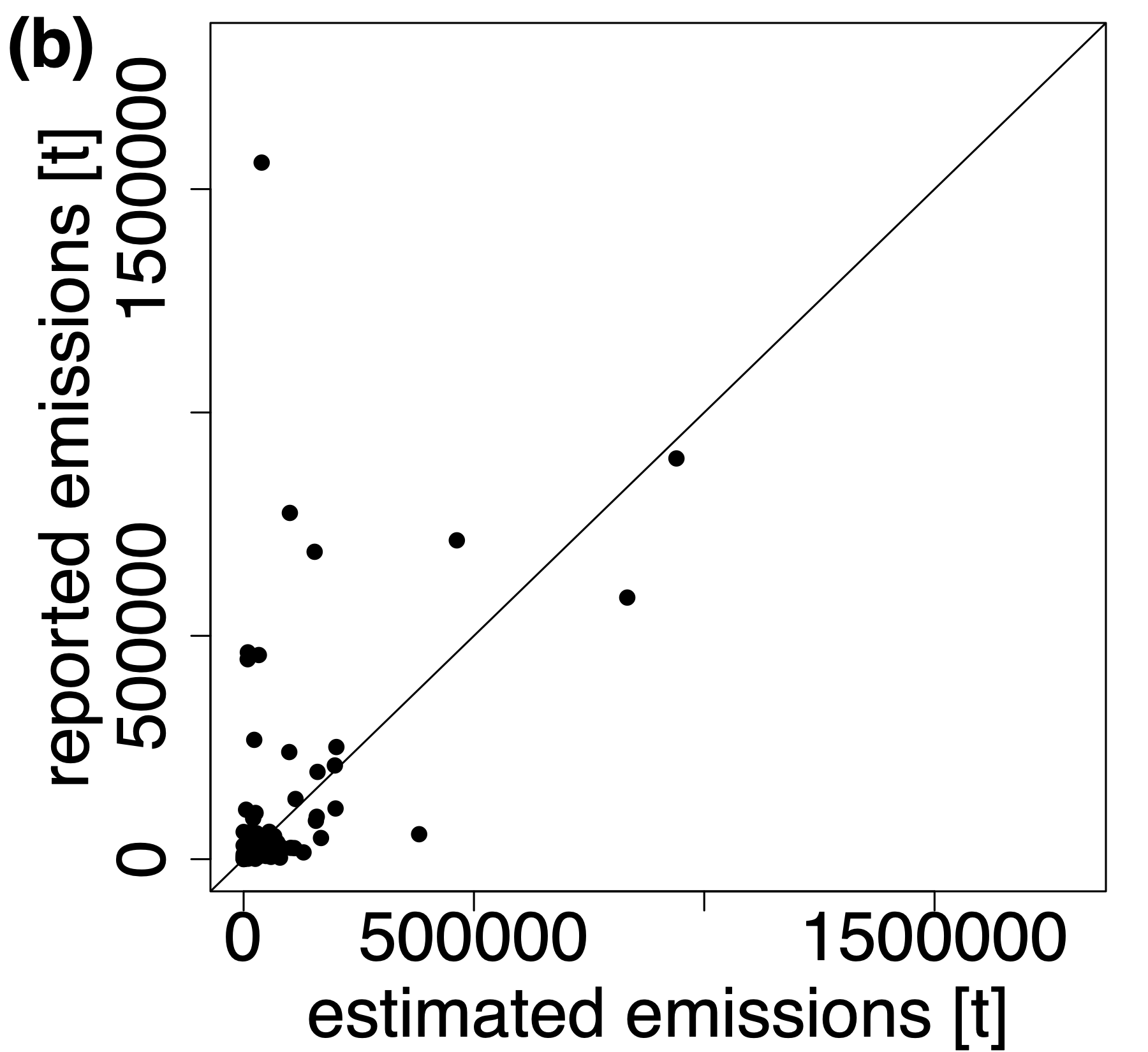}
  	\caption{\textbf{Comparison of estimated emissions with reported emissions of 106 EU ETS I firms.} The figure shows estimated emissions on $x$-axes against reported CO$_2$ emissions on $y$-axes in tonnes of 106 firms that are listed in EU ETS I. Panels \textbf{(a)} and \textbf{(b)} show log-log and linear scales respectively. Magnitude of estimated emissions matches for most firms in the network. For 60 firms our method underestimates reported emissions (dots above the diagonal). 
    The correlation coefficient for the log ETS I emissions and log estimated emissions is 0.61, for linear scales the correlation is lower, 0.22.
     Overall, reported emissions of 106 firms are equal to $16$Mt and their estimated emissions from oil and gas purchases are equal to $6$Mt.
    } 
	\label{figSI_scatterplot_emissions}       

\end{figure}

\begin{table}[!ht]
    \centering
    \begin{tabular}{|l||l|l|l|l|l|}
    \hline
        NACE1 & \# non-emitters & \# emitters & emissions [t] & emissions per firm [t/firm] \\ \hline \hline
        A & 3 130 & 6 768 & 915 670
        & 135
        \\ \hline
        B & 170 & 196 & 211 225 & 1 078 \\ \hline
        C & 10 009 & 17 814 & 3 931 904 & 221\\ \hline
        D & 1 225 & 705 & 1 760 219 & 2 497 \\ \hline
        E & 408 & 867 & 418 148 & 482 \\ \hline
        F & 15 548 & 23 848 & 924 086 & 39 \\ \hline
        G & 35 702 & 35 518 & 1 498 434 & 42 \\ \hline
        H & 3 082 & 8 201 & 3 372 103 & 411 \\ \hline
        I & 6 702 & 8 416 & 236 681 & 28 \\ \hline
        J & 14 592 & 3 005 & 161 914 & 54 \\ \hline
        K & 3 299 & - & - & - \\ \hline
        L & 13 138 & 10 084 & 452 781 & 45 \\ \hline
        M & 32 607 & 10 727 & 145 707 & 14 \\ \hline
        N & 9 232 & 5 453 & 298 580 & 55 \\ \hline
        O & 98 & 82 & 3 022 & 37\\ \hline
        P & 2 176 & 605 & 20 629 & 34 \\ \hline
        Q & 2 228 & 543 & 9 383 & 17 \\ \hline
        R & 3 829 & 1 765 & 60 085 & 34 \\ \hline
        S & 2 795 & 1 871 & 64 724 & 35 \\ \hline
        T & 1 & 0 & 0 & - \\ \hline
        U & 0 & 1 & 52 & 52 \\ \hline
        Z & 64 767 & 49 316 & 3 983 990 & 81 \\ \hline
    \end{tabular}
    \caption{\textbf{Summary of estimated emissions across NACE 1 industry sectors.} 
    \textbf{Legend:} \textbf{A} Agriculture, Forestry $\&$ Fishing,
  \textbf{B} Mining $\&$ Quarrying,
  \textbf{C} Manufacturing,
  \textbf{D} Electricity, Gas, Steam, Air Conditioning,
  \textbf{E} Water Supply, Sewerage, Waste,
  \textbf{F} Construction,
  \textbf{G} Wholesale $\&$ Retail Trade,
  \textbf{H} Transportation $\&$ Storage,
  \textbf{I} Accommodation $\&$ Food Service,
  \textbf{J} Information $\&$ Communication.,
  \textbf{K} Financial $\&$ Insurance Activities,
  \textbf{L} Real Estate Activities,
  \textbf{M-U} Other Service Activities,
  \textbf{Z} Undefined}
    \label{tab:emissions_nace1}
\end{table}

\begin{figure}[t]
	\centering
	\includegraphics[scale=.1, keepaspectratio]{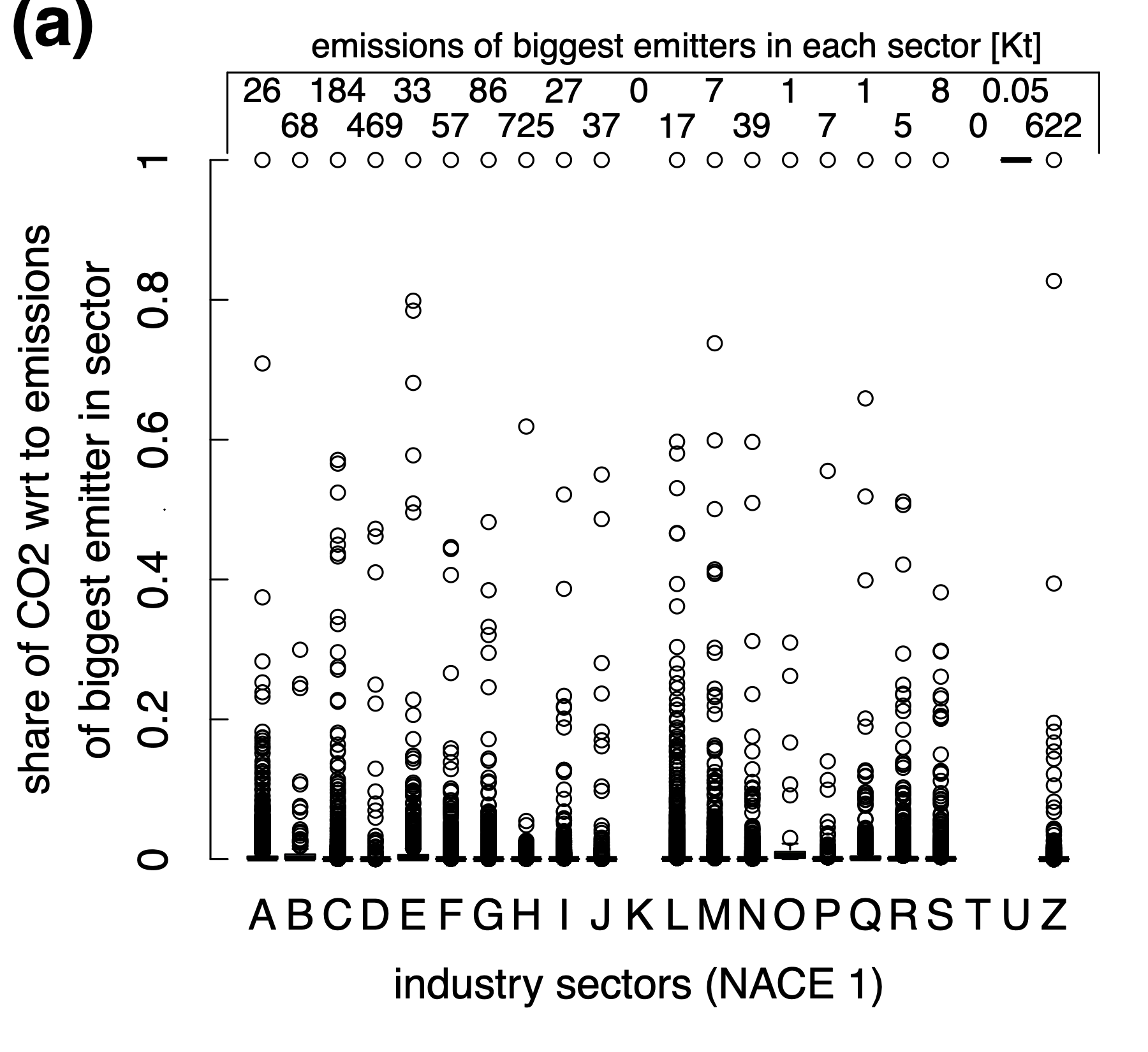}
	\includegraphics[scale=.1, keepaspectratio]{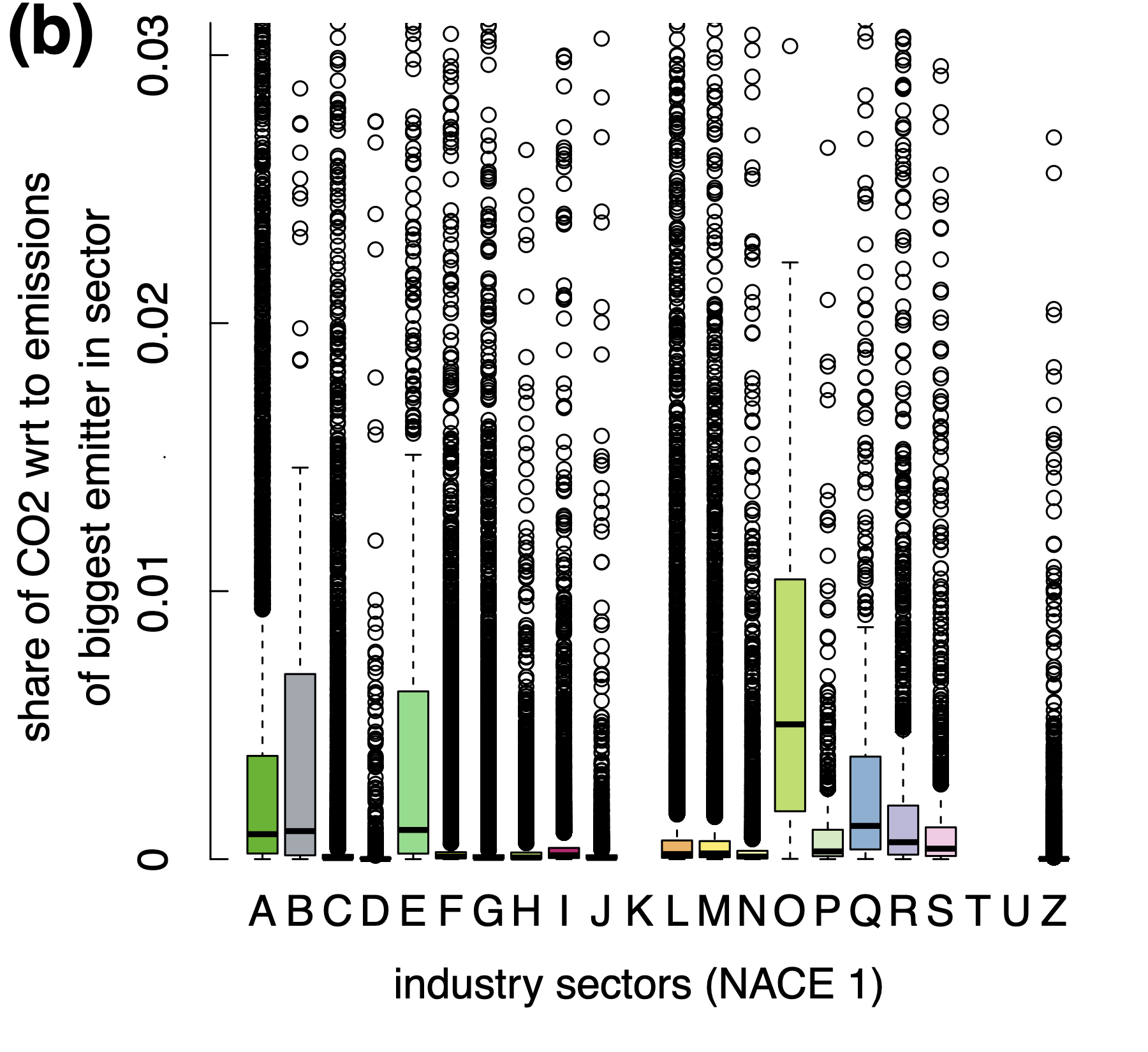}
  	\caption{\textbf{Distributions of firm-level estimated emissions across NACE1 sectors.} Panel \textbf{(a)} and \textbf{(b)} has NACE1 industry sectors on the $x$-axis with respective boxplots of firm-level emissions in each of the sectors presented in both panels. $y$-axis on the panel \textbf{(a)} ranges from 0 to 1, while the panel \textbf{(b)} shows the same plot zoomed for bottom $3\%$. Emissions in each of the sectors are weighted with respect to the biggest emitter in that sector. Thus, dots in panel (a) aligned at 1 represent biggest emitters in every sector with their emissions in Kt presented at the top of the plot. For example, the most polluting firm in sector A emits estimated $26$Kt of CO2 and the second biggest in this sector emits $0.7*26=18.2$Kt. Sector K has no emissions because it is excluded from the estimation method. Sector T has only 1 firm 0 emissions, and there is only 1 emitter in sector U with 50 tonnes of CO$_2$. We see that emissions from firms within sectors are heterogeneous, with several large emitters in each sector, while the majority emit less than $2\%$ of the emissions of the largest emitter in their respective sector. \textbf{Legend:} \textbf{A} Agriculture, Forestry $\&$ Fishing,
  \textbf{B} Mining $\&$ Quarrying,
  \textbf{C} Manufacturing,
  \textbf{D} Electricity, Gas, Steam, Air Conditioning,
  \textbf{E} Water Supply, Sewerage, Waste,
  \textbf{F} Construction,
  \textbf{G} Wholesale $\&$ Retail Trade,
  \textbf{H} Transportation $\&$ Storage,
  \textbf{I} Accommodation $\&$ Food Service,
  \textbf{J} Information $\&$ Communication.,
  \textbf{K} Financial $\&$ Insurance Activities,
  \textbf{L} Real Estate Activities,
  \textbf{M-U} Other Service Activities,
  \textbf{Z} Undefined  } 
	\label{figSI_boxplots_emissions}       

\end{figure}

\newpage
\clearpage

\section{Details on direct and supply chain contagion adjusted output and banking system losses}\label{SI_loss_comparison_table}

Appendix \ref{SI_loss_comparison_table} summarizes the detailed economic losses incurred by firms and banks for important carbon price scenarios described in the main text, namely 10 EUR/t, 45 EUR/t (EU ETS II cap), 100 EUR/t, 200 EUR/t (upper bound of CO\textsubscript{2} prices consistent with 2 degree warming in 2030. 
Table \ref{SI_tab_loss_comparison_per_co2price} summarize the results presented in the main text and gives the amplification sector for supply chain contagion in the optimistic and pessimistic scenarios, both, for firm output losses and bank equity losses and across different CO\textsubscript{2} price scenarios.

\begin{table}[H]
\centering
\begin{tabular}{lccccccccc}
\textbf{Scenario} & \textbf{10 EUR/t} & \textbf{45 EUR/t} & \textbf{100 EUR/t} & \textbf{200 EUR/t} \\
\toprule
\addlinespace
\textbf{Firm Direct Output Losses} (incl. cost pass-through) & 0.20 \% & 1.31 \% & 2.29 \% & 3.77 \% \\
\addlinespace
\textbf{Firm Output Losses Optimistic SC Contagion} &  1.66 \% &  5.27 \% &  8.26 \% &  12.29 \% \\
\addlinespace
\textbf{Firm Output Losses Pessimistic SC Contagion} & 2.75 \% &  53.43 \% &  53.94 \% &  54.86 \% \\
\hline
\addlinespace
\textbf{Amplification of Output Losses Optimistic} &  8.30 &  4.02 & 3.61 &  3.26 \\
\addlinespace
\textbf{Amplification of Output Losses Pessimistic} &  13.75 &  40.79 &  23.55 &  14.55 \\
\addlinespace
\midrule
\addlinespace
\textbf{Bank Equity Direct Losses} (incl. cost pass-through) &  0.27 \% &  1.21 \% & 2.21 \% &  4.72 \% \\
\addlinespace
\textbf{Bank Equity Losses Optimistic SC Contagion} &  1.08 \% & 2.68 \% &  5.46 \% &  9.09 \% \\
\addlinespace
\textbf{Bank Equity Losses Pessimistic SC Contagion} &  1.58 \% &  42.59 \% &  44.57 \% &  46.80 \% \\
\hline
\addlinespace
\textbf{Amplification of Financial Losses Optimistic} &  4.00 &  2.21 &  2.47 &  1.93 \\
\addlinespace
\textbf{Amplification of Financial Losses Pessimistic} &  5.85 &  35.20 &  20.17 &  9.91 \\
\end{tabular}
\caption{Firms' Output and Bank Equity Losses for different carbon prices for direct losses and losses adjusted for supply chain contagion for the optimistic and pessimistic scenarios. Results correspond to Fig. \ref{fig4_fin_prod_loss}, with costs pass-through.}\label{SI_tab_loss_comparison_per_co2price}
\end{table}

\newpage
\clearpage

\section{Bank exposures to firms affected by carbon pricing and CPRS firms}\label{appendixD_bank_exposures_and_cprs}

\begin{figure}[ht]
	\centering
	\includegraphics[scale=.13, keepaspectratio]{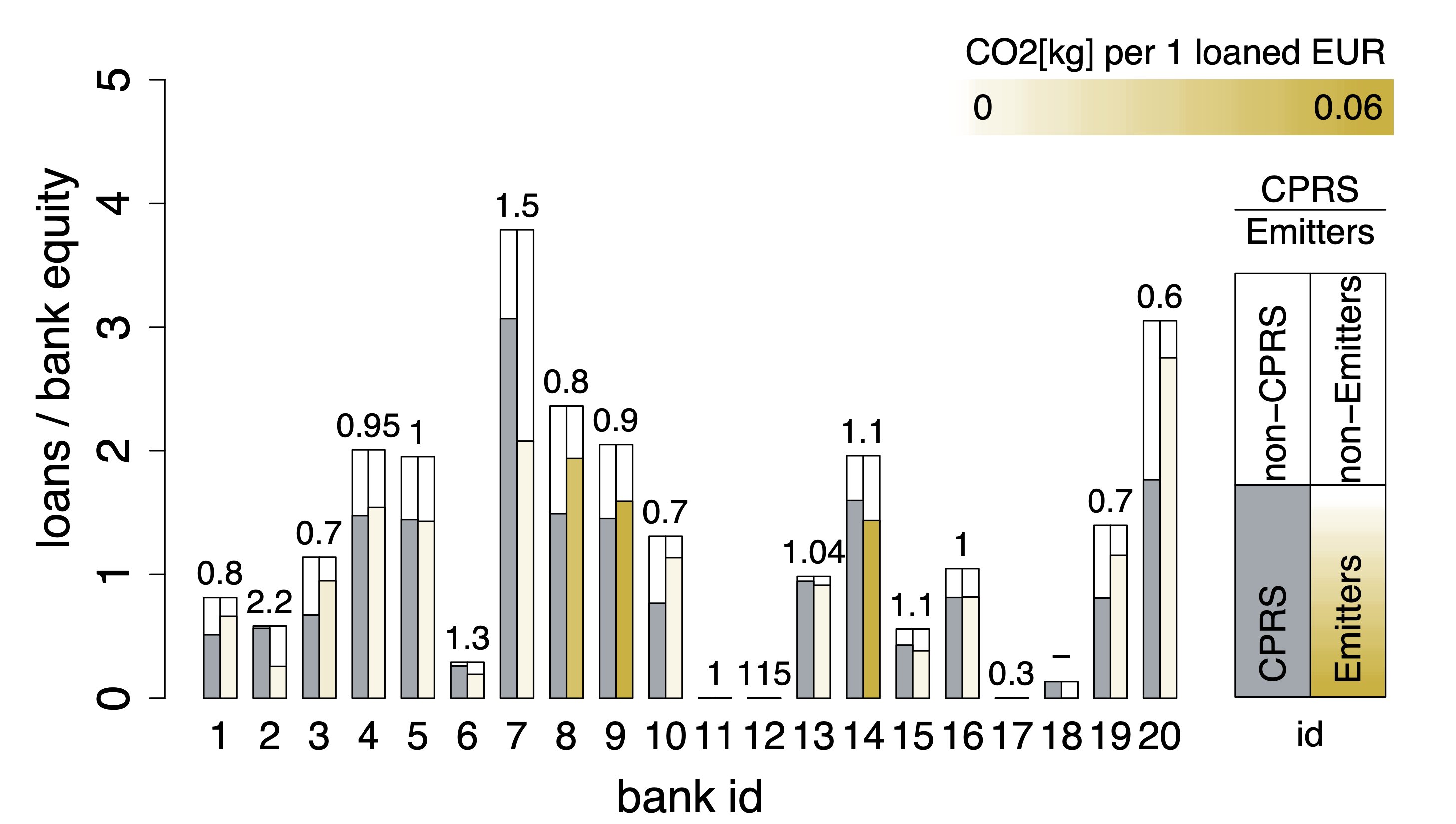}

  	\caption{\textbf{Exposure of banks to carbon risky assets based on estimated emissions and state of the art climate transition risk taxonomy CPRS.} Every couple of bars belongs to a bank with id number on the $x$-axis. The height of the couple is identical, and it indicates exposure of a bank to firms divided by its equity (CET1),  $\frac{1}{e_k}\sum_i B_{ik}$. The CPRS and emissions based methods are used to identify climate risky loans of every bank. The left hand side bars show results of CPRS based approach, where a height of the grey bar refers to volume of loans to firms from climate policy relevant NACE 4 sectors. The remaining share of loans (white space above the grey bar) is to firms from sectors that are classified as non-CPRS. Bars on the right-hand side show results obtained from the emissions-based approach. Yellow bar shows banks' loans to clients with newly estimated emissions (loans of firms with reported emissions are not included). Additionally, shade of the yellow color indicates CO$_2$[kg] per 1 loaned EUR of a bank ranging from $0$ to $0.06$ kg/EUR (from light to dark shades). Far beyond this range are banks number 9 and 14 with 10 kg/EUR and 24 kg/EUR respectively, which makes them the most carbon intensive. They are followed by banks number $8$ with $0.06$ kg/EUR and $3, 10, 1, 15, 4, 7, 5, 13, 20$ (in decreasing order) with carbon intensities between $0.009$ and $0.001$ kg/EUR. Carbon intensities of all the remaining banks are less than $0.001$ kg/EUR, and bank number $18$ has no emitting clients. 
    A number above a bar shows the degree to which the CPRS approach misestimates climate risk, calculated as the ratio of loans to CPRS clients versus loans to emitting clients.
   For banks number 2, 6, 7, 12, 13, 14, 15 this number is bigger than 1 (grey bars are higher than respective yellow bars), it means that CPRS taxonomy overestimates climate risks stemming from emitting clients. Grey bars of banks number 1, 3, 4, 8, 9, 10, 17, 19 and 20 are smaller than respective yellow bars, which means that sector based approach underestimates climate risks. Both approaches give same risk assessment for banks number 5, 11 and 16. 
   Here we see that the emissions-based risk assessment method not only allows us to identify climate-risky loans, but, unlike the CPRS approach, it can also assess the carbon intensity of those loans and, consequently, of banks.
} 
	\label{fig2_two_risk_approaches}       

\end{figure}

Here we assess how financially exposed the banking system is to the introduction of carbon pricing from lending to firms. We analyse the largest $m=20$ banks reporting CET1 capital data (totaling EUR 13 bn) and all their commercial loans to $56,595$ Hungarian firms (totaling EUR 24 bn), for details on the data see Section Data. 

We calculate for each bank the amount of loans outstanding to firms affected by carbon pricing (ETS II) and the carbon intensity of its loan portfolio defined as CO$_2$ emissions in kg per euro of lending. To compare our novel exposure measures we calculate  the amount of loans banks lent to firms in climate policy relevant sectors (CPRS) \citep{battiston2022nace}. 
Figure \ref{fig2_two_risk_approaches} shows every bank ($x$-axis) as a composite-bar where height represents the bank's commercial loan portfolio value as a fraction of CET1 capital ($y$-axis). The fractions of loans to bank equity strongly varies across banks, indicating different business models across banks. 
The composite-bar is dived into two sub-bars. The height of the grey sub-bar (left) denotes the value of loans lent to firms in CPRS as fraction of capital, whereas the colored sub-bar (right) denotes the value of loans lent to firms with positive carbon emissions. The color insensitivity of the right sub-bar corresponds to the carbon intensity of a bank's loan portfolio; dark orange means high CO$_2$ emissions in kg per euro of loan portfolio, white means zero kg CO$_2$ per euro. Higher values mean that banks are likely to be affected stronger by carbon pricing (everything else equal).
Overall, we clearly see that a large share of banks' loan portfolios will be affected by CO$_2$ pricing, according to both CO$_2$ emission based and CPRS based exposure estimates. However, for individual banks the exposure estimates differ by factors of up to 3.3 (bank 17) --- the number above each composite-bar gives the ratio of CPRS based and CO$_2$ emission based exposures; values smaller than 1 indicate underestimation of exposures by CPRS. 
In Fig.\ref{figSI_loans_NACE4}(a) we further confirm that NACE 4 sectors can not capture the carbon intensity of firms, hence, CPRS assigns zero risks to 263 sectors (out of 537 with loans in our dataset) that account for 23\% of total loan volume. Conversely, 12 climate policy-relevant sectors, representing 9\% of the loan volume, have no estimated emissions.
The CO$_2$ emission intensity of loan portfolios varies substantially across banks --- emissions in portfolios of banks' 8,9 and 14 are particularly high. Overall, our findings illustrate that using CO$_2$ emission data at the firm-level is crucial for assessing banks carbon pricing exposures accurately. Additionally, in Fig.\ref{figSI_cprs_emit_loans} we show climate risky and non-risky loans based on two climate risk assessment methods distributed across NACE1 sectors.


\begin{figure}[t]
	\centering
	\includegraphics[scale=.1, keepaspectratio]{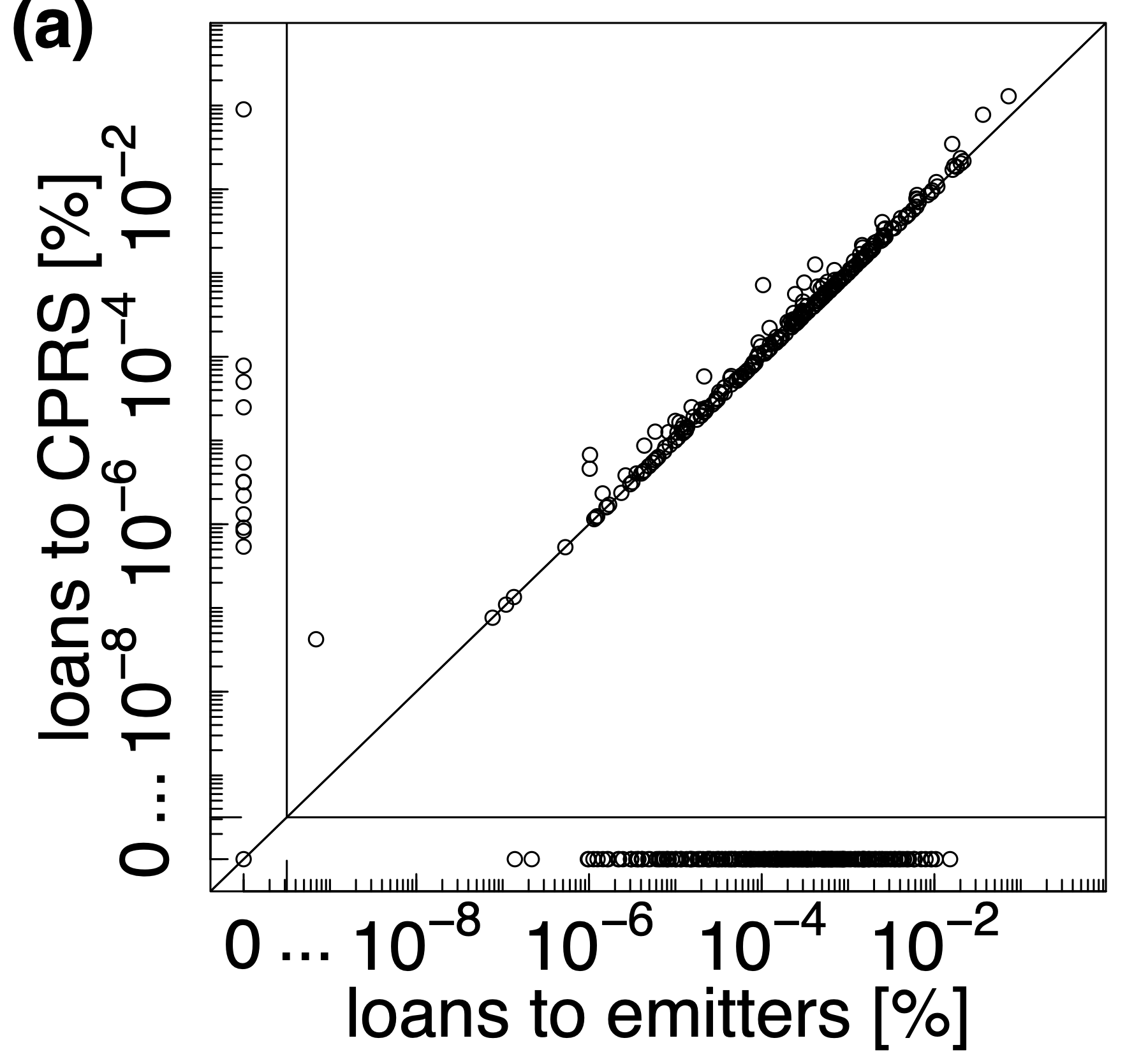}
	\includegraphics[scale=.1, keepaspectratio]{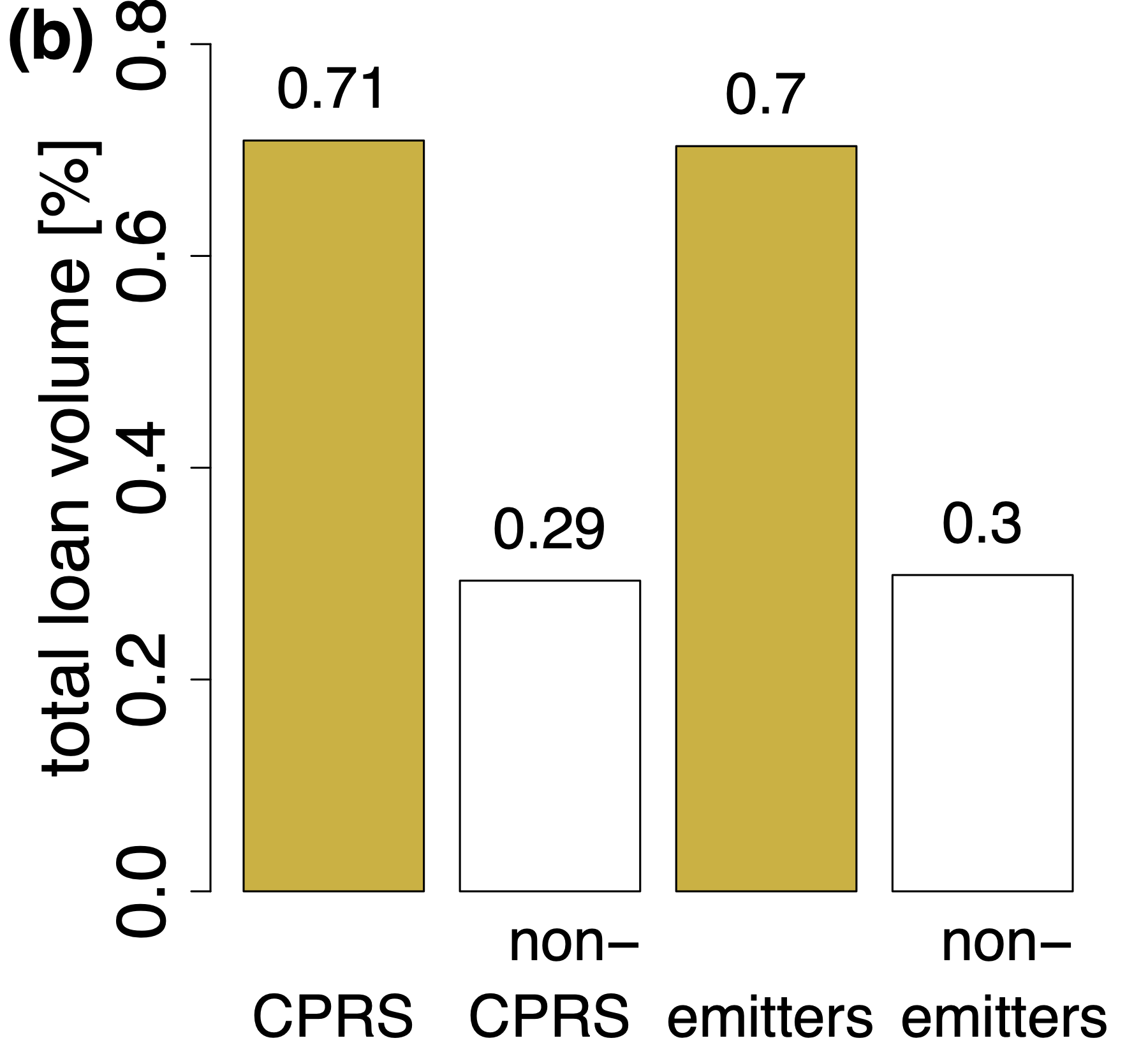}
  	\caption{\textbf{Climate risky loans in NACE 4 sectors based on emissions and CPRS taxonomy.} The log-log scatterplot in panel \textbf{(a)} shows loans of emitters in NACE 4 sectors (black circles) on the $x$-axis and loans of firms categorized by the CPRS taxonomy on the $y$-axis. There is $600$ NACE 4 sectors in total in this plot. If a circle lies on the diagonal it means that all firms with loans in the respective NACE 4 sector are emitters, and both methods lead to the same result. In our case $78$ NACE 4 sectors are on the diagonal. If a circle lies above the diagonal, it means that not all firms with loans in respective sector are emitters, and CPRS method overestimates risks ($184$ sectors). Circles that have zero $x$-coordinate indicate sectors that have no emitting firms, but based on the CPRS taxonomy are climate risk relevant ($12$ sectors). These sectors are A2.3.0, C13.9.3, C28.9.5, D35.1.2, H50.2.0, K.64.2.0, K64.9.2, K64.9.9, K66.1.9, K66.2.1, K66.2.2, K66.2.9. Note that we omit sector K-Financial activities from our emissions estimation method, hence it has no estimated emissions. 
   Contrary to it, circles that have zero $y$-coordinate denote sectors that are climate non-risky according to the CPRS taxonomy, but given the estimated emissions contain emitting firms ($263$ sectors). For example, sectors like C10.1.2, C10.3.9, C10.6.1, C25.1.1, C.25.6.2, G46.1.1, G46.2.1, G46.7.5, G46.9.0, G47.1.1. Firms in the remaining 63 sectors have no loans in our dataset and are denoted by a circle at zero-zero coordinate.   Panel \textbf{(b)} shows aggregated results from the two methods.  The first bar in yellow labeled as CPRS shows total loan volume, $71\%$, of firms that belong to climate policy relevant sectors. It is equal to a sum over $y$-axis coordinates of circles in non-zero regions in panel \textbf{(a)}. Similarly, the third bar shows total loan volume, $70\%$, of newly emitting firms and is equal to sum of $x$-coordinates of circles in no-zero region of panel \textbf{(a)}. Bar in white color labeled as non-CPRS is equal to loan volume, $29\%$, of firms that are not climate policy relevant based on the CPRS taxonomy, and is equal to the sum of $y$-coordinates of circles with zero $x$-coordinate. Finally, the last bar in white color labeled as non-emitters shows total loan volume, $30\%$, of non-emitting firms. It is equal to the sum of $x$-coordinates of circles with zero $y$-coordinates in panel \textbf{(a)}. These results show that on aggregated level two risk assessment methods lead to very similar results --- around $70\%$ of loan volume is climate risky and remaining $30\%$ of loans is not climate risky. However, on the NACE 4 level CPRS method can misestimate emissions-related direct exposure of banks. 
   }

	\label{figSI_loans_NACE4}       

\end{figure}

\begin{figure}[t]
	\centering
	\includegraphics[scale=.1, keepaspectratio]{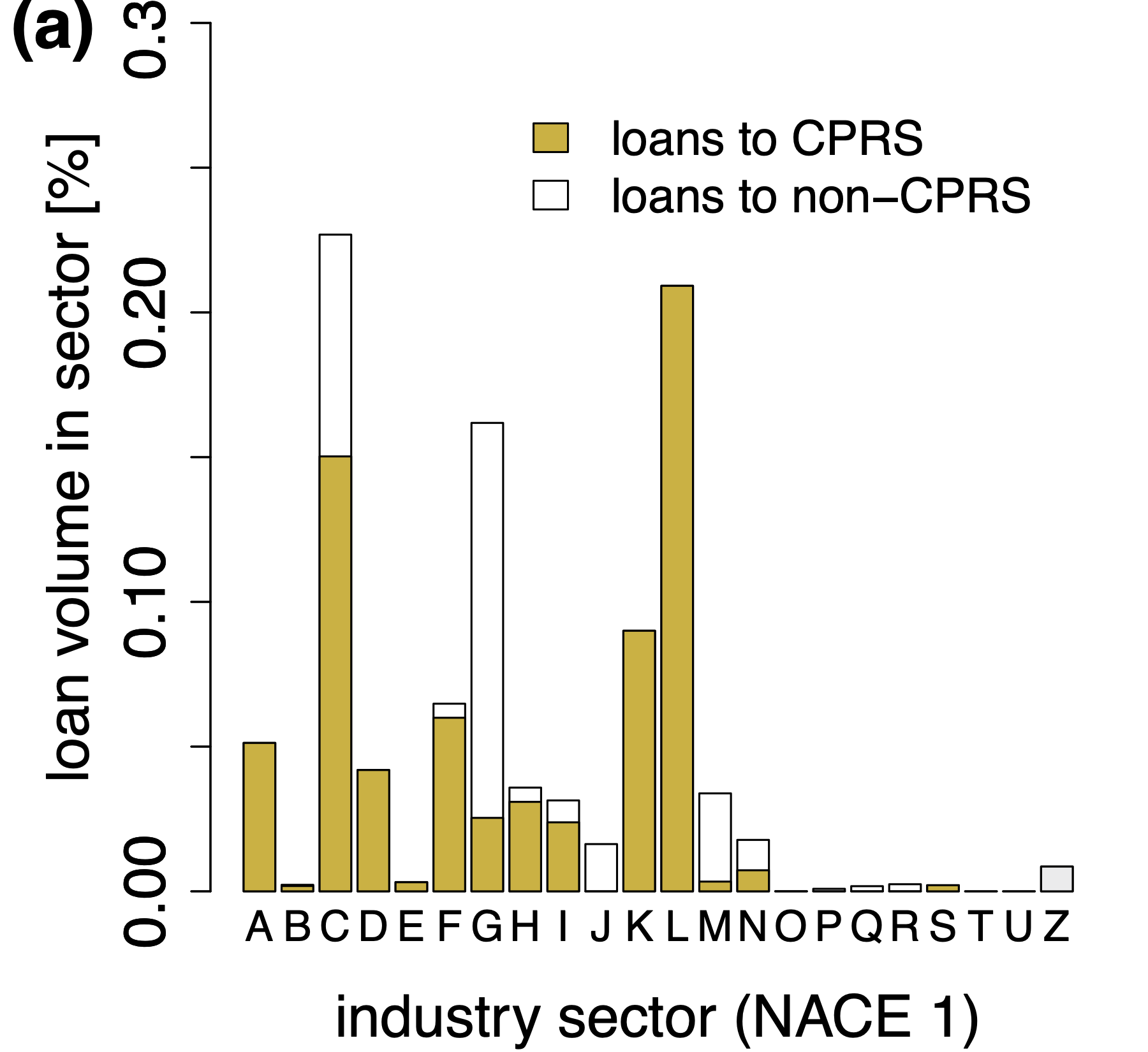}
	\includegraphics[scale=.1, keepaspectratio]{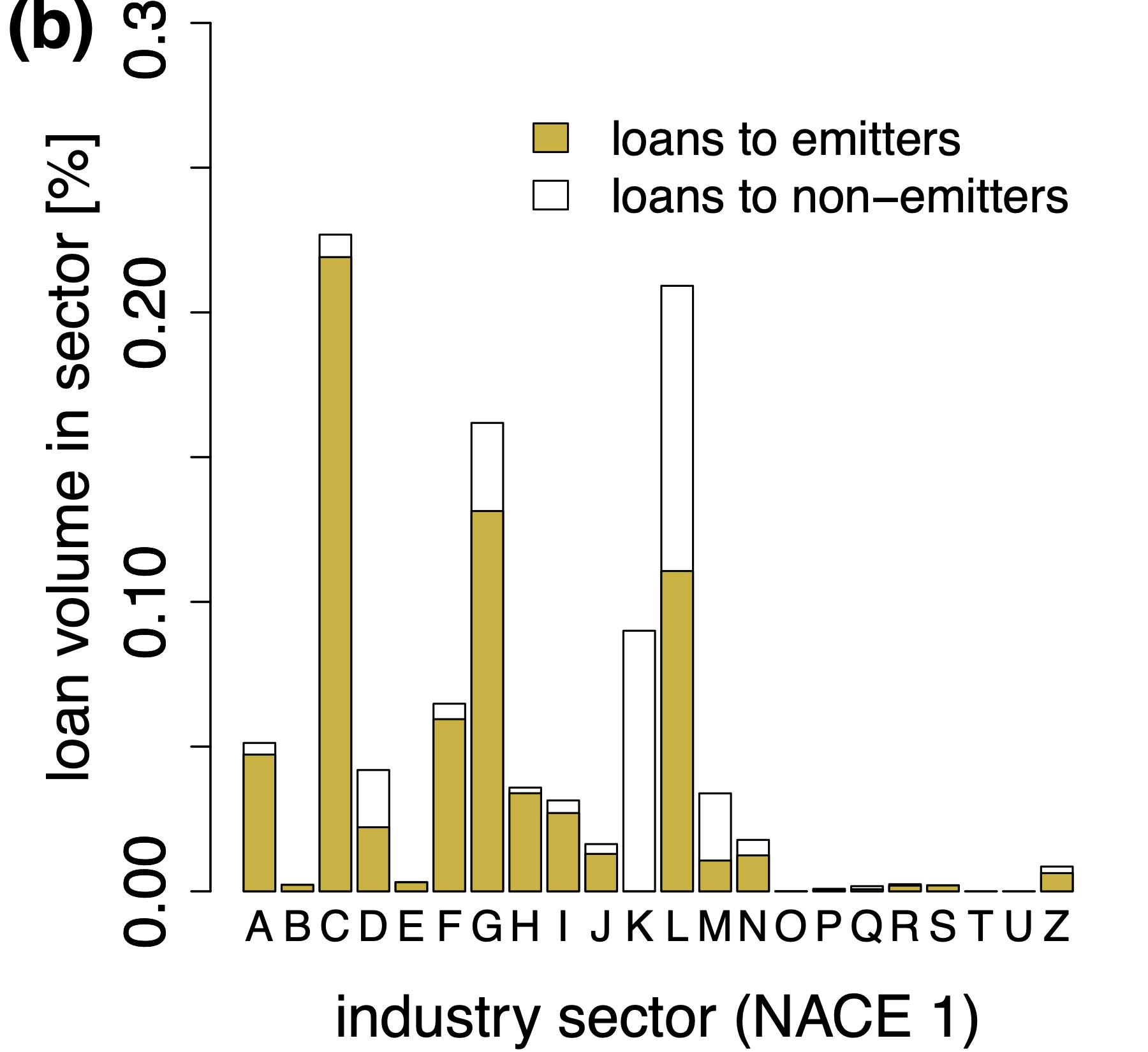}
  	\caption{\textbf{Climate risky and non-risky loans based on two climate risk assessment methods across NACE 1 sectors.} Both panels have NACE 1-level sectors on the $x$-axis and the percentage share of total loan volume in these sectors on the $y$-axis. The heights of the bars, which indicate the total loan volume of firms within each sector, agree in both panels and add up to 1. Yellow and white colors indicate loans to climate risky and non-risky firms respectively. In panel \textbf{(a)} we present results based on the CPRS classification of climate relevant and non-relevant sectors, thus yellow color denotes loans to the CPRS firms and white color -- to non-CPRS firms . In panel \textbf{(b)} we present results based on estimated emissions of banks' clients, where loans to emitters are in yellow color and loans to non-emitting clients are in white. For many sectors two methods yield different risk assessments. For example, in sectors C, Manufacturing, and G, Wholesale $\&$ Retail, CPRS approach underestimates risks by $31\%$ and $81\%$ respectively. Conversely, it overestimates climate risks in sector L, Real Estate Activities,  by $90\%$. Note that we excluded sector K from our emission estimation method and it has no emissions by design. 
   \textbf{Legend:} \textbf{A} Agriculture, Forestry $\&$ Fishing,
  \textbf{B} Mining $\&$ Quarrying,
  \textbf{C} Manufacturing,
  \textbf{D} Electricity, Gas, Steam, Air Conditioning,
  \textbf{E} Water Supply, Sewerage, Waste,
  \textbf{F} Construction,
  \textbf{G} Wholesale $\&$ Retail Trade,
  \textbf{H} Transportation $\&$ Storage,
  \textbf{I} Accommodation $\&$ Food Service,
  \textbf{J} Information $\&$ Communication.,
  \textbf{K} Financial $\&$ Insurance Activities,
  \textbf{L} Real Estate Activities,
  \textbf{M-U} Other Service Activities,
  \textbf{Z} Undefined }  
	\label{figSI_cprs_emit_loans}       

\end{figure}

\clearpage

\newpage
\section{Additional details on systemic risky firms and losses from supply chain contagion}\label{AppendixF_ad_info_systemic_risky_firms}

In the pessimistic GL scenario the absence of even a single essential input can halt the entire production process of a firm, regardless of the availability of other inputs. This explains why the contagion losses are substantially higher with the GL production function compared to the Linear.
As discussed in Section \ref{step_4} Methods Step 4, extensive losses can occur from contagion with the GL function when a so-called \textit{systemically important firm (SIF)} fails. In the scenario with carbon price of $30$ EUR/t, we observe failure of one systemically important firm, see Fig.\ref{figSI_fsri_esri_costs_pass_through}(a). This firm belongs to NACE 4 sector \textit{H52.2.1 - Service
activities incidental to land transportation}, and its estimated emissions are equal to 725 Kt.  As we show in Fig.\ref{figSI_boxplots_emissions}, this is the biggest emitter in the transportation sector. Note that our model does not allow for firms adjusting their prices during the propagation of supply network shocks, even though customers of transport firms might be willing to pay higher prices to ensure the transportation of their goods. Additional high systemic risk firms may fail at higher prices, but as they all trigger the same contagion (see \cite{diem2022quantifying}), the resulting losses overlap, leading to no further significant jumps. 
Notably, if the carbon costs pass-through mechanism is omitted from our model systemically important firm fails already in the 10 EUR/t scenario, see Fig.\ref{figSI_fsri_esri_costs_pass_through}(b). This means that the costs pass-through mitigates the effects of direct stress on SI firm by distributing it across downstream firms within the network.

Notably, if the carbon costs pass-through mechanism is omitted from our model systemically important firm fails already in the 10 EUR/t scenario, see Fig.\ref{figSI_fsri_esri_costs_pass_through}(b). This means that the costs pass-through mitigates the effects of direct stress on SI firm by distributing it across downstream firms within the network.

To ensure the robustness of our results wrt. details of the network structure, we conduct climate stress tests on modified supply chain networks. Specifically, we remove all links in the network that fall below a certain threshold. We apply two thresholds: EUR 25,000, which reduces the network to 203,592 firms, retaining 91\% of the original sales volume, and EUR 250,000, which leaves 51,618 firms, preserving 74\% of the original network's out-strength. For other characteristics of the thresholded networks see Tab.\ref{tab:thresholded_W}. The results in Fig.\ref{figSI_fsri_esri_costs_pass_through}(b) show that, qualitatively, the financial losses remain similar to those observed in the full network. In other words, the systemically important firm that fails in the 30 EUR/t scenario on the full network still triggers significant contagion in the sub-networks at the same price level. However, the nature of this contagion changes slightly. Notably, as Fig.\ref{figSI_thr_ntw_prod_fin_loss} shows, the L-Real Estate Activities sector experiences significantly lower losses. While it accounted for nearly 17\% of total financial losses in the full network, it contributes to less than 3\% of the total equity losses in the sub-network. This indicates that removing smaller links has disrupted certain connections that would have otherwise spread the contagion.

\begin{figure}[h]
	\centering
	\includegraphics[scale=.1, keepaspectratio]{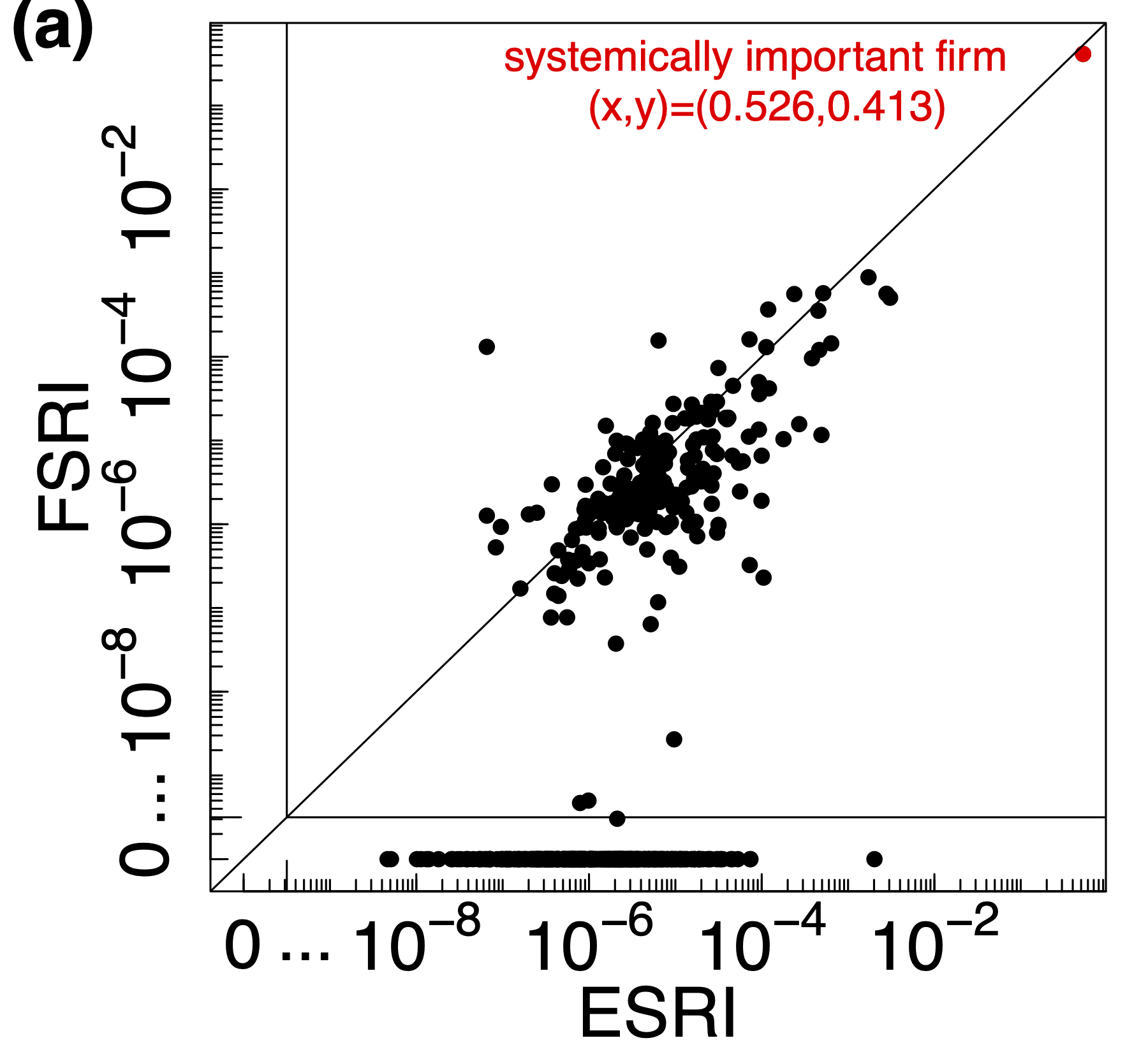}
	\includegraphics[scale=.1, keepaspectratio]{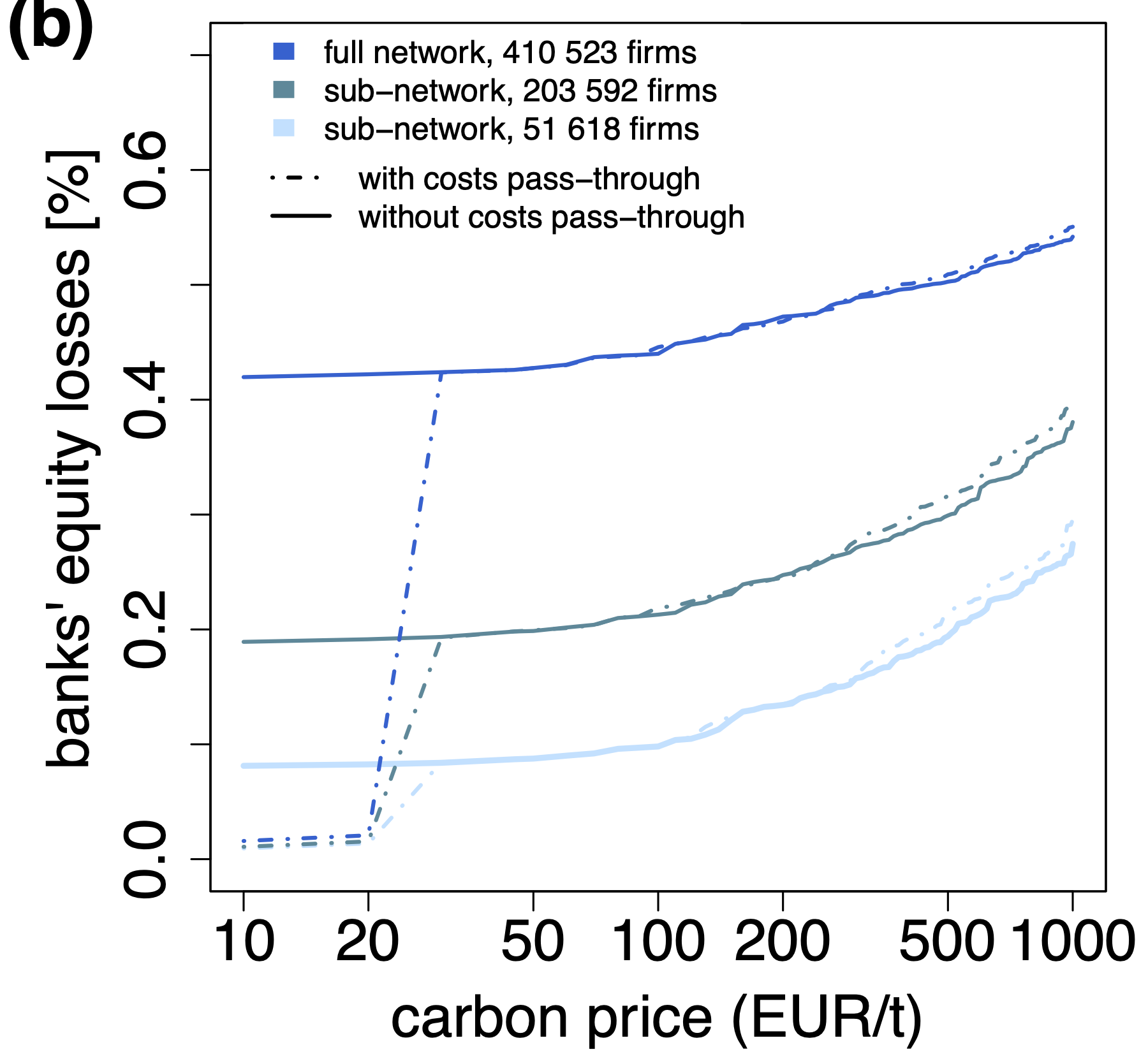}

  	\caption{Panel \textbf{(a)} shows Economic ($x$-axis) and Financial ($y$-axis) Systemic Risk Indices (ESRI and FSRI) of firms that default in $30$ EUR/t scenario on the full network. Axes are log-scaled. Firms with no financial systemic risk are plotted in the separate area at the bottom of the figure. One systemically important firm (red dot) fails in this scenario. Other firms have very small economic and financial systemic risks, as they don't trigger any substantial SCN contagion. This means, that the high losses in $30$ EUR/t scenario presented in Fig.\ref{fig4_fin_prod_loss}\textbf{(a)}, stem from the contagion triggered by a failure of one systemically risky firm. This firm belongs to the NACE 4 digit sector \textit{52.2.1 - Service activities incidental to land transportation}. It is the biggest emitter in the transportation sector with net estimated emissions of 725 Kt. Panel \textbf{(b)} shows climate stress testing results with the SCN contagion involving General Leontief production function with and without costs pass-through mechanism. $x$-axis denotes carbon prices and the $y$-axis shows banks' equity losses. The upper (dark blue) dot-dashed line corresponds to the main results presented in Fig.\ref{fig4_fin_prod_loss}\textbf{(a)}. The solid line that overlaps it shows results of the stress testing without cost pass-through mechanism. It means, that firms with estimated emissions have to cover their carbon costs by own profit without passing it on their buyers. In such case, the systemic risk of a firm materializes already in 10 EUR/t scenario. This means, that the carbon costs pass through softens the impacts of the direct stress by distributing it across firms in the network. Additionally, we show results of the stress test on smaller networks created by dropping links smaller than particular thresholds (see section \ref{sec_data}.Data and Tab.\ref{tab:thresholded_W}). Lines in the middle (dark green) and at the bottom (light blue) show results for network with links bigger than EUR 25 000 and 250 000 respectively. We see, that qualitatively results remain the same, but are shifted down. The systemically important firm fails in 10 EUR/t scenario without costs pass-through and in 45 EUR/t scenario with costs pass-through. The losses are, however, substantially lower even though overall out-strength of sub-networks drops by 9 percentage points for the network with 200K firms, and by 36 percentage points for the network with 50K firms (see Tab.\ref{tab:thresholded_W}). The systemic losses on the sub-networks are relatively small in comparison with losses on the full network, as some links spreading the contagion on the latter are not present in the former mentioned (see Fig.\ref{figSI_thr_ntw_prod_fin_loss}).
     }  
	\label{figSI_fsri_esri_costs_pass_through}       

\end{figure}

\begin{figure}[ht]
	\centering
	\includegraphics[scale=.1, keepaspectratio]{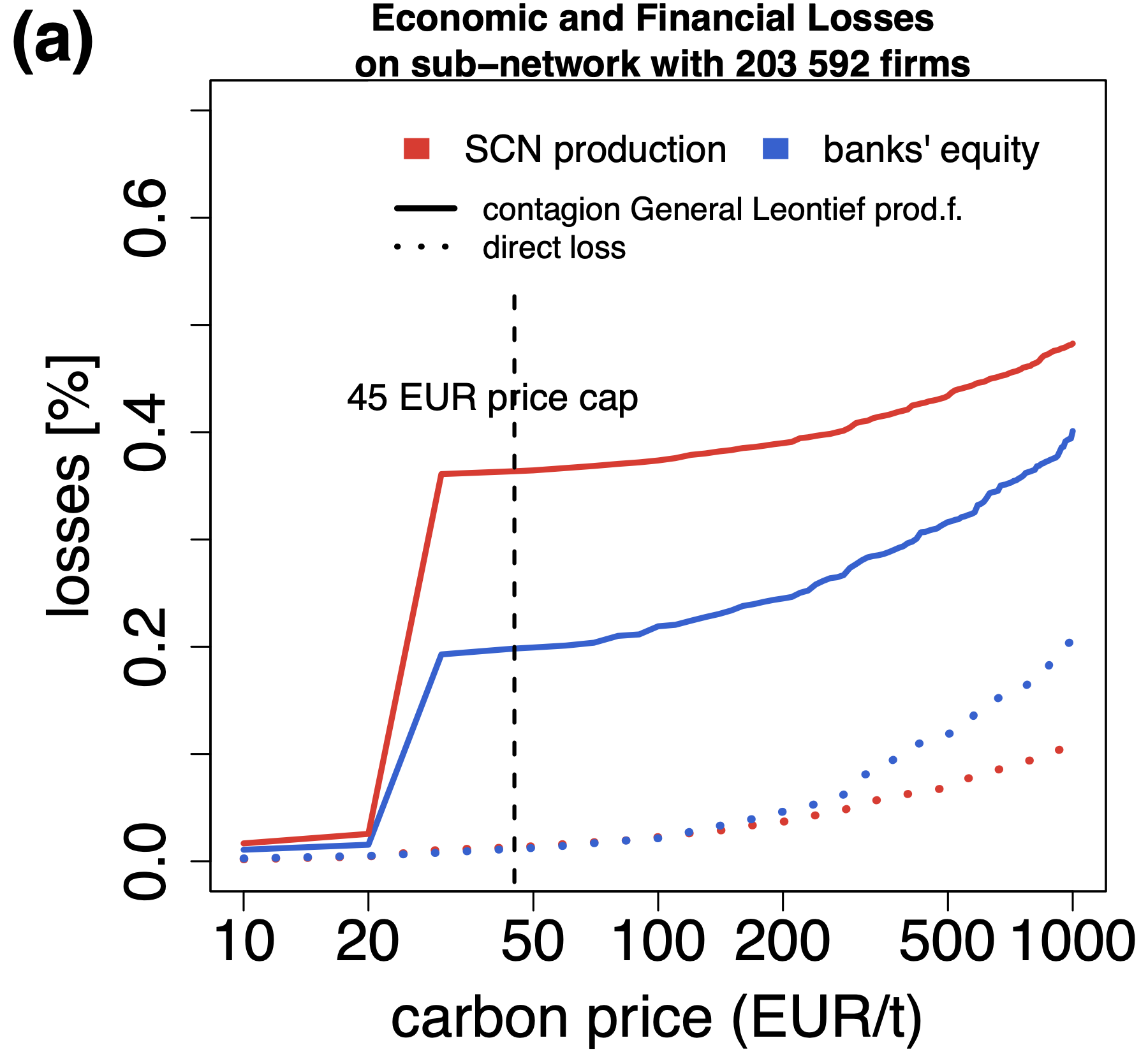}
	\includegraphics[scale=.1, keepaspectratio]{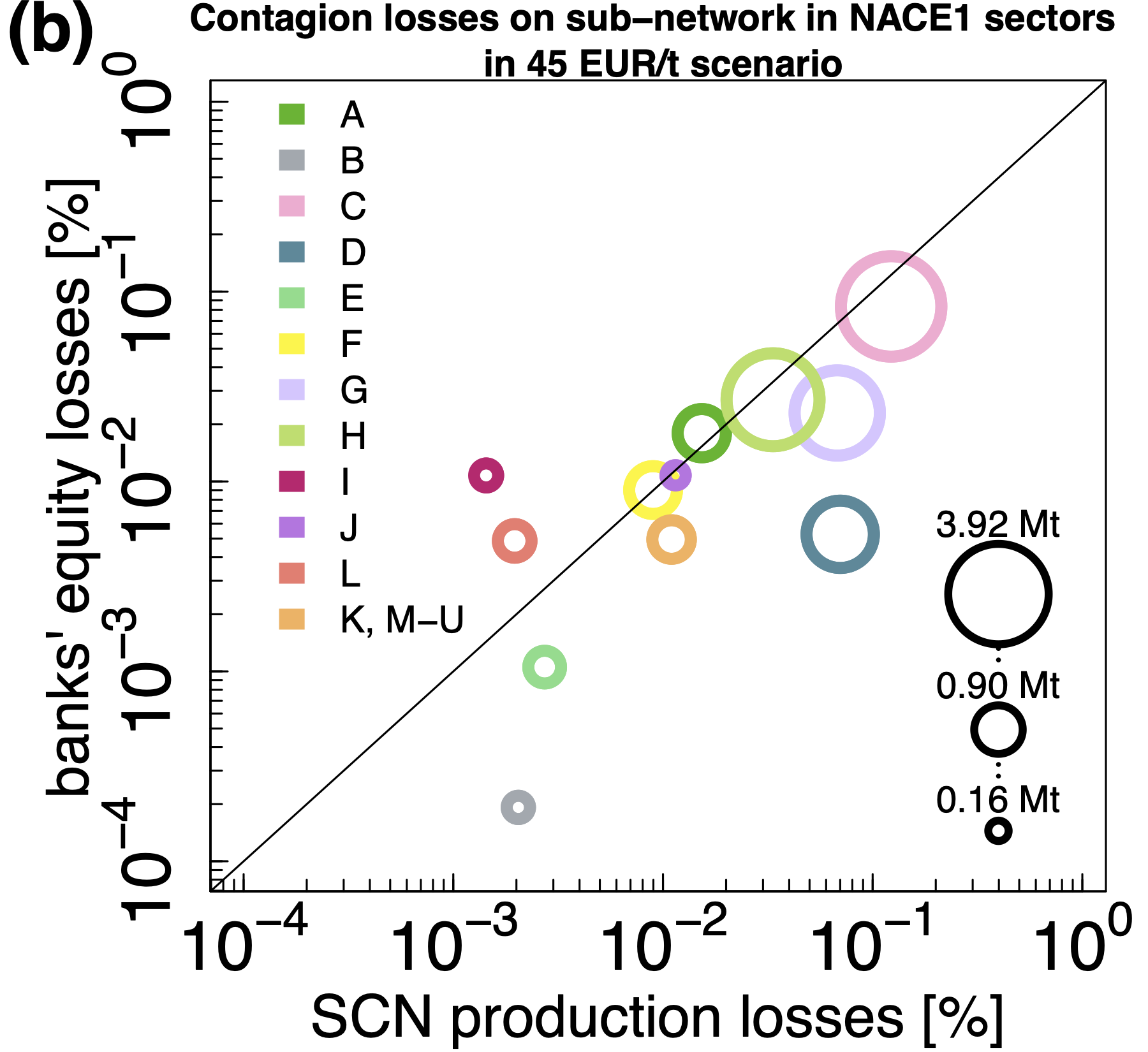}

  	\caption{\textbf{Direct and supply chain contagion adjusted losses of firms and banks after the carbon price shock on sub-network.} These results are produced in the similar was as the ones in Fig.\ref{fig4_fin_prod_loss}.  The sub-network, $W^s$, with 203 592 firms is derived from $W$ by removing links smaller than EUR $25$K (see Tab.\ref{tab:thresholded_W}). In panel \textbf{(a)} $x$-axis shows carbon prices and the $y$-axis shows losses of the supply chain network (SCN) and of the banks, in red and blue colors respectively, quantified across 100 scenarios. The $x$-axis is log-scaled. Dashed vertical line indicates EU ETS II price cap of $45$ EUR/t.  The direct production losses, $\Lambda^\textrm{dir}$ and the respective loan write offs of banks, $\mathcal{L}^\textrm{dir}$ are denoted by the dotted red and blue lines respectively. The contagion-adjusted losses involving General Leontief production function incurred by the SC network, $\Lambda^\textrm{GL}$, and by banks, $\mathcal{L}^\textrm{GL}$, are given by the red and blue solid lines. The results are qualitatively similar to the results obtained on the full network, with the systemic risk materializing in the 30 EUR/t scenario. At this price the sub-network SCN, $W^s$, suffers contagion-adjusted losses $\Lambda^\textrm{GL}=36\%$,  and $\mathcal{L}^\textrm{GL}=19\%$. 
    Panel \textbf{(b)} shows production, $\Lambda^\textrm{GL}$, ($x$-axis) and financial, $\mathcal{L}^\textrm{GL}$, ($y$-axis) losses at $45$ EUR/t scenario disaggregated to NACE1 sectors. The scatterplot is in log-log scale. Each circle denotes a sector and its size indicates estimated carbon emissions. Note, that emissions here are slightly smaller than in Fig.\ref{fig4_fin_prod_loss}, as many firms are dropped. Colors distinguish sectors A-J and L, and sectors K and M-U are aggregated into one category (orange circle). Production and financial losses are in the same units as in panel \textbf{(a)}. Thus all $x$-coordinates add up to $0.36$ and all $y$-coordinates add up to $0.19$. We see, that in this case the biggest production and financial contagion-adjusted losses stem from sectors C-Manufacturing, H-Transportation and G-Wholesale, while sector L-Real Estate Activities remains relatively unaffected. Note, that on the full network it accounted for almost $17\%$ of the total financial losses, while on the sub-network it causes less than $3\%$ of the banks' equity losses. This demonstrates, that by applying threshold on network's links we drop some important connections, that transmit the shock. In this case - to the real estate sector.
    } 
	\label{figSI_thr_ntw_prod_fin_loss}       
\end{figure}

\clearpage
\newpage

\section{Bank-level equity losses for 45EUR/t pessimistic scenario} \label{appendix_bank_losses_45_pessimistic}

\begin{figure}[ht]
	\centering
	\includegraphics[scale=.1, keepaspectratio]{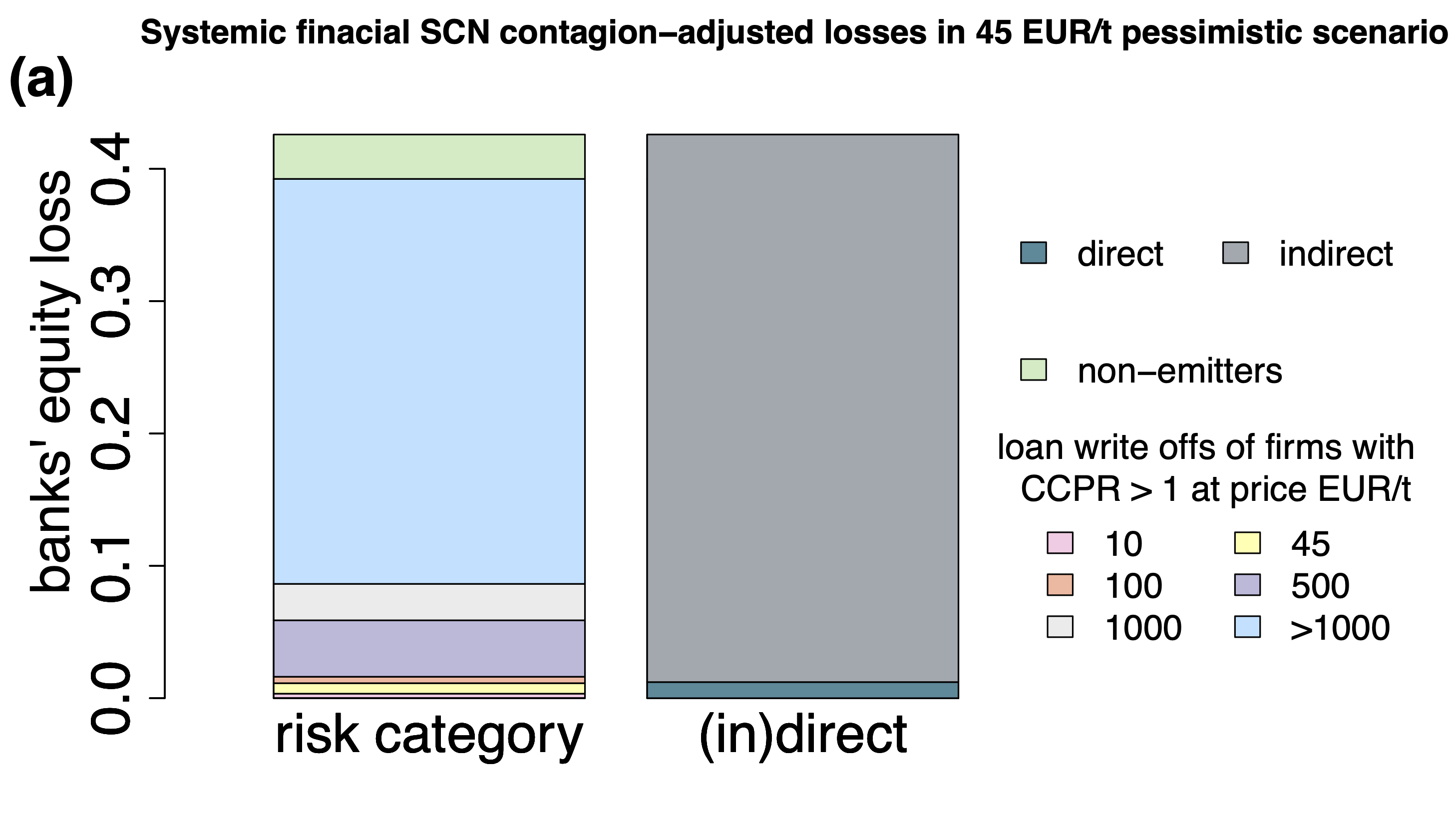}
	\includegraphics[scale=.1, keepaspectratio]{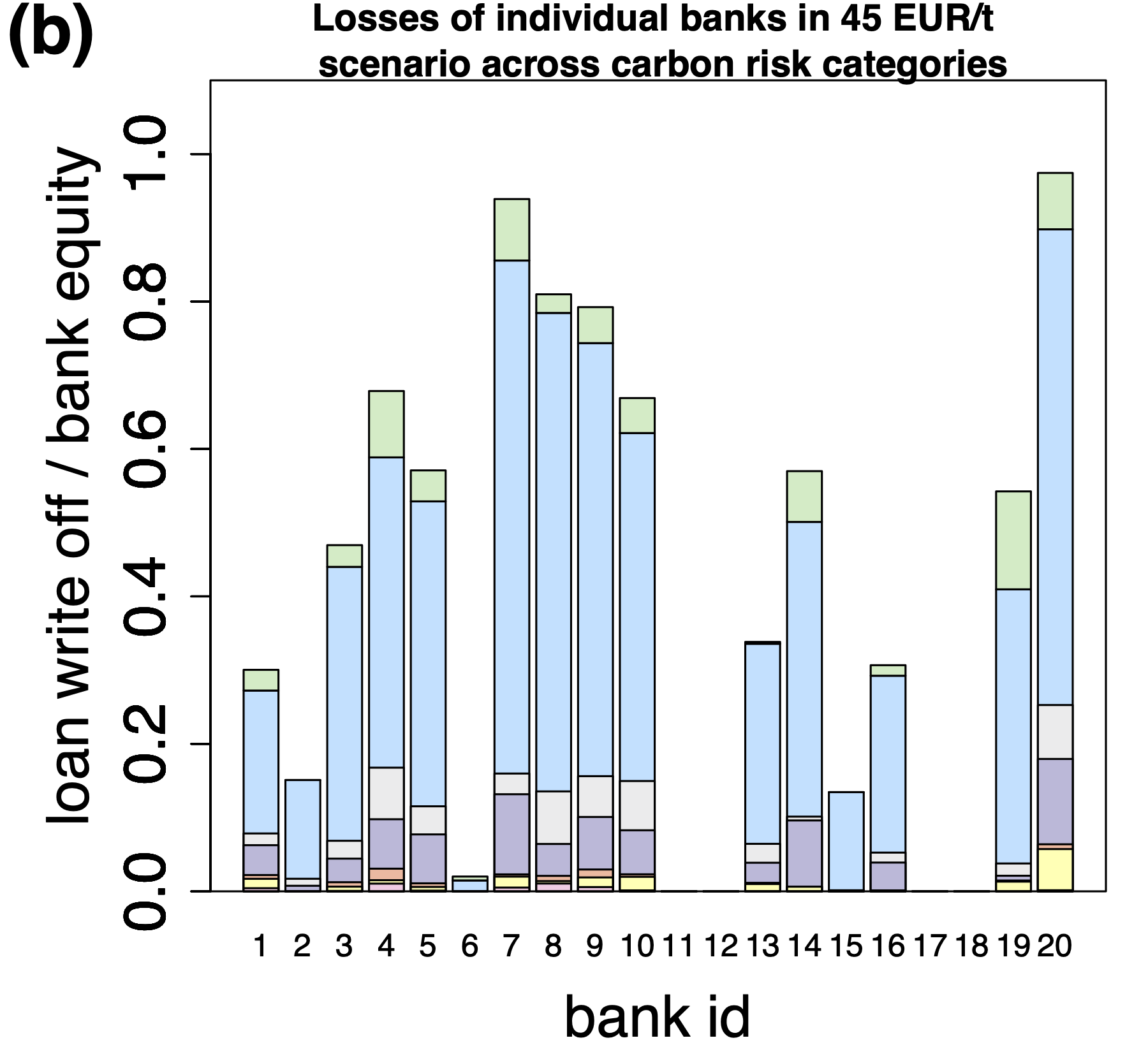}
	\includegraphics[scale=.1, keepaspectratio]{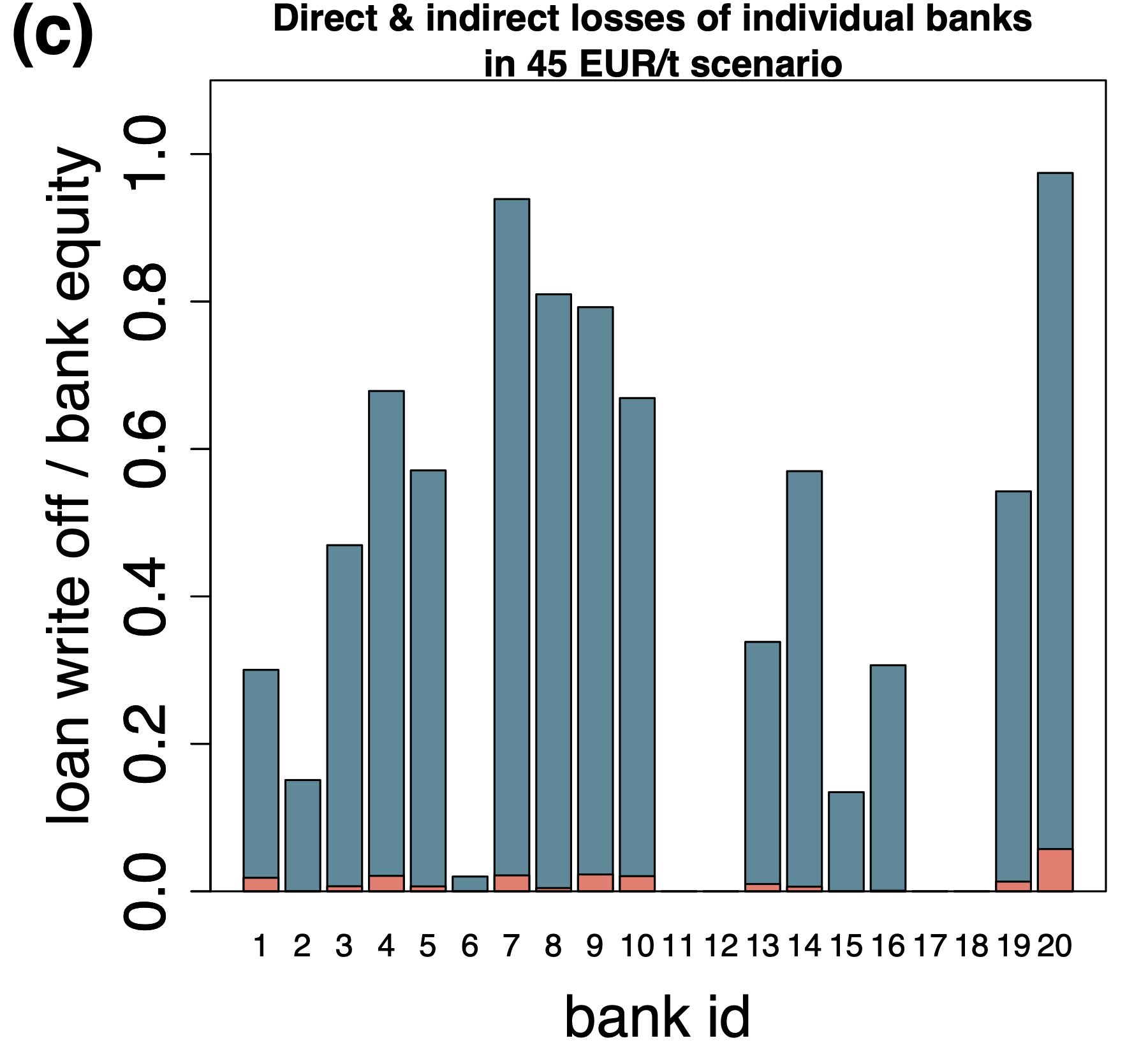}
    \caption{ \textbf{Supply chain network contagion-adjusted financial system losses in 45 EUR/t pessimistic scenario.} Panel (a) shows two identical bars, representing 42\% bank-equity losses, as in Fig.\ref{fig4_fin_prod_loss} (blue solid line) at 45 EUR/t. The two bars are disaggregated by carbon costs-to-profit ratio (CCPR)-induced risk categories of loans on the left-hand side, and by loans defaulting directly (dark green, due to high carbon costs) and indirectly (gray, due to production losses) on the right-hand side. Direct financial losses account for only 1\% of bank equity losses, while indirect losses are substantially higher due to the failure of a systemically important firm. The CCPR-induced risk categories reveal the carbon risks of loans that default in the 45 EUR/t pessimistic scenario. The colors indicate the risk categories of firms, based on the CCPR ratio, which reflects the carbon price at which each firm would become unprofitable when carbon costs exceed profits at a given price level. Specifically, we show 10, 45, 100, 500 and 1000 EUR/t in pink, yellow, red, purple and grey colors respectively. Firms that would only become unprofitable at prices above 1000 EUR/t are shown in blue, while non-emitting firms are categorized in green. Panels (b) and (c) show these contagion-adjusted losses in 45 EUR/t pessimistic scenario additionally disaggregated to 20 banks. $x$-axes show ids of 20 banks and $y$-axes show contagion-adjusted losses of these banks with respect to their own equities.  }

	\label{figSI_45_eur_ccpr_risk}       

\end{figure}

In Fig.\ref{figSI_45_eur_ccpr_risk} we study the systemic financial losses in 45 EUR/t pessimistic scenario in detail (see the caption for description of the figure). The carbon costs-to-profit ratio carbon risk categories classify loans (firms) to several groups (various colors in the left bar the plot of panel (a)) each of them indicating at which carbon price, $\pi$, a firm , $i$, would default on its loan, i.e. CCPR$_i(\pi)>1$. In the presented scenario, the loans of directly defaulting firms and loans in risk categories 10 and 45 combined coincide, as both represent loans to firms with a CCPR ratio greater than 1 at a carbon price of 45 EUR/t. The disaggregation of indirect losses indicates that the majority of loans defaulting from supply chain contagion originate from firms with minimal climate risk. The light-blue segment represents loans to firms that would remain profitable even at a carbon price of 1000 EUR/t. Nevertheless, the supply chain contagion initiated by a carbon shock of just 45 EUR/t imposes additional stress on these firms, leading to their indirect default. Additionally, 3 percentage points of indirect losses stem from non-emitting firms. This implies that carbon risk of firms based solely on firm-level information given by the CCPR doesn't fully reflect firms' exposure to the climate transition risk. Supply chain contagion can amplify carbon risk of emitting firms, passing it on to non-emitting firms or those with low carbon risk. In such cases, a firm's credit risk is impacted by production losses resulting from climate-related supply chain disruptions. Furthermore, the financial viability of firms may also be threatened by increased production costs due to higher supplier prices driven by carbon costs. Now, we analyze the 45 EUR/t scenario, in which a systemically important firm fails, from the perspective of individual banks. Specifically, we are interested in how much the banks' direct losses are amplified and whether these losses originate from firms with low or high climate transition risk. We demonstrate these bank-level losses in 
panels (b) and (c) of Fig.\ref{figSI_45_eur_ccpr_risk}. $x$-axes show ids of 20 banks and $y$-axes show contagion-adjusted losses of these banks with respect to their own equities. Contagion-adjusted losses of banks number 7 and 20 are equal to almost $100\%$ of own equities. Banks number 4, 5, 8, 9, 10, 14 and 19 write off loans worth more than $50\%$ of own equities. The rest of the banks have losses smaller than $40\%$ of own equities. Sizes of loan write offs across banks are heterogeneous, which is given by their different equity sizes. Panel (b) shows bank-level losses categorized by carbon risk of defaulted firms. Majority of banks incur losses from their emitting as well as non-emitting clients. On average, $7\%$ of banks' losses stem from the non-emitters. Panel (c) shows bank-level contagion-adjusted losses disaggregated to direct, $\mathcal{L}^\textrm{dir}_k$, and indirect, $\mathcal{L}^\textrm{indir}_k$, write offs. Banks number 2, 6, 11, 12, 15, 16, 17 and 18 have no direct losses, but they experience indirect losses. The indirect losses are substantially higher than the direct ones. Average amplification factor over 13 banks that incur direct losses is $80$.

\clearpage
\newpage

\section{NACE 1 sector contagion adjusted production and financial losses in 45 EUR/t scenario}
\label{appendix_45_N1_losses}

To gain a clearer understanding of the supply chain contagion propagation, we present the systemic production and financial losses across industry sectors. In particular, we look at a scenario in which a systemically important firm fails. Given that the EU ETS II carbon price is not expected to exceed 45 EUR/t until 2030, we focus on this scenario as it represents the worst-case outcome when the policy is implemented. 

Figure \ref{figSI_fin_prod_loss_N1} (a) shows production and financial losses across various carbon price shock scenarios as in Fig.\ref{fig4_fin_prod_loss}. Panels \textbf{(b)}, \textbf{(c)} and \textbf{(d)} show production, $\Lambda^\textrm{dir}$,
   $\Lambda^\textrm{L}$,
   $\Lambda^\textrm{GL}$, ($x$-axes) and financial, 
   $\mathcal{L}^\textrm{dir}$,
   $\mathcal{L}^\textrm{L}$,
   $\mathcal{L}^\textrm{GL}$, ($y$-axes) losses respectively at $45$ EUR/t scenario disaggregated to NACE 1 sectors. Both axes are log-scaled. Recall that the GL production function is associated with the pessimistic substitution and the Linear production function - with optimistic substitution. Circles indicate industries A-J and L. Industries that incur small losses, M-U and K, are aggregated into one circle (orange). Size of a circle indicates the amount of estimated emissions presented in Fig.\ref{fig1_emissions}\textbf{(a)} in yellow color. Production and financial losses are in the same units as in panel \textbf{(a)}. Thus all $x$-coordinates add up to $0.01$ and all $y$-coordinates add up to $0.01$ in \textbf{(b)}, --- to $0.5$ and $0.03$ in \textbf{(c)}, and --- to  $0.53$ and $0.42$ in \textbf{(d)}. We can see that the direct production and financial losses to individual sectors in panel \textbf{(b)} are minimal across most of the sectors. The biggest losses are suffered by sectors G, Wholesale $\&$ Retail, C, Manufacturing, and H, Transportation $\&$ Storage. In panels \textbf{(c)} and \textbf{(d)} presented losses are amplified by the contagion within and across the sectors. Thus, C, G and H incur additional shock from the cascade, which further seriously affects sectors L, Real Estate Activities, F, Construction and D, Electricity.

This means, that banks are exposed to climate transition risk not only from clients in big emitting sectors C, H, D or G, but also from firms in sectors with relatively small emission shares like sectors L and F. Sectors C, G and L are very likely to propagate shocks to banks, as more than a half of the entire loan volume is to firms in these sectors (Fig.\ref{figSI_cprs_emit_loans}\textbf{(b)}), while, as the biggest emitters, sectors C, H, G and D are most likely to incur direct losses that are in turn propagated. Indeed, industries C, G and H experience high direct losses, but D is almost not affected. This means that most of the firms in the latter don't default directly as they have sufficient profit buffers to cover their emission costs. Conversely, firms in the transportation sector experience a high number of defaults. This sector incurs losses comparable to losses of C, but its total loan volume is 6 times smaller (see Fig.\ref{figSI_cprs_emit_loans}\textbf{(b)})).

Propagation of losses through the production network is highly dependent on the network's structure -- pairwise links between firms. Contagion-adjusted losses spread from the initially failed firms not only to other firms within the same sector but also to firms in other sectors. For example, sector H sells/provides $23\%$ of its produce/services to sector C, and $20\%$ -- to G, see Tab.\ref{tab_SI_NACE1_buy_sup_out}. That is how numerous direct failures in H can propagate shock downstream to C and G. Similarly, sector L purchases $7\%$, $14\%$, $15\%$ and $19\%$ of its inputs from sectors C, D, F and G respectively. Sector F, in turn, gets $15\%$ of its own inputs from sector C and $34\%$ -- from sector G, see Tab.\ref{tab_SI_NACE1_buy_sup_in}, which can be a reason of higher contagion losses in F. However, these numbers provide only a rough indication of how a shock may propagate. The final magnitude of losses and the sectors affected also depend significantly on the production functions used. As previously mentioned, the General Leontief function in the cascade can create bottlenecks in production processes, greatly amplifying the direct losses. A comparison of panels \textbf{(c)} (with the Linear production function) and \textbf{(d)} (with the General Leontief production function) shows that the relative positions of the industries remain consistent, except for four sectors: L, I, J, and D. The use of the General Leontief (GL) function in the cascade shifts these sectors into a higher notional carbon-risk category. Notably, the previously almost unaffected non-emitting sector L, Real Estate Activities, causes loss equal to $7\%$ of total bank equity, which accounts for $18\%$ of the contagion-adjusted losses to the financial system in this scenario. Sectors I, J and D together incur losses of $6\%$ of banks' equity.

\begin{figure}[t]
	\centering
	\includegraphics[scale=.1, keepaspectratio]{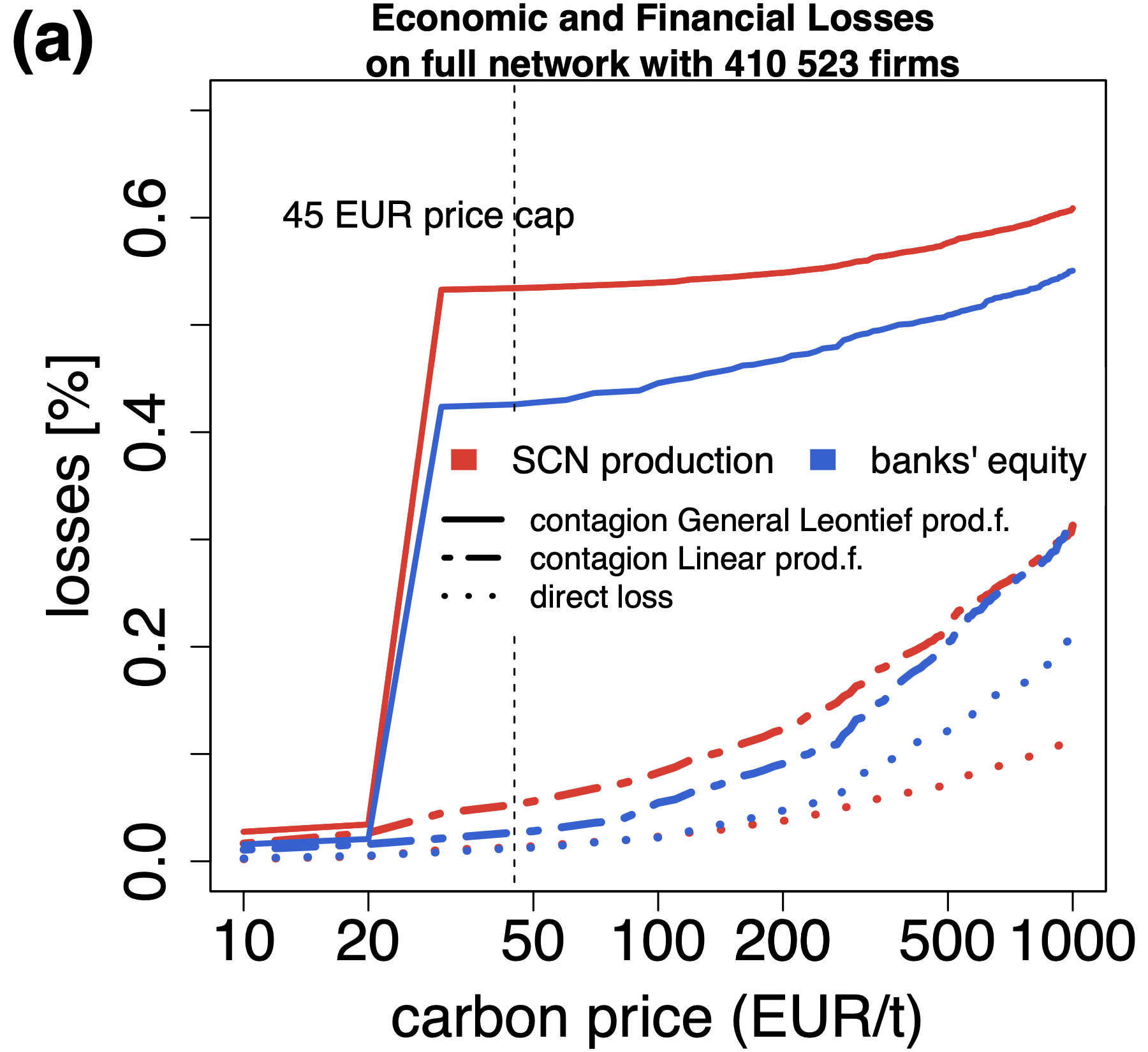}
	\includegraphics[scale=.1, keepaspectratio]{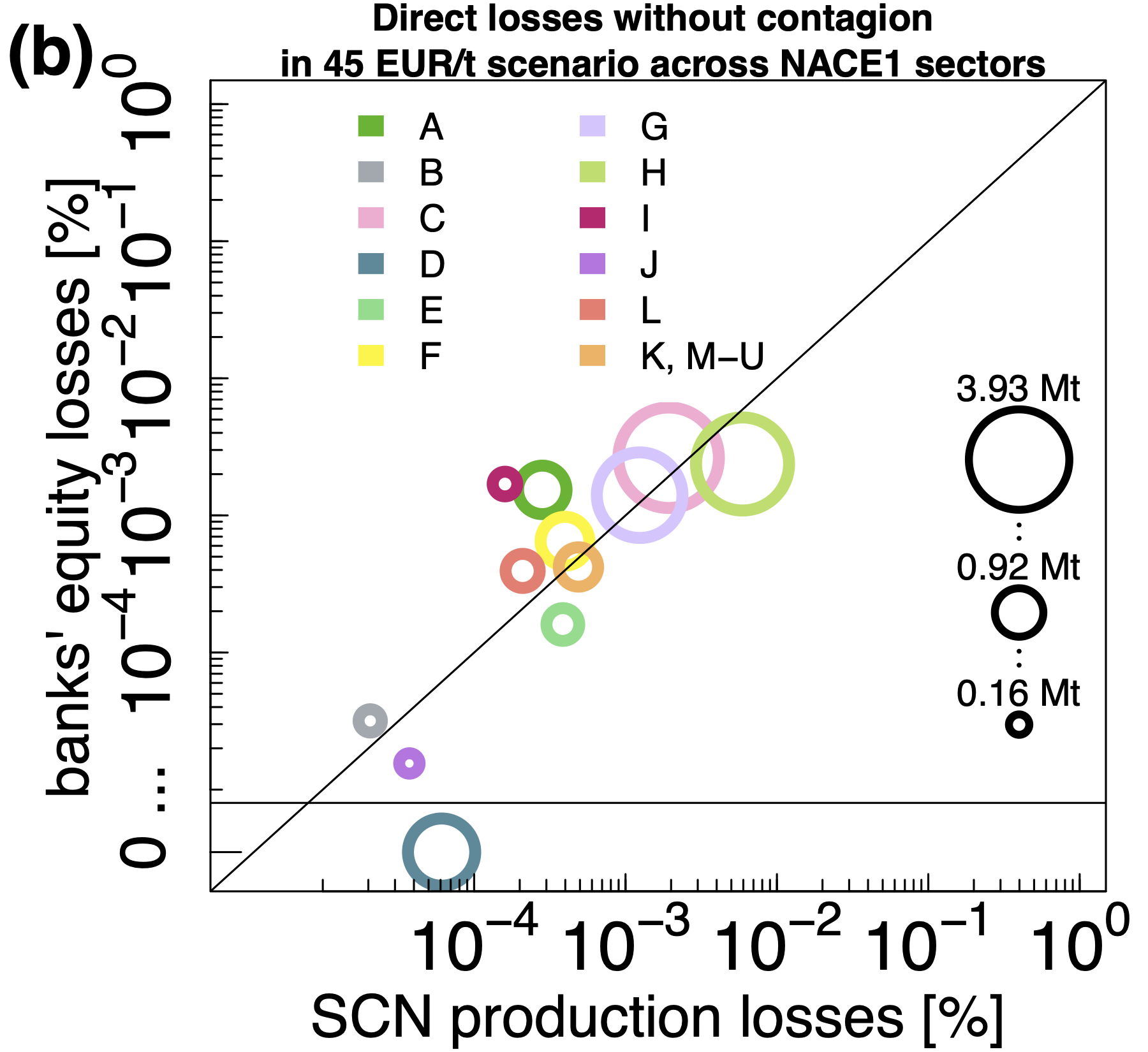}
 	\includegraphics[scale=.1, keepaspectratio]{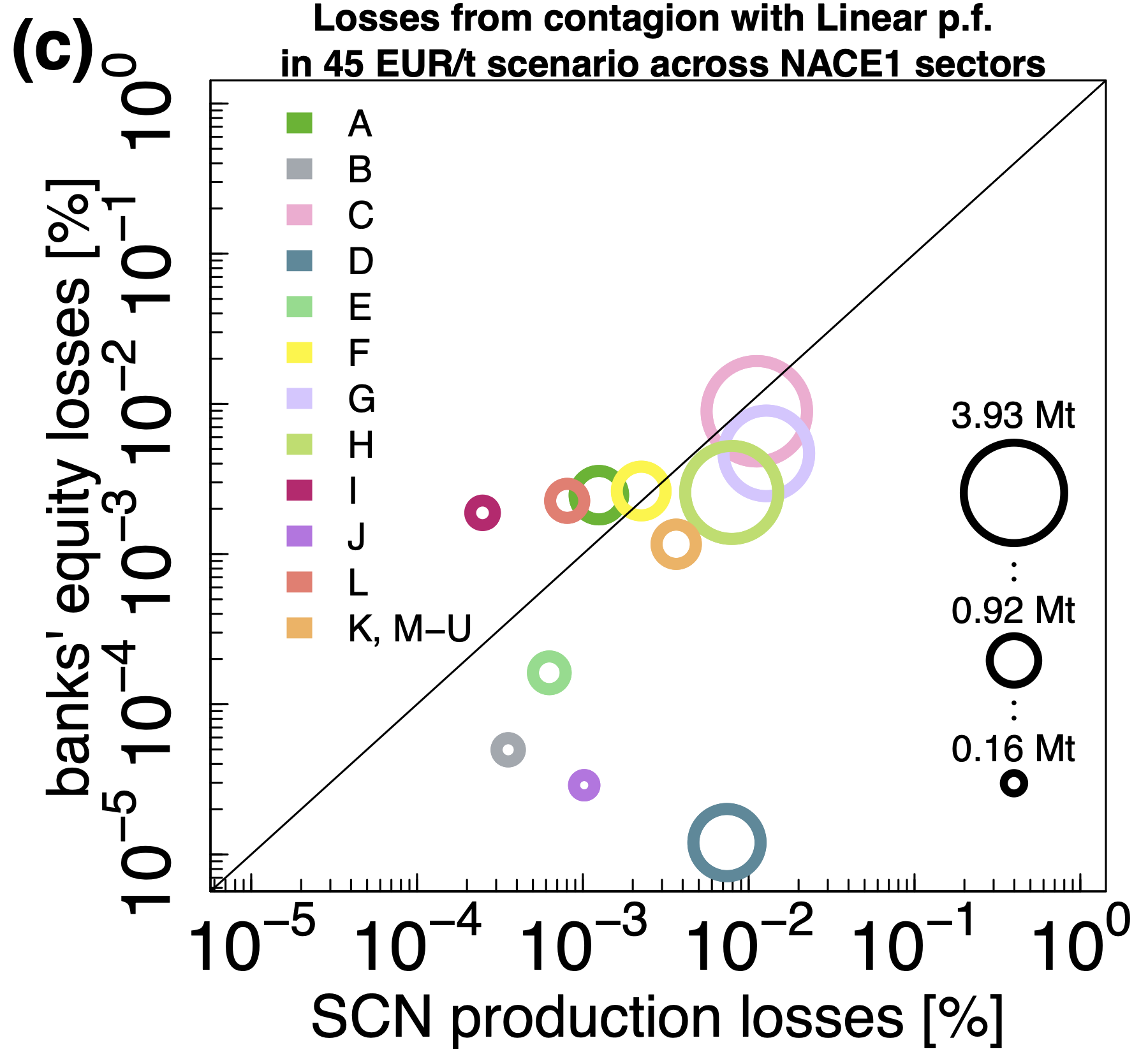}
	\includegraphics[scale=.1, keepaspectratio]{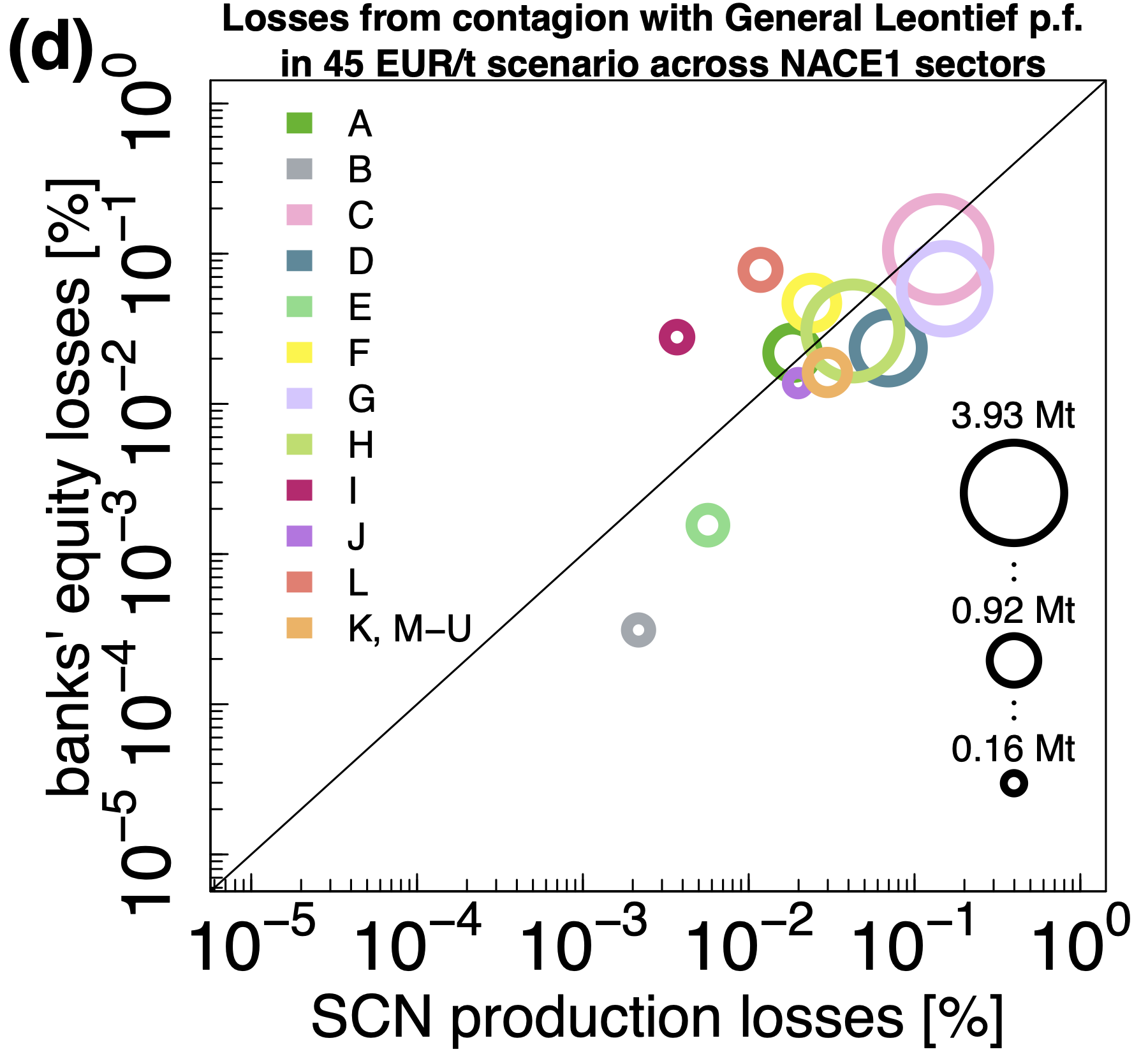}

  	\caption{\textbf{Direct and supply chain contagion adjusted losses of firms and banks after the carbon price shock.} Panel \textbf{(a)} shows carbon prices on the log-scaled $x$-axis and losses of the supply chain network (SCN) and of the banks on the $y$-axis, as in Fig.\ref{fig4_fin_prod_loss}. Red color indicates SCN losses, while blue color denotes total banks' equity losses quantified across 100 scenarios. The direct production losses, $\Lambda^\textrm{dir}$ and the respective loan write offs of banks, $\mathcal{L}^\textrm{dir}$ are denoted by the dotted red and blue lines respectively. The contagion-adjusted losses involving Linear production function incurred by the SC network, $\Lambda^\textrm{L}$, and by banks, $\mathcal{L}^\textrm{L}$, are denoted by dot-dashed lines in red and blue respectively. Finally, the contagion-adjusted losses involving General Leontief production function incurred by the SC network, $\Lambda^\textrm{GL}$, and by banks, $\mathcal{L}^\textrm{GL}$, are given by the red and blue solid lines. The SCN contagion amplifies losses caused by the direct defaults of firms, with abrupt jump at $30$ EUR/t for GL. At this price level, the direct losses, $\Lambda^\textrm{dir}=1\%$ and $\mathcal{L}^\textrm{dir}=0.8\%$, get amplified by the production cascade -- leading to the contagion-adjusted production losses $\Lambda^\textrm{GL}=53\%$ and $\Lambda^\textrm{L}=4\%$, and the respective financial system contagion-adjusted banks' equity losses are equal to $\mathcal{L}^\textrm{GL}=42\%$ and $\mathcal{L}^\textrm{L}=2\%$. The huge amplification in simulations involving GL production function is caused by a failure of systemically important firm in the given scenario (see Fig.\ref{figSI_fsri_esri_costs_pass_through}\textbf{(a)}). Dashed vertical line indicates EU ETS II price cap of $45$ EUR/t. 
   Panels \textbf{(b)}, \textbf{(c)} and \textbf{(d)} show production, $\Lambda^\textrm{dir}$,
   $\Lambda^\textrm{L}$,
   $\Lambda^\textrm{GL}$, ($x$-axes) and financial, 
   $\mathcal{L}^\textrm{dir}$,
   $\mathcal{L}^\textrm{L}$,
   $\mathcal{L}^\textrm{GL}$, ($y$-axes) losses respectively at $45$ EUR/t scenario disaggregated to NACE1 sectors. These scatterplots are in log-log scale. Each circle denotes a sector and its size indicates estimated carbon emissions from Fig.\ref{fig1_emissions}. Colors distinguish sectors A-J and L, and sectors K and M-U are aggregated into one category (orange circle). Production and financial losses are in the same units as in panel \textbf{(a)}. The biggest direct and indirect production and financial losses are caused by sectors C, G and H. Additionally, sectors L and D incur high financial and production losses respectively from the cascade with the GL production function.
    } 
	\label{figSI_fin_prod_loss_N1}       
\end{figure}

\begin{table}[!ht]
    \centering
    \begin{tabular} {l||m{0.02\textwidth}|m{0.02\textwidth}|m{0.02\textwidth}|m{0.02\textwidth}|m{0.02\textwidth}|m{0.02\textwidth}|m{0.02\textwidth}|m{0.02\textwidth}|m{0.02\textwidth}|m{0.02\textwidth}|m{0.02\textwidth}|m{0.02\textwidth}|m{0.02\textwidth}|m{0.02\textwidth}|m{0.02\textwidth}|m{0.02\textwidth}|m{0.02\textwidth}|m{0.02\textwidth}|m{0.02\textwidth}|m{0.02\textwidth}|m{0.02\textwidth}|m{0.02\textwidth}||l}

        & \textbf{A} & \textbf{B} & \textbf{C} & \textbf{D} & \textbf{E} & \textbf{F} & \textbf{G} & \textbf{H} & \textbf{I} & \textbf{J} & \textbf{K} & \textbf{L} & \textbf{M} & \textbf{N} & \textbf{O} & \textbf{P} & \textbf{Q} & \textbf{R} & \textbf{S} & \textbf{T} & \textbf{U} &\textbf{Z} & $s^{\textrm{out}}$ \\ \hline
        \hline
        \textbf{A} & 23 &.& 30 &.&.& 1 & 23 & 1 &.&.&.&.&.& 1 &.&.&.&.&.&.&.& 20 & 5  \\ \hline
        \textbf{B} &.& 10 & 63 & 3 &.& 8 & 6 & 3 &.&.&.&.& 1 & 1 &.&.&.&.&.&.&.& 5 & 0.9 \\ \hline
        \textbf{C} & 2 &.& 31 & 1 &.& 3 & 30 & 3 & 1 & 1 & 1 &.& 1 & 1 &.&.&.&.&.&.&.& 25 & 57 \\ \hline
        \textbf{D} & 1 &.& 14 & 60 & 1 &.& 2 & 2 & 1 &.& 3 & 2 & 1 & 1 &.&.&.&.&.&.&.& 13 & 35 \\ \hline
        \textbf{E} & 1 &.& 16 & 4 & 28 & 5 & 7 & 1 & 2 & 1 &.& 3 & 2 & 21 &.&.&.& 1 & 1 &.&.& 8 & 2 \\ \hline
        \textbf{F} & 1 &.& 17 & 4 & 1 & 32 & 10 & 4 & 2 & 3 &.& 7 & 5 & 4 &.&.&.& 1 &.&.&.& 8 & 9 \\ \hline
        \textbf{G} & 3 &.& 14 &.&.& 6 & 48 & 3 & 3 & 2 & 2 & 1 & 2 & 2 &.&.&.&.&.&.&.& 13 & 71 \\ \hline
        \textbf{H} & 1 &.& 23 & 4 & 1 & 3 & 20 & 32 &.& 1 & 4 &.& 2 & 1 &.&.&.&.&.&.&.& 8 & 13 \\ \hline
        \textbf{I} & 1 &.& 7 &.&.& 5 & 11 & 3 & 22 & 3 &.& 3 & 2 & 9 &.& 3 & 1 & 2 & 1 &.&.& 28 & 1 \\ \hline
        \textbf{J} &.&.& 8 & 4 &.& 2 & 13 & 5 & 1 & 38 & 1 & 1 & 14 & 3 &.&.&.& 1 & 1 &.&.& 8 & 7 \\ \hline
        \textbf{K} & 1 &.& 4 & 43 &.& 3 & 10 & 17 & 1 & 1 & 2 & 1 & 2 & 6 &.&.&.&.&.&.&.& 8 & 4 \\ \hline
        \textbf{L} & 1 &.& 14 & 1 & 1 & 5 & 29 & 5 & 5 & 6 & 1 & 11 & 6 & 3 &.&.&.& 1 & 1 &.&.& 9 & 5 \\ \hline
        \textbf{M} & 1 &.& 19 & 5 & 1 & 7 & 18 & 4 & 2 & 7 & 1 & 5 & 17 & 4 &.&.&.& 1 &.&.&.& 6 & 10 \\ \hline
        \textbf{N} & 1 &.& 24 & 4 & 1 & 7 & 16 & 5 & 2 & 4 & 1 & 4 & 5 & 17 &.&.&.& 1 &.&.&.& 8 & 8 \\ \hline
        \textbf{O} & 1 &.& 28 & 2 & 1 & 8 & 5 & 1 & 1 & 23 & 1 & 6 & 8 & 4 &.&.&.& 1 &.&.&.& 11 & 0.05 \\ \hline
        \textbf{P} & 2 &.& 17 & 1 &.& 5 & 22 & 4 & 2 & 12 & 1 & 4 & 10 & 6 &.& 4 &.& 3 & 1 &.&.& 5 & 0.2 \\ \hline
        \textbf{Q} & 3 &.& 16 &.& 1 & 3 & 27 & 2 & 2 & 10 &.& 3 & 9 & 6 &.& 1 & 3 & 3 & 1 &.&.& 8 & 0.09 \\ \hline
        \textbf{R} &.&.& 8 & 1 &.& 5 & 8 & 1 & 3 & 11 & 2 & 3 & 5 & 10 &.&.&.& 26 & 1 &.&.& 16 & 0.9 \\ \hline
        \textbf{S} & 1 &.& 16 & 1 & 1 & 4 & 18 & 2 & 9 & 16 &.& 3 & 4 & 6 &.&.& 1 & 2 & 6 &.&.& 9 & 0.5 \\ \hline
        \textbf{T} &.&.&.&.&.&.&.&.&.&.&.&.&.&.&.&.&.&.&.&.&.&.&.\\ \hline
        \textbf{U} &.&.& 16 &.& 4 & 4 & 33 & 4 & 11 & 1 &.& 14 & 3 & 2 &.&.&.& 1 & 2 &.&.& 5 & * \\ \hline
        \textbf{Z} & 3 &.& 29 & 2 & 1 & 5 & 28 & 9 & 1 & 3 & 1 & 1 & 3 & 2 &.&.&.&.&.&.&.& 13 & 25 \\ \hline
    \end{tabular}
    \caption{ \textbf{Percentage share of sales from NACE1 to NACE1 level industry sectors.} Rows and columns (A-Z) denote suppliers and buyers aggregated to NACE1 industry sectors. The last column, $s^\textrm{out}$, is in billions of EUR and adds up to EUR 254bn, which is the total output of the supply chain network, $W$. Total output of sector U denoted by $*$ is equal to EUR 200 000. Elements of the table are equal to sales from supplier in a row to its buyers in columns with respect to its total output, $s^\textrm{out}$, in the last column. For example, sector A sells $23\%$ of its output to other companies in sector A, $30\%$ to firms in sector C, and $23\%$ to sector G. Each row adds up to $100\%$ over all columns A-Z.}
    \label{tab_SI_NACE1_buy_sup_out}
\end{table}

\begin{table}[!ht]
    \centering
    \begin{tabular}{l||m{0.02\textwidth}|m{0.02\textwidth}|m{0.02\textwidth}|m{0.02\textwidth}|m{0.02\textwidth}|m{0.02\textwidth}|m{0.02\textwidth}|m{0.02\textwidth}|m{0.02\textwidth}|m{0.02\textwidth}|m{0.02\textwidth}|m{0.02\textwidth}|m{0.02\textwidth}|m{0.02\textwidth}|m{0.02\textwidth}|m{0.02\textwidth}|m{0.02\textwidth}|m{0.02\textwidth}|m{0.02\textwidth}|m{0.02\textwidth}|m{0.02\textwidth}|m{0.02\textwidth}|}

        ~ & \textbf{A} & \textbf{B} & \textbf{C} & \textbf{D} & \textbf{E} & \textbf{F} & \textbf{G} & \textbf{H} & \textbf{I} & \textbf{J} & \textbf{K} & \textbf{L} & \textbf{M} & \textbf{N} & \textbf{O} & \textbf{P} & \textbf{Q} & \textbf{R} & \textbf{S} & \textbf{T} & \textbf{U} &\textbf{Z}  \\ \hline
        \hline
        \textbf{A} & 19 & 1 & 3 &.&.&.& 2 &.&.&.&.&.&.& 1 &.& 2 &.& 1 &.&.& 3 & 3 \\ \hline
        \textbf{B} &.& 22 & 1 &.&.& 1 &.&.&.&.&.&.&.&.&.&.&.&.&.&.&.&.\\ \hline
        \textbf{C} & 19 & 31 & 34 & 2 & 11 & 15 & 24 & 12 & 10 & 9 & 9 & 7 & 9 & 5 & 11 & 6 & 8 & 5 & 13 & 5 & 9 & 38 \\ \hline
        \textbf{D} & 4 & 6 & 10 & 80 & 14 & 1 & 1 & 5 & 6 & 2 & 26 & 14 & 2 & 3 & 5 & 5 & 4 & 6 & 8 &.&.& 12 \\ \hline
        \textbf{E} &.&.& 1 &.& 25 & 1 &.&.& 1 &.&.& 1 &.& 6 &.& 1 & 1 & 1 & 1 &.&.&.\\ \hline
        \textbf{F} & 2 & 6 & 3 & 2 & 7 & 22 & 1 & 3 & 4 & 3 & 1 & 15 & 7 & 7 & 14 & 3 & 7 & 5 & 4 &.& 1 & 2 \\ \hline
        \textbf{G} & 39 & 13 & 19 & 1 & 16 & 34 & 49 & 16 & 46 & 17 & 34 & 19 & 18 & 25 & 24 & 23 & 34 & 16 & 29 & 78 & 19 & 25 \\ \hline
        \textbf{H} & 1 & 9 & 6 & 2 & 4 & 3 & 4 & 31 & 1 & 2 & 12 & 1 & 4 & 3 & 1 & 1 & 4 & 1 & 3 &.&.& 3 \\ \hline
        \textbf{I} &.&.&.&.&.&.&.&.& 6 &.&.& 1 &.& 2 &.& 16 & 5 & 2 & 3 &.& 1 & 1 \\ \hline
        \textbf{J} &.&.& 1 & 1 & 1 & 1 & 1 & 3 & 2 & 35 & 2 & 2 & 13 & 4 & 8 & 5 & 5 & 5 & 8 &.& 1 & 1 \\ \hline
        \textbf{K} & 1 & 1 &.& 7 & 1 & 1 & 1 & 5 & 1 & 1 & 2 & 1 & 1 & 4 &.& 1 & 1 &.& 1 &.&.& 1 \\ \hline
        \textbf{L} &.& 1 & 1 &.& 2 & 2 & 2 & 2 & 7 & 4 & 1 & 13 & 4 & 3 & 3 & 6 & 5 & 5 & 4 &.&.& 1 \\ \hline
        \textbf{M} & 2 & 3 & 4 & 2 & 6 & 6 & 3 & 3 & 5 & 10 & 3 & 11 & 25 & 6 & 10 & 9 & 6 & 9 & 6 &.& 4 & 2 \\ \hline
        \textbf{N} & 1 & 3 & 4 & 1 & 4 & 4 & 2 & 3 & 4 & 4 & 2 & 8 & 5 & 21 & 8 & 6 & 4 & 9 & 5 &.& 59 & 2 \\ \hline
        \textbf{O} &.&.&.&.&.&.&.&.&.&.&.&.&.&.& 1 &.&.&.&.&.&.&.\\ \hline
        \textbf{P} &.&.&.&.&.&.&.&.&.&.&.&.&.&.&.& 4 &.& 1 &.&.&.&.\\ \hline
        \textbf{Q} &.&.&.&.&.&.&.&.&.&.&.&.&.&.&.&.& 2 &.&.&.&.&.\\ \hline
        \textbf{R} &.&.&.&.&.&.&.&.& 1 & 1 &.& 1 & 1 & 1 &.& 1 & 1 & 21 & 2 &.&.&.\\ \hline
        \textbf{S} &.&.&.&.&.&.&.&.& 1 & 1 &.&.&.&.&.& 1 & 3 & 1 & 5 &.&.&.\\ \hline
        \textbf{T} &.&.&.&.&.&.&.&.&.&.&.&.&.&.&.&.&.&.&.&.&.&.\\ \hline
        \textbf{U} &.&.&.&.&.&.&.&.&.&.&.&.&.&.&.&.&.&.&.&.&.&.\\ \hline
        \textbf{Z} & 10 & 4 & 14 & 2 & 9 & 9 & 10 & 16 & 5 & 9 & 6 & 6 & 10 & 9 & 12 & 8 & 10 & 11 & 9 & 17 & 2 & 8 \\ \hline\hline
        $s^\textrm{in}$ & 6 & 0.4 & 52 & 26 & 2 & 13 & 70 & 13 & 4 & 7 & 4 & 4 & 7 & 6 &* & 0.2 & 0.2 & 1 & 0.6 &*&*& 37 \\ 
    \end{tabular}
        \caption{ \textbf{Percentage share of purchases of NACE 1 from NACE 1 level industry sectors.} Rows and columns (A-Z) denote suppliers and buyers aggregated to NACE1 industry sectors. The last row, $s^\textrm{in}$, is in billions of EUR and adds up to EUR 254bn, which is the total volume of purchases in the supply chain network, $W$. Total purchases of sectors O,T and U denoted by $*$ are equal to EUR 32m, 91 000 and 304 000. Elements of the table are equal to purchases of buyers in a column from its suppliers in rows with respect to its total input, $s^\textrm{in}$, in the last column. For example, sector A buys $19\%$ of its input from other companies in sector A, $19\%$ from firms in sector C, and $39\%$ from sector G. Each column adds up to $100\%$ over all columns A-Z.}
    \label{tab_SI_NACE1_buy_sup_in}
\end{table}

\clearpage
\newpage

\section{Policy implications and model limitations}\label{SI_limitations_and_policy}

\textbf{Policy implications.}
Our estimates show that a rapid introduction of relatively high carbon prices of around 200 EUR/t (approx. upper price bound compliant with 2 degrees warming) involves substantial transition risks as economic losses could be substantially amplified through supply chain contagion. 

In principle, the currently scheduled introduction of EU ETS II carbon pricing with an initial cap of prices at a relatively low level --- allowing for some degree of adaptation --- before allowing for higher prices by 2030 seems to be a reasonable compromise between managing transition risk and reaching climate warming goals. However, decarbonization has to happen fast \cite{fankhauser2022meaning} and should feature minimal economic costs, hence, the exact implementation matters. 

The currently scheduled implementation --- jumping from 0 EUR/t to 45 EUR/t in 2025 and from 45 EUR/t to no cap in 2030 --- seems to be suboptimal as it allows for two rapid and potentially large price shocks. In particular, in 2030 there is the inherent potential of a large jump in prices, when considering that the 2 degree warming compatible CO\textsubscript{2} price could be as high as 200 EUR/t and even higher when aiming for 1.5 degree compatibility. Additionally the size of the jump is a priori not easily predictable and, hence, constitutes substantial uncertainty for firms. If firms underestimate the actual price jump in practice they might adapt too slowly and face difficulties. The non-linear increase of economic losses --- (shown in Fig. \ref{fig4_fin_prod_loss}), especially when considering indirect losses from supply chain contagion --- suggests that small price shocks with a recovery period in between would lead to smaller economic losses than few, but large and unpredictable, shocks that could lead to a substantial number of direct defaults and potentially high SC contagion effects.
Considering this, a better alternative would be starting carbon pricing as soon as possible with a low price cap, that increases each year by fixed percentage. The increase needs to be large enough to achieve sufficient decarbonization, but small enough to avoid a large number of defaults at a given time. Suitable climate stress testing models like the one presented here and model extensions can help policy makers to balance this trade-off and improve the implementation quality. Naturally, a late rapid implementation of high prices seems to bear the most substantial risks. 

Further, policy makers and central banks should monitor the timely adaptation of high systemic risk firms to avoid unnecessary transition risks. \\

\textbf{Limitations.}
Several limitations need to be addressed in future research. First, we need a unified model for carbon cost pass through and supply chain contagion that also considers price impacts on profits and supply chain rewiring through reduced sales. Second, we assume that the network of 2022 will be in still in place 2027, i.e., that firms do not adapt, which is unrealistic. We need models that simulate realistic supply network rewiring up to 2027 / 2030 given a set of carbon price expectations. Here financial system transition risks only originate from commercial loans, ignoring other important channels such as price impacts on stocks and bonds.

We discuss specific limitations in more detail.

\begin{enumerate}

         \item \textbf{Emissions estimation}    We implicitly assume a uniform price of gas on the one hand and oil-related products on the other hand, as we distribute emissions based on the relative in-strength firms display towards gas and oil distributors. Since smaller consumers generally pay higher prices for energy, particularly for gas, we may underestimate emissions for large consumers and overestimate them for small consumers. Additionally, oil is a heterogeneous product with various derivatives such as gasoline, heating oil, naphtha etc. These oil products are traded at different prices, which is not accounted for when aggregating the transaction values of different oil-product providing sectors. Additionally, results of estimation are highly dependent on the network data, i.e. the number of firms and links between them. As the final firm-level emissions are proportional to relative in-strengths which depend on the number of gas or oil buyers, then 10 or 100 of them will make a big difference. The 2022 VAT dataset we use was not subject to any reporting transaction thresholds, and we can be confident that the majority of firm-level transactions within the country is covered. To compare, the 2019 dataset with reporting threshold contains only around $60\%$ of firms in 2022 and the method would be misleading in that case.

    \item \textbf{Business and investors expectations}
            We assume that by 2027, the year the EU ETS II policy is set to be implemented, Hungary will remain in the same state as described by the 2022 empirical data. This is equivalent to an assumption that firms and investors are skeptical about the policy and make no preparations, such as investing in green technologies and infrastructure or adjusting their business models to mitigate its effects.
            
    \item \textbf{Price pass-through}
            In our model, we assume that the demand for oil and gas is inelastic, meaning firms' consumption of these resources remains unchanged despite rising emission prices. Further, we employ a naive carbon cost pass-through mechanism that relies on downstream bargaining power, estimated based on the NACE 4 industry size of a firm. While this is a simplified approach to the complex processes at play in an open economy with competition, Figure \ref{figSI_fsri_esri_costs_pass_through}\textbf{(b)} shows that even in this simple form, it demonstrates how passing costs downstream reduces direct pressure on firms and, consequently, mitigates losses for banks.
            
    \item \textbf{Loss given default}
            In our calculation of bank losses, we assume a uniform loss given default (LGD) of 100\% for all defaulted firms. This parameter is calibrated to represent the maximum potential losses in our climate stress test. We selected this value because our results are scalable with the LGD, making it easy to infer outcomes for lower values. Additionally, we do not account for possible measures such as installment postponements, firm restructuring in case of financial difficulties, or the use of collateral by banks.
        \item \textbf{Interbank- and other financial contagion.} Here we focus on contagion effects in the real economy, but we neglect contagion among financial institutions, which has been amply studied \cite{glasserman2016contagion, diem2020minimal}. 
        \item \textbf{International supply chain relations.} Even though the current data set is remarkably granular and complete for supply chain links within Hungary, it does not contain international supply chain links. These links can be important to transmit contagion internationally \cite{chakraborty2024inequality}, and hence we miss transition risks imported from abroad. 
        \item \textbf{Products and establishments.} For multi-product and multi-establishment \cite{diem2024supply, inoue2024disruption} firms we can not assess emissions of individual products and establishments potentially mis-estimate cost impacts from carbon pricing. 
\end{enumerate}

\clearpage
\newpage

\section{Descriptive statistics of firms' variables} \label{appendixC_data}

\begin{table}[!ht]
    \centering
    \begin{tabular}{|l||l|l|l|l|l|l|l|l|}
    \hline
        ~ & \makecell{\textbf{equity} \\ ($z$)} & \makecell{\textbf{liquidity} \\($a$)} & \makecell{\textbf{profit}\\ ($p$)} & \makecell{\textbf{net profit} \\ ($\mathcal{P})$} & \makecell{\textbf{ret. earn.}\\ ($\zeta$)} & \makecell{\textbf{emis.} \\ ($\mathcal{E}$)} & \makecell{\textbf{loans} \\ ($B$)}  \\ \hline
        \textbf{mean} & 401 863 & 199 236 & 128 229 & 31 249 & 43 101 &  73 & 31 667  \\ \hline       
        \textbf{std} & 22 744 742 & 9 602 207 & 6 010 017 & 866 538 & 1 945 682 & 7 296 & 1 512 935  \\ \hline \hline
        \textbf{1\%}  & -53 084 & -40 343 & -41 209 & -53 993 & -66 904 & 0 & 0  \\ \hline
        \textbf{5\%} & -7 272 & -945 & -3 173 & -8 898 & -9 225 & 0 & 0  \\ \hline
        \textbf{25\%} & 3 246 & 2 654 & 884 & -218 & -166 & 0 & 0  \\ \hline
        \textbf{50\%} & 12 728 & 9 911 & 8 092 & 1 524 & 1 308 & 0 & 0 \\ \hline
        \textbf{75\%} & 51 222 & 37 610 & 28 905 & 10 676 & 9 723 & 3 & 0  \\ \hline
        \textbf{95\%} & 496 127 & 305 920 & 225 605 & 103 555 & 102 292 & 66 & 42 857 \\ \hline
        \textbf{99\%} & 2 986 192 & 1 767 827 & 1 355 390 & 534 038 & 564 299 & 536 & 325 833  \\ \hline
    \end{tabular}
    \caption{Descriptive statistics of financial variables and estimated emissions of 410 523 firms in the supply chain network (in $10^3$ 2022 HUF). }
    \label{tab:fin_var_410K}
\end{table}

\begin{table}[!ht]
    \centering
    \begin{tabular}{|l||l|l|l|l|l|l|l|l|}
    \hline
        ~ & \makecell{\textbf{equity} \\ ($z$)} & \makecell{\textbf{liquidity} \\($a$)} & \makecell{\textbf{profit}\\ ($p$)} & \makecell{\textbf{net profit} \\ ($\mathcal{P})$} & \makecell{\textbf{ret. earn.}\\ ($\zeta$)} & \makecell{\textbf{emis.} \\ ($\mathcal{E}$)} & \makecell{\textbf{loans} \\ ($B$)}  \\ \hline \hline
        \textbf{mean} & 370 656 & 217 995 & 178 393 & 55 528 & 59 366 & 79 & 36 654  \\ \hline
        \textbf{std} & 18 092 327 & 9 937 983 & 7 256 430 & 707 844 & 1 335 054 & 8 327 & 1 812 510 \\ \hline \hline
        \textbf{1\%} & 10 & 0 & 0 & 0 & -73 & 0 & 0  \\ \hline
        \textbf{5\%}  & 2 108 & 1 050 & 503 & 96 & 27 & 0 & 0\\ \hline
        \textbf{25\%}  & 7 277 & 5 761 & 5 800 & 1 228 & 952 & 0 & 0 \\ \hline
        \textbf{50\%}  & 21 434 & 17 084 & 16 380 & 5 354 & 4 566 & 0 & 0 \\ \hline
        \textbf{75\%}  & 72 627 & 54 161 & 47 055 & 20 739 & 18 699 & 5 & 0 \\ \hline
        \textbf{95\%}  & 618 018 & 392 965 & 340 094 & 161 029 & 151 610 & 92 & 57 232  \\ \hline
        \textbf{99\%}  & 3 294 995 & 1 993 394 & 1 823 218 & 735 342 & 715 481 & 688 & 400 000  \\ \hline
    \end{tabular}
    \caption{Descriptive statistics of financial variables and estimated emissions of 299 830 firms in the supply chain network (in $10^3$ 2022 HUF).}
    \label{tab:fin_var_299K}
\end{table}

\clearpage
\newpage

\section{Details on estimating firms' carbon emissions from supply chain data} \label{appendixA_emis}

\begin{figure*}[h]
	\centering
	\includegraphics[scale=.1, keepaspectratio]{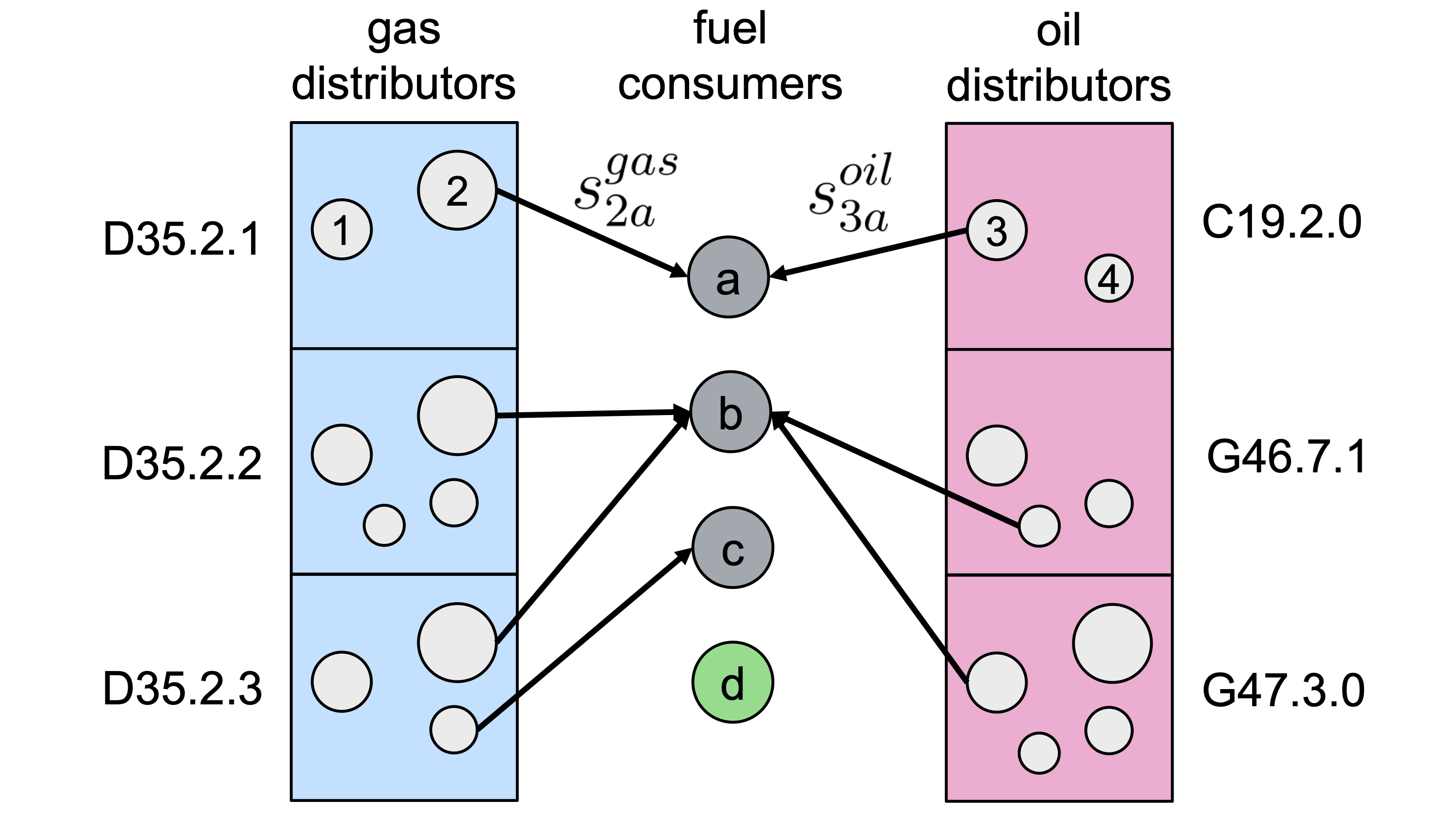}
  	\caption{\textbf{Schematic view of emission estimation method.} Circles in the figure indicate firms. In particular, firms that belong to gas distributing industries D35.2.1, D35.2.2 and D35.2.3 are placed in blue area, while firms that we treat as the oil distributors from C19.2.0, G46.7.1 and G47.3.0 are in the pink area. Firms \textit{a,b,c,d} in the middle of the figure represent the remaining firms - potential consumers of oil and gas. To estimate oil-related emissions, we identify all in-links of non-distributing firms from oil distributors, for example $s^{oil}_{3a}$. We sum over all these in-links of a firm and divide the result by total out-strength of the oil distributors to find a relative oil consumption of each firm with respect to others. Proportionally to these shares we distribute total oil-related emissions $\mathcal{E}^{oil}$ of the commercial sector in the country to firms. In the same way we find gas-related emissions of firms. The final emissions are obtained as the sum of emissions from oil and gas consumption. Thus, firms $a$ and $b$ have oil- as well as gas-related emissions. Firm $c$ has only emissions from gas utilization, while firm $d$ has no estimated emissions.}

	\label{figSI_emissions_method}       

\end{figure*}

\clearpage
\newpage

\section{Details on convergence of the cost pass through mechanism}\label{SI_cost_pass_through}

The convergence of the iteration is ensured by the use of a sub-stochastic matrix. To demonstrate this, we can rewrite the first equation of eq.(\ref{eq:cpt}) as $c(k+1)=[W_w\circ\mu]^T \cdot c(k)$, where the matrix in the brackets is sub-stochastic. A sub-stochastic matrix is defined as one in which the row sums are less than or equal to 1. This property is satisfied here because each row is scaled by 
$\mu_i \in [0,1]$. The spectral radius of a matrix—defined as the largest absolute value of its eigenvalues—determines the convergence behavior. If the spectral radius is less than 1, the iteration converges to the zero vector (note, that the spectral radius is unchanged by transposing the matrix). From the \textit{Gershgorin circle theorem}, we know that the eigenvalues of a square matrix are bounded by the row sums of the matrix. This implies that the eigenvalues of a row sub-stochastic matrix cannot exceed 1. However, for convergence to the zero vector, the spectral radius must be strictly less than 1. The market share $\mu$ is defined at the NACE 4 level, making it highly unlikely that many firms have a market share of exactly 1. In our dataset of 410,523 firms, only 21 firms have $\mu_i=1$, which accounts for less than 0.01\% of the total. Consequently, only 21 rows of the matrix have sums exactly equal to 1. This small subset of rows contributes minimally to the eigenvalues near or on the unit circle. The dominant behavior of the matrix is instead governed by the remaining 410,502 rows, where the row sums are strictly less than 1. These rows induce decay over iterations, ensuring that the spectral radius is less than 1 and driving the convergence to the zero vector.


\end{widetext}


\begin{thebibliography}{65}%
\makeatletter
\providecommand \@ifxundefined [1]{%
 \@ifx{#1\undefined}
}%
\providecommand \@ifnum [1]{%
 \ifnum #1\expandafter \@firstoftwo
 \else \expandafter \@secondoftwo
 \fi
}%
\providecommand \@ifx [1]{%
 \ifx #1\expandafter \@firstoftwo
 \else \expandafter \@secondoftwo
 \fi
}%
\providecommand \natexlab [1]{#1}%
\providecommand \enquote  [1]{``#1''}%
\providecommand \bibnamefont  [1]{#1}%
\providecommand \bibfnamefont [1]{#1}%
\providecommand \citenamefont [1]{#1}%
\providecommand \href@noop [0]{\@secondoftwo}%
\providecommand \href [0]{\begingroup \@sanitize@url \@href}%
\providecommand \@href[1]{\@@startlink{#1}\@@href}%
\providecommand \@@href[1]{\endgroup#1\@@endlink}%
\providecommand \@sanitize@url [0]{\catcode `\\12\catcode `\$12\catcode `\&12\catcode `\#12\catcode `\^12\catcode `\_12\catcode `\%12\relax}%
\providecommand \@@startlink[1]{}%
\providecommand \@@endlink[0]{}%
\providecommand \url  [0]{\begingroup\@sanitize@url \@url }%
\providecommand \@url [1]{\endgroup\@href {#1}{\urlprefix }}%
\providecommand \urlprefix  [0]{URL }%
\providecommand \Eprint [0]{\href }%
\providecommand \doibase [0]{https://doi.org/}%
\providecommand \selectlanguage [0]{\@gobble}%
\providecommand \bibinfo  [0]{\@secondoftwo}%
\providecommand \bibfield  [0]{\@secondoftwo}%
\providecommand \translation [1]{[#1]}%
\providecommand \BibitemOpen [0]{}%
\providecommand \bibitemStop [0]{}%
\providecommand \bibitemNoStop [0]{.\EOS\space}%
\providecommand \EOS [0]{\spacefactor3000\relax}%
\providecommand \BibitemShut  [1]{\csname bibitem#1\endcsname}%
\let\auto@bib@innerbib\@empty
\bibitem [{\citenamefont {{Kyoto Protocol}}(1997)}]{protocol1997kyoto}%
  \BibitemOpen
  \bibfield  {author} {\bibinfo {author} {\bibnamefont {{Kyoto Protocol}}},\ }\bibfield  {title} {\bibinfo {title} {Kyoto protocol},\ }\href@noop {} {\bibfield  {journal} {\bibinfo  {journal} {UNFCCC Website. Available online: https://unfccc.int/resource/docs/convkp/kpeng.pdf (accessed on 9 Sep. 2022)}\ } (\bibinfo {year} {1997})}\BibitemShut {NoStop}%
\bibitem [{\citenamefont {{Paris Agreement}}(2015)}]{agreement2015paris}%
  \BibitemOpen
  \bibfield  {author} {\bibinfo {author} {\bibnamefont {{Paris Agreement}}},\ }\bibfield  {title} {\bibinfo {title} {Paris agreement},\ }in\ \href@noop {} {\emph {\bibinfo {booktitle} {Report of the Conference of the Parties to the United Nations Framework Convention on Climate Change (21st Session, 2015: Paris). Retrived December}}},\ Vol.~\bibinfo {volume} {4}\ (\bibinfo {organization} {HeinOnline},\ \bibinfo {year} {2015})\ p.\ \bibinfo {pages} {2017}\BibitemShut {NoStop}%
\bibitem [{\citenamefont {Mercure}\ \emph {et~al.}(2018)\citenamefont {Mercure}, \citenamefont {Pollitt}, \citenamefont {Vi{\~n}uales}, \citenamefont {Edwards}, \citenamefont {Holden}, \citenamefont {Chewpreecha}, \citenamefont {Salas}, \citenamefont {Sognnaes}, \citenamefont {Lam},\ and\ \citenamefont {Knobloch}}]{mercure2018macroeconomic}%
  \BibitemOpen
  \bibfield  {author} {\bibinfo {author} {\bibfnamefont {J.-F.}\ \bibnamefont {Mercure}}, \bibinfo {author} {\bibfnamefont {H.}~\bibnamefont {Pollitt}}, \bibinfo {author} {\bibfnamefont {J.~E.}\ \bibnamefont {Vi{\~n}uales}}, \bibinfo {author} {\bibfnamefont {N.~R.}\ \bibnamefont {Edwards}}, \bibinfo {author} {\bibfnamefont {P.~B.}\ \bibnamefont {Holden}}, \bibinfo {author} {\bibfnamefont {U.}~\bibnamefont {Chewpreecha}}, \bibinfo {author} {\bibfnamefont {P.}~\bibnamefont {Salas}}, \bibinfo {author} {\bibfnamefont {I.}~\bibnamefont {Sognnaes}}, \bibinfo {author} {\bibfnamefont {A.}~\bibnamefont {Lam}},\ and\ \bibinfo {author} {\bibfnamefont {F.}~\bibnamefont {Knobloch}},\ }\bibfield  {title} {\bibinfo {title} {Macroeconomic impact of stranded fossil fuel assets},\ }\href@noop {} {\bibfield  {journal} {\bibinfo  {journal} {Nature climate change}\ }\textbf {\bibinfo {volume} {8}},\ \bibinfo {pages} {588} (\bibinfo {year} {2018})}\BibitemShut {NoStop}%
\bibitem [{\citenamefont {Lamperti}\ \emph {et~al.}(2019)\citenamefont {Lamperti}, \citenamefont {Bosetti}, \citenamefont {Roventini},\ and\ \citenamefont {Tavoni}}]{lamperti2019public}%
  \BibitemOpen
  \bibfield  {author} {\bibinfo {author} {\bibfnamefont {F.}~\bibnamefont {Lamperti}}, \bibinfo {author} {\bibfnamefont {V.}~\bibnamefont {Bosetti}}, \bibinfo {author} {\bibfnamefont {A.}~\bibnamefont {Roventini}},\ and\ \bibinfo {author} {\bibfnamefont {M.}~\bibnamefont {Tavoni}},\ }\bibfield  {title} {\bibinfo {title} {The public costs of climate-induced financial instability},\ }\href@noop {} {\bibfield  {journal} {\bibinfo  {journal} {Nature Climate Change}\ }\textbf {\bibinfo {volume} {9}},\ \bibinfo {pages} {829} (\bibinfo {year} {2019})}\BibitemShut {NoStop}%
\bibitem [{\citenamefont {on~Climate Change~(IPCC)}(2022)}]{IPCC_2022}%
  \BibitemOpen
  \bibfield  {author} {\bibinfo {author} {\bibfnamefont {I.~P.}\ \bibnamefont {on~Climate Change~(IPCC)}},\ }\bibinfo {title} {Mitigation pathways compatible with 1.5°c in the context of sustainable development},\ in\ \href@noop {} {\emph {\bibinfo {booktitle} {Global Warming of 1.5°C: IPCC Special Report on Impacts of Global Warming of 1.5°C above Pre-industrial Levels in Context of Strengthening Response to Climate Change, Sustainable Development, and Efforts to Eradicate Poverty}}}\ (\bibinfo  {publisher} {Cambridge University Press},\ \bibinfo {year} {2022})\ p.\ \bibinfo {pages} {93–174}\BibitemShut {NoStop}%
\bibitem [{\citenamefont {Bressan}\ \emph {et~al.}(2024)\citenamefont {Bressan}, \citenamefont {{D}uranovi{\'c}}, \citenamefont {Monasterolo},\ and\ \citenamefont {Battiston}}]{bressan2024asset}%
  \BibitemOpen
  \bibfield  {author} {\bibinfo {author} {\bibfnamefont {G.}~\bibnamefont {Bressan}}, \bibinfo {author} {\bibfnamefont {A.}~\bibnamefont {{D}uranovi{\'c}}}, \bibinfo {author} {\bibfnamefont {I.}~\bibnamefont {Monasterolo}},\ and\ \bibinfo {author} {\bibfnamefont {S.}~\bibnamefont {Battiston}},\ }\bibfield  {title} {\bibinfo {title} {Asset-level assessment of climate physical risk matters for adaptation finance},\ }\href@noop {} {\bibfield  {journal} {\bibinfo  {journal} {Nature Communications}\ }\textbf {\bibinfo {volume} {15}},\ \bibinfo {pages} {5371} (\bibinfo {year} {2024})}\BibitemShut {NoStop}%
\bibitem [{\citenamefont {Battiston}\ \emph {et~al.}(2017)\citenamefont {Battiston}, \citenamefont {Mandel}, \citenamefont {Monasterolo}, \citenamefont {Sch{\"u}tze},\ and\ \citenamefont {Visentin}}]{battiston2017climate}%
  \BibitemOpen
  \bibfield  {author} {\bibinfo {author} {\bibfnamefont {S.}~\bibnamefont {Battiston}}, \bibinfo {author} {\bibfnamefont {A.}~\bibnamefont {Mandel}}, \bibinfo {author} {\bibfnamefont {I.}~\bibnamefont {Monasterolo}}, \bibinfo {author} {\bibfnamefont {F.}~\bibnamefont {Sch{\"u}tze}},\ and\ \bibinfo {author} {\bibfnamefont {G.}~\bibnamefont {Visentin}},\ }\bibfield  {title} {\bibinfo {title} {A climate stress-test of the financial system},\ }\href@noop {} {\bibfield  {journal} {\bibinfo  {journal} {Nature Climate Change}\ }\textbf {\bibinfo {volume} {7}},\ \bibinfo {pages} {283} (\bibinfo {year} {2017})}\BibitemShut {NoStop}%
\bibitem [{\citenamefont {Campiglio}\ \emph {et~al.}(2018)\citenamefont {Campiglio}, \citenamefont {Dafermos}, \citenamefont {Monnin}, \citenamefont {Ryan-Collins}, \citenamefont {Schotten},\ and\ \citenamefont {Tanaka}}]{campiglio2018climate}%
  \BibitemOpen
  \bibfield  {author} {\bibinfo {author} {\bibfnamefont {E.}~\bibnamefont {Campiglio}}, \bibinfo {author} {\bibfnamefont {Y.}~\bibnamefont {Dafermos}}, \bibinfo {author} {\bibfnamefont {P.}~\bibnamefont {Monnin}}, \bibinfo {author} {\bibfnamefont {J.}~\bibnamefont {Ryan-Collins}}, \bibinfo {author} {\bibfnamefont {G.}~\bibnamefont {Schotten}},\ and\ \bibinfo {author} {\bibfnamefont {M.}~\bibnamefont {Tanaka}},\ }\bibfield  {title} {\bibinfo {title} {Climate change challenges for central banks and financial regulators},\ }\href@noop {} {\bibfield  {journal} {\bibinfo  {journal} {Nature climate change}\ }\textbf {\bibinfo {volume} {8}},\ \bibinfo {pages} {462} (\bibinfo {year} {2018})}\BibitemShut {NoStop}%
\bibitem [{\citenamefont {Battiston}\ \emph {et~al.}(2021)\citenamefont {Battiston}, \citenamefont {Monasterolo}, \citenamefont {Riahi},\ and\ \citenamefont {van Ruijven}}]{battiston2021accounting}%
  \BibitemOpen
  \bibfield  {author} {\bibinfo {author} {\bibfnamefont {S.}~\bibnamefont {Battiston}}, \bibinfo {author} {\bibfnamefont {I.}~\bibnamefont {Monasterolo}}, \bibinfo {author} {\bibfnamefont {K.}~\bibnamefont {Riahi}},\ and\ \bibinfo {author} {\bibfnamefont {B.~J.}\ \bibnamefont {van Ruijven}},\ }\bibfield  {title} {\bibinfo {title} {Accounting for finance is key for climate mitigation pathways},\ }\href@noop {} {\bibfield  {journal} {\bibinfo  {journal} {Science}\ }\textbf {\bibinfo {volume} {372}},\ \bibinfo {pages} {918} (\bibinfo {year} {2021})}\BibitemShut {NoStop}%
\bibitem [{\citenamefont {Dincer}\ and\ \citenamefont {Rosen}(1999)}]{dincer1999energy}%
  \BibitemOpen
  \bibfield  {author} {\bibinfo {author} {\bibfnamefont {I.}~\bibnamefont {Dincer}}\ and\ \bibinfo {author} {\bibfnamefont {M.~A.}\ \bibnamefont {Rosen}},\ }\bibfield  {title} {\bibinfo {title} {Energy, environment and sustainable development},\ }\href@noop {} {\bibfield  {journal} {\bibinfo  {journal} {Applied energy}\ }\textbf {\bibinfo {volume} {64}},\ \bibinfo {pages} {427} (\bibinfo {year} {1999})}\BibitemShut {NoStop}%
\bibitem [{\citenamefont {{D}irective ({EU})~2003/87/{EC}}(2003)}]{directive2003ETSI}%
  \BibitemOpen
  \bibfield  {author} {\bibinfo {author} {\bibnamefont {{D}irective ({EU})~2003/87/{EC}}},\ }\bibfield  {title} {\bibinfo {title} {of the {E}uropean {P}arliament and of the {C}ouncil of 13 october 2003 establishing a scheme for greenhouse gas emission allowance trading within the {C}ommunity and amending {C}ouncil {D}irective 96/61/{EC}},\ }\href@noop {} {\bibfield  {journal} {\bibinfo  {journal} {Official Journal of the European Union}\ }\textbf {\bibinfo {volume} {L}},\ \bibinfo {pages} {32} (\bibinfo {year} {2003})}\BibitemShut {NoStop}%
\bibitem [{\citenamefont {{D}irective ({EU})~2023/959}(2023)}]{directive2023ETSII}%
  \BibitemOpen
  \bibfield  {author} {\bibinfo {author} {\bibnamefont {{D}irective ({EU})~2023/959}},\ }\bibfield  {title} {\bibinfo {title} {of the {E}uropean {P}arliament and of the {C}ouncil of 10 may 2023 amending {D}irective 2003/87/{EC} establishing a system for greenhouse gas emission allowance trading within the {U}nion and {D}ecision ({EU}) 2015/1814 concerning the establishment and operation of a market stability reserve for the {U}nion greenhouse gas emission trading system},\ }\href@noop {} {\bibfield  {journal} {\bibinfo  {journal} {Official Journal of the European Union}\ }\textbf {\bibinfo {volume} {L}},\ \bibinfo {pages} {134} (\bibinfo {year} {2023})}\BibitemShut {NoStop}%
\bibitem [{\citenamefont {{R}egulation ({EU})~2023/956}(2023)}]{regulation2023CBAM}%
  \BibitemOpen
  \bibfield  {author} {\bibinfo {author} {\bibnamefont {{R}egulation ({EU})~2023/956}},\ }\bibfield  {title} {\bibinfo {title} {of the {E}uropean {P}arliament and of the {C}ouncil of 10 may 2023 establishing a {C}arbon {B}order {A}djustment {M}echanism},\ }\href@noop {} {\bibfield  {journal} {\bibinfo  {journal} {Official Journal of the European Union}\ }\textbf {\bibinfo {volume} {L}},\ \bibinfo {pages} {52} (\bibinfo {year} {2023})}\BibitemShut {NoStop}%
\bibitem [{\citenamefont {{R}egulation ({EU})~2023/955}(2023)}]{regulation2023SCF}%
  \BibitemOpen
  \bibfield  {author} {\bibinfo {author} {\bibnamefont {{R}egulation ({EU})~2023/955}},\ }\bibfield  {title} {\bibinfo {title} {{of the European Parliament and of the Council of 10 May 2023 establishing a Social Climate Fund and amending Regulation (EU) 2021/1060}},\ }\href@noop {} {\bibfield  {journal} {\bibinfo  {journal} {Official Journal of the European Union}\ }\textbf {\bibinfo {volume} {L}},\ \bibinfo {pages} {1} (\bibinfo {year} {2023})}\BibitemShut {NoStop}%
\bibitem [{\citenamefont {Stechemesser}\ \emph {et~al.}(2024)\citenamefont {Stechemesser}, \citenamefont {Koch}, \citenamefont {Mark}, \citenamefont {Dilger}, \citenamefont {Kl{\"o}sel}, \citenamefont {Menicacci}, \citenamefont {Nachtigall}, \citenamefont {Pretis}, \citenamefont {Ritter}, \citenamefont {Schwarz} \emph {et~al.}}]{stechemesser2024climate}%
  \BibitemOpen
  \bibfield  {author} {\bibinfo {author} {\bibfnamefont {A.}~\bibnamefont {Stechemesser}}, \bibinfo {author} {\bibfnamefont {N.}~\bibnamefont {Koch}}, \bibinfo {author} {\bibfnamefont {E.}~\bibnamefont {Mark}}, \bibinfo {author} {\bibfnamefont {E.}~\bibnamefont {Dilger}}, \bibinfo {author} {\bibfnamefont {P.}~\bibnamefont {Kl{\"o}sel}}, \bibinfo {author} {\bibfnamefont {L.}~\bibnamefont {Menicacci}}, \bibinfo {author} {\bibfnamefont {D.}~\bibnamefont {Nachtigall}}, \bibinfo {author} {\bibfnamefont {F.}~\bibnamefont {Pretis}}, \bibinfo {author} {\bibfnamefont {N.}~\bibnamefont {Ritter}}, \bibinfo {author} {\bibfnamefont {M.}~\bibnamefont {Schwarz}}, \emph {et~al.},\ }\bibfield  {title} {\bibinfo {title} {Climate policies that achieved major emission reductions: Global evidence from two decades},\ }\href@noop {} {\bibfield  {journal} {\bibinfo  {journal} {Science}\ }\textbf {\bibinfo {volume} {385}},\ \bibinfo {pages} {884} (\bibinfo {year} {2024})}\BibitemShut {NoStop}%
\bibitem [{\citenamefont {Fierro}\ \emph {et~al.}(2024)\citenamefont {Fierro}, \citenamefont {Reissl}, \citenamefont {Lamperti}, \citenamefont {Campiglio}, \citenamefont {Drouet}, \citenamefont {Emmerling}, \citenamefont {Kremer},\ and\ \citenamefont {Tavoni}}]{fierro2024macro}%
  \BibitemOpen
  \bibfield  {author} {\bibinfo {author} {\bibfnamefont {L.~E.}\ \bibnamefont {Fierro}}, \bibinfo {author} {\bibfnamefont {S.}~\bibnamefont {Reissl}}, \bibinfo {author} {\bibfnamefont {F.}~\bibnamefont {Lamperti}}, \bibinfo {author} {\bibfnamefont {E.}~\bibnamefont {Campiglio}}, \bibinfo {author} {\bibfnamefont {L.}~\bibnamefont {Drouet}}, \bibinfo {author} {\bibfnamefont {J.}~\bibnamefont {Emmerling}}, \bibinfo {author} {\bibfnamefont {E.}~\bibnamefont {Kremer}},\ and\ \bibinfo {author} {\bibfnamefont {M.}~\bibnamefont {Tavoni}},\ }\bibfield  {title} {\bibinfo {title} {Macro-financial transition risks along mitigation pathways: evidence from a hybrid agent-based integrated assessment model},\ }\href@noop {} {\bibfield  {journal} {\bibinfo  {journal} {preprint}\ }\textbf {\bibinfo {volume} {doi.org/10.21203/rs.3.rs-5111841/v1}} (\bibinfo {year} {2024})}\BibitemShut {NoStop}%
\bibitem [{\citenamefont {Acharya}\ \emph {et~al.}(2023)\citenamefont {Acharya}, \citenamefont {Berner}, \citenamefont {Engle}, \citenamefont {Jung}, \citenamefont {Stroebel}, \citenamefont {Zeng},\ and\ \citenamefont {Zhao}}]{acharya2023climate}%
  \BibitemOpen
  \bibfield  {author} {\bibinfo {author} {\bibfnamefont {V.~V.}\ \bibnamefont {Acharya}}, \bibinfo {author} {\bibfnamefont {R.}~\bibnamefont {Berner}}, \bibinfo {author} {\bibfnamefont {R.}~\bibnamefont {Engle}}, \bibinfo {author} {\bibfnamefont {H.}~\bibnamefont {Jung}}, \bibinfo {author} {\bibfnamefont {J.}~\bibnamefont {Stroebel}}, \bibinfo {author} {\bibfnamefont {X.}~\bibnamefont {Zeng}},\ and\ \bibinfo {author} {\bibfnamefont {Y.}~\bibnamefont {Zhao}},\ }\bibfield  {title} {\bibinfo {title} {Climate stress testing},\ }\href@noop {} {\bibfield  {journal} {\bibinfo  {journal} {Annual Review of Financial Economics}\ }\textbf {\bibinfo {volume} {15}},\ \bibinfo {pages} {291} (\bibinfo {year} {2023})}\BibitemShut {NoStop}%
\bibitem [{\citenamefont {Allen}\ \emph {et~al.}(2020)\citenamefont {Allen}, \citenamefont {Dees}, \citenamefont {Caicedo~Graciano}, \citenamefont {Chouard}, \citenamefont {Clerc}, \citenamefont {de~Gaye}, \citenamefont {Devulder}, \citenamefont {Diot}, \citenamefont {Lisack}, \citenamefont {Pegoraro} \emph {et~al.}}]{allen2020climate}%
  \BibitemOpen
  \bibfield  {author} {\bibinfo {author} {\bibfnamefont {T.}~\bibnamefont {Allen}}, \bibinfo {author} {\bibfnamefont {S.}~\bibnamefont {Dees}}, \bibinfo {author} {\bibfnamefont {C.~M.}\ \bibnamefont {Caicedo~Graciano}}, \bibinfo {author} {\bibfnamefont {V.}~\bibnamefont {Chouard}}, \bibinfo {author} {\bibfnamefont {L.}~\bibnamefont {Clerc}}, \bibinfo {author} {\bibfnamefont {A.}~\bibnamefont {de~Gaye}}, \bibinfo {author} {\bibfnamefont {A.}~\bibnamefont {Devulder}}, \bibinfo {author} {\bibfnamefont {S.}~\bibnamefont {Diot}}, \bibinfo {author} {\bibfnamefont {N.}~\bibnamefont {Lisack}}, \bibinfo {author} {\bibfnamefont {F.}~\bibnamefont {Pegoraro}}, \emph {et~al.},\ }\bibfield  {title} {\bibinfo {title} {Climate-related scenarios for financial stability assessment: An application to france},\ }\href@noop {} {\bibfield  {journal} {\bibinfo  {journal} {Banque de France Working Paper}\ } (\bibinfo {year} {2020})}\BibitemShut {NoStop}%
\bibitem [{\citenamefont {Vermeulen}\ \emph {et~al.}(2021)\citenamefont {Vermeulen}, \citenamefont {Schets}, \citenamefont {Lohuis}, \citenamefont {K{\"o}lbl}, \citenamefont {Jansen},\ and\ \citenamefont {Heeringa}}]{vermeulen2021heat}%
  \BibitemOpen
  \bibfield  {author} {\bibinfo {author} {\bibfnamefont {R.}~\bibnamefont {Vermeulen}}, \bibinfo {author} {\bibfnamefont {E.}~\bibnamefont {Schets}}, \bibinfo {author} {\bibfnamefont {M.}~\bibnamefont {Lohuis}}, \bibinfo {author} {\bibfnamefont {B.}~\bibnamefont {K{\"o}lbl}}, \bibinfo {author} {\bibfnamefont {D.-J.}\ \bibnamefont {Jansen}},\ and\ \bibinfo {author} {\bibfnamefont {W.}~\bibnamefont {Heeringa}},\ }\bibfield  {title} {\bibinfo {title} {The heat is on: A framework for measuring financial stress under disruptive energy transition scenarios},\ }\href@noop {} {\bibfield  {journal} {\bibinfo  {journal} {Ecological Economics}\ }\textbf {\bibinfo {volume} {190}},\ \bibinfo {pages} {107205} (\bibinfo {year} {2021})}\BibitemShut {NoStop}%
\bibitem [{\citenamefont {Guth}\ \emph {et~al.}(2021)\citenamefont {Guth}, \citenamefont {Hesse}, \citenamefont {K{\"o}nigswieser}, \citenamefont {Krenn}, \citenamefont {Lipp}, \citenamefont {Neudorfer}, \citenamefont {Schneider}, \citenamefont {Weiss} \emph {et~al.}}]{guth2021oenb}%
  \BibitemOpen
  \bibfield  {author} {\bibinfo {author} {\bibfnamefont {M.}~\bibnamefont {Guth}}, \bibinfo {author} {\bibfnamefont {J.}~\bibnamefont {Hesse}}, \bibinfo {author} {\bibfnamefont {C.}~\bibnamefont {K{\"o}nigswieser}}, \bibinfo {author} {\bibfnamefont {G.}~\bibnamefont {Krenn}}, \bibinfo {author} {\bibfnamefont {C.}~\bibnamefont {Lipp}}, \bibinfo {author} {\bibfnamefont {B.}~\bibnamefont {Neudorfer}}, \bibinfo {author} {\bibfnamefont {M.}~\bibnamefont {Schneider}}, \bibinfo {author} {\bibfnamefont {P.}~\bibnamefont {Weiss}}, \emph {et~al.},\ }\bibfield  {title} {\bibinfo {title} {Oenb climate risk stress test--modeling a carbon price shock for the austrian banking sector},\ }\href@noop {} {\bibfield  {journal} {\bibinfo  {journal} {Financial Stability Report}\ }\textbf {\bibinfo {volume} {42}},\ \bibinfo {pages} {27} (\bibinfo {year} {2021})}\BibitemShut {NoStop}%
\bibitem [{\citenamefont {Roncoroni}\ \emph {et~al.}(2021)\citenamefont {Roncoroni}, \citenamefont {Battiston}, \citenamefont {Escobar-Farf{\'a}n},\ and\ \citenamefont {Martinez-Jaramillo}}]{roncoroni2021climate}%
  \BibitemOpen
  \bibfield  {author} {\bibinfo {author} {\bibfnamefont {A.}~\bibnamefont {Roncoroni}}, \bibinfo {author} {\bibfnamefont {S.}~\bibnamefont {Battiston}}, \bibinfo {author} {\bibfnamefont {L.~O.}\ \bibnamefont {Escobar-Farf{\'a}n}},\ and\ \bibinfo {author} {\bibfnamefont {S.}~\bibnamefont {Martinez-Jaramillo}},\ }\bibfield  {title} {\bibinfo {title} {Climate risk and financial stability in the network of banks and investment funds},\ }\href@noop {} {\bibfield  {journal} {\bibinfo  {journal} {Journal of Financial Stability}\ }\textbf {\bibinfo {volume} {54}},\ \bibinfo {pages} {100870} (\bibinfo {year} {2021})}\BibitemShut {NoStop}%
\bibitem [{\citenamefont {Sever}\ and\ \citenamefont {Perez-Archila}(2021)}]{sever2021climate}%
  \BibitemOpen
  \bibfield  {author} {\bibinfo {author} {\bibfnamefont {C.}~\bibnamefont {Sever}}\ and\ \bibinfo {author} {\bibfnamefont {M.}~\bibnamefont {Perez-Archila}},\ }\href@noop {} {\emph {\bibinfo {title} {Climate-Related Stress Testing: Transition Risk in Colombia}}}\ (\bibinfo  {publisher} {International Monetary Fund},\ \bibinfo {year} {2021})\BibitemShut {NoStop}%
\bibitem [{\citenamefont {L{\"o}schenbrand}\ \emph {et~al.}(2024)\citenamefont {L{\"o}schenbrand}, \citenamefont {Maier}, \citenamefont {Millischer},\ and\ \citenamefont {Resch}}]{loschenbrand2024credit}%
  \BibitemOpen
  \bibfield  {author} {\bibinfo {author} {\bibfnamefont {S.}~\bibnamefont {L{\"o}schenbrand}}, \bibinfo {author} {\bibfnamefont {M.}~\bibnamefont {Maier}}, \bibinfo {author} {\bibfnamefont {L.}~\bibnamefont {Millischer}},\ and\ \bibinfo {author} {\bibfnamefont {F.}~\bibnamefont {Resch}},\ }\bibfield  {title} {\bibinfo {title} {Credit risk where it’s due: Carbon pricing and firm defaults},\ }\href@noop {} {\bibfield  {journal} {\bibinfo  {journal} {Available at SSRN 4572907}\ } (\bibinfo {year} {2024})}\BibitemShut {NoStop}%
\bibitem [{\citenamefont {Alogoskoufis}\ \emph {et~al.}(2021)\citenamefont {Alogoskoufis}, \citenamefont {Dunz}, \citenamefont {Emambakhsh}, \citenamefont {Hennig}, \citenamefont {Kaijser}, \citenamefont {Kouratzoglou}, \citenamefont {Mu{\~n}oz}, \citenamefont {Parisi},\ and\ \citenamefont {Salleo}}]{alogoskoufis2021ecb}%
  \BibitemOpen
  \bibfield  {author} {\bibinfo {author} {\bibfnamefont {S.}~\bibnamefont {Alogoskoufis}}, \bibinfo {author} {\bibfnamefont {N.}~\bibnamefont {Dunz}}, \bibinfo {author} {\bibfnamefont {T.}~\bibnamefont {Emambakhsh}}, \bibinfo {author} {\bibfnamefont {T.}~\bibnamefont {Hennig}}, \bibinfo {author} {\bibfnamefont {M.}~\bibnamefont {Kaijser}}, \bibinfo {author} {\bibfnamefont {C.}~\bibnamefont {Kouratzoglou}}, \bibinfo {author} {\bibfnamefont {M.~A.}\ \bibnamefont {Mu{\~n}oz}}, \bibinfo {author} {\bibfnamefont {L.}~\bibnamefont {Parisi}},\ and\ \bibinfo {author} {\bibfnamefont {C.}~\bibnamefont {Salleo}},\ }\href@noop {} {\emph {\bibinfo {title} {ECB economy-wide climate stress test: Methodology and results}}},\ \bibinfo {number} {281}\ (\bibinfo  {publisher} {ECB Occasional Paper},\ \bibinfo {year} {2021})\BibitemShut {NoStop}%
\bibitem [{\citenamefont {Hallegatte}\ \emph {et~al.}(2022)\citenamefont {Hallegatte}, \citenamefont {Lipinsky}, \citenamefont {Morales}, \citenamefont {Oura}, \citenamefont {Ranger}, \citenamefont {Regelink},\ and\ \citenamefont {Reinders}}]{hallegatte2022bank}%
  \BibitemOpen
  \bibfield  {author} {\bibinfo {author} {\bibfnamefont {S.}~\bibnamefont {Hallegatte}}, \bibinfo {author} {\bibfnamefont {M.~F.}\ \bibnamefont {Lipinsky}}, \bibinfo {author} {\bibfnamefont {P.}~\bibnamefont {Morales}}, \bibinfo {author} {\bibfnamefont {M.~H.}\ \bibnamefont {Oura}}, \bibinfo {author} {\bibfnamefont {N.}~\bibnamefont {Ranger}}, \bibinfo {author} {\bibfnamefont {M.~G.~J.}\ \bibnamefont {Regelink}},\ and\ \bibinfo {author} {\bibfnamefont {H.~J.}\ \bibnamefont {Reinders}},\ }\href@noop {} {\emph {\bibinfo {title} {Bank stress testing of physical risks under climate change macro scenarios: typhoon risks to the Philippines}}}\ (\bibinfo  {publisher} {International Monetary Fund},\ \bibinfo {year} {2022})\BibitemShut {NoStop}%
\bibitem [{\citenamefont {Lepore}\ and\ \citenamefont {Fernando}(2023)}]{LeporeFernando2023}%
  \BibitemOpen
  \bibfield  {author} {\bibinfo {author} {\bibfnamefont {C.}~\bibnamefont {Lepore}}\ and\ \bibinfo {author} {\bibfnamefont {R.}~\bibnamefont {Fernando}},\ }\href@noop {} {\emph {\bibinfo {title} {Global Economic Impacts of Physical Climate Risks}}},\ \bibinfo {type} {Tech. Rep.}\ (\bibinfo {year} {2023})\ \bibinfo {note} {volume and DOI not provided}\BibitemShut {NoStop}%
\bibitem [{\citenamefont {{European Environmental Agency}}(2024)}]{EEA2024}%
  \BibitemOpen
  \bibfield  {author} {\bibinfo {author} {\bibnamefont {{European Environmental Agency}}},\ }\href {https://doi.org/10.2800/8671471} {\emph {\bibinfo {title} {European Climate Risk Assessment}}},\ EEA Report 01/2024\ (\bibinfo  {publisher} {Publications Office of the European Union},\ \bibinfo {address} {Luxembourg},\ \bibinfo {year} {2024})\BibitemShut {NoStop}%
\bibitem [{\citenamefont {{World Bank Group}}(nd)}]{WorldBankClimatePortal}%
  \BibitemOpen
  \bibfield  {author} {\bibinfo {author} {\bibnamefont {{World Bank Group}}},\ }\href@noop {} {\bibinfo {title} {Climate knowledge portal: Country profiles}},\ \bibinfo {howpublished} {\url{https://climateknowledgeportal.worldbank.org/country-profiles}} (\bibinfo {year} {n.d.}),\ \bibinfo {note} {accessed: 2024-12-01}\BibitemShut {NoStop}%
\bibitem [{\citenamefont {{Basel Committee on Banking Supervision}}(2021{\natexlab{a}})}]{baselClimateMeth}%
  \BibitemOpen
  \bibfield  {author} {\bibinfo {author} {\bibnamefont {{Basel Committee on Banking Supervision}}},\ }\href@noop {} {\bibinfo {title} {{{C}limate-related financial risks – measurement methodologies.}}} (\bibinfo {year} {{2021}}{\natexlab{a}})\BibitemShut {NoStop}%
\bibitem [{\citenamefont {{Basel Committee on Banking Supervision}}(2021{\natexlab{b}})}]{baselClimateRiskTransmission}%
  \BibitemOpen
  \bibfield  {author} {\bibinfo {author} {\bibnamefont {{Basel Committee on Banking Supervision}}},\ }\href@noop {} {\bibinfo {title} {{{C}limate-related risk drivers and their transmission channels.}}} (\bibinfo {year} {{2021}}{\natexlab{b}})\BibitemShut {NoStop}%
\bibitem [{\citenamefont {{Basel Committee on Banking Supervision}}(2022)}]{baselClimatePrinciples}%
  \BibitemOpen
  \bibfield  {author} {\bibinfo {author} {\bibnamefont {{Basel Committee on Banking Supervision}}},\ }\href@noop {} {\bibinfo {title} {{{P}rinciples for the effective management and supervision of climate-related financial risks.}}} (\bibinfo {year} {{2022}})\BibitemShut {NoStop}%
\bibitem [{\citenamefont {Duprez}\ and\ \citenamefont {Magerman}(2018)}]{duprez2018price}%
  \BibitemOpen
  \bibfield  {author} {\bibinfo {author} {\bibfnamefont {C.}~\bibnamefont {Duprez}}\ and\ \bibinfo {author} {\bibfnamefont {G.}~\bibnamefont {Magerman}},\ }\href@noop {} {\emph {\bibinfo {title} {Price updating in production networks}}},\ \bibinfo {type} {Tech. Rep.}\ (\bibinfo  {institution} {NBB Working Paper},\ \bibinfo {year} {2018})\BibitemShut {NoStop}%
\bibitem [{\citenamefont {NGFS}(2024)}]{NGFS_emission_data}%
  \BibitemOpen
  \bibfield  {author} {\bibinfo {author} {\bibnamefont {NGFS}},\ }\href@noop {} {\bibinfo {title} {{Improving Greenhouse Gas Emissions Data}}} (\bibinfo {year} {2024})\BibitemShut {NoStop}%
\bibitem [{\citenamefont {{European Parliament and Council of the European Union}}(2022)}]{EU2022CSRD}%
  \BibitemOpen
  \bibfield  {author} {\bibinfo {author} {\bibnamefont {{European Parliament and Council of the European Union}}},\ }\href {https:/p/eur-lex.europa.eu/legal-content/EN/TXT/?uri=CELEX%3A32022L2464} {\bibinfo {title} {Directive ({EU}) 2022/2464 of the european parliament and of the council of 14 december 2022 on corporate sustainability reporting}} (\bibinfo {year} {2022}),\ \bibinfo {note} {official Journal of the European Union, L 322, 16.12.2022, pp. 15-42}\BibitemShut {NoStop}%
\bibitem [{\citenamefont {Battiston}\ \emph {et~al.}(2020)\citenamefont {Battiston}, \citenamefont {Guth}, \citenamefont {Monasterolo}, \citenamefont {Neudorfer}, \citenamefont {Pointner} \emph {et~al.}}]{battiston2020austrian}%
  \BibitemOpen
  \bibfield  {author} {\bibinfo {author} {\bibfnamefont {S.}~\bibnamefont {Battiston}}, \bibinfo {author} {\bibfnamefont {M.}~\bibnamefont {Guth}}, \bibinfo {author} {\bibfnamefont {I.}~\bibnamefont {Monasterolo}}, \bibinfo {author} {\bibfnamefont {B.}~\bibnamefont {Neudorfer}}, \bibinfo {author} {\bibfnamefont {W.}~\bibnamefont {Pointner}}, \emph {et~al.},\ }\bibfield  {title} {\bibinfo {title} {Austrian banks’ exposure to climate-related transition risk},\ }\href@noop {} {\bibfield  {journal} {\bibinfo  {journal} {Austrian National Bank Financial Stability Report}\ }\textbf {\bibinfo {volume} {40}},\ \bibinfo {pages} {31} (\bibinfo {year} {2020})}\BibitemShut {NoStop}%
\bibitem [{\citenamefont {Kosztowniak}(2023)}]{kosztowniak2023climate}%
  \BibitemOpen
  \bibfield  {author} {\bibinfo {author} {\bibfnamefont {A.}~\bibnamefont {Kosztowniak}},\ }\bibfield  {title} {\bibinfo {title} {Climate policy relevant sectors in the polish commercial banks},\ }\href@noop {} {\bibfield  {journal} {\bibinfo  {journal} {Central European Review of Economics \& Finance}\ }\textbf {\bibinfo {volume} {42}},\ \bibinfo {pages} {50} (\bibinfo {year} {2023})}\BibitemShut {NoStop}%
\bibitem [{\citenamefont {Battiston}\ \emph {et~al.}(2023)\citenamefont {Battiston}, \citenamefont {Mandel}, \citenamefont {Monasterolo},\ and\ \citenamefont {Roncoroni}}]{battiston2023climate}%
  \BibitemOpen
  \bibfield  {author} {\bibinfo {author} {\bibfnamefont {S.}~\bibnamefont {Battiston}}, \bibinfo {author} {\bibfnamefont {A.}~\bibnamefont {Mandel}}, \bibinfo {author} {\bibfnamefont {I.}~\bibnamefont {Monasterolo}},\ and\ \bibinfo {author} {\bibfnamefont {A.}~\bibnamefont {Roncoroni}},\ }\bibfield  {title} {\bibinfo {title} {Climate credit risk and corporate valuation},\ }\href@noop {} {\bibfield  {journal} {\bibinfo  {journal} {Available at SSRN 4124002}\ } (\bibinfo {year} {2023})}\BibitemShut {NoStop}%
\bibitem [{\citenamefont {Diem}\ \emph {et~al.}(2024{\natexlab{a}})\citenamefont {Diem}, \citenamefont {Borsos}, \citenamefont {Reisch}, \citenamefont {Kert{\'e}sz},\ and\ \citenamefont {Thurner}}]{diem2024estimating}%
  \BibitemOpen
  \bibfield  {author} {\bibinfo {author} {\bibfnamefont {C.}~\bibnamefont {Diem}}, \bibinfo {author} {\bibfnamefont {A.}~\bibnamefont {Borsos}}, \bibinfo {author} {\bibfnamefont {T.}~\bibnamefont {Reisch}}, \bibinfo {author} {\bibfnamefont {J.}~\bibnamefont {Kert{\'e}sz}},\ and\ \bibinfo {author} {\bibfnamefont {S.}~\bibnamefont {Thurner}},\ }\bibfield  {title} {\bibinfo {title} {Estimating the loss of economic predictability from aggregating firm-level production networks},\ }\href@noop {} {\bibfield  {journal} {\bibinfo  {journal} {PNAS nexus}\ }\textbf {\bibinfo {volume} {3}},\ \bibinfo {pages} {pgae064} (\bibinfo {year} {2024}{\natexlab{a}})}\BibitemShut {NoStop}%
\bibitem [{\citenamefont {Barrot}\ and\ \citenamefont {Sauvagnat}(2016)}]{barrot2016input}%
  \BibitemOpen
  \bibfield  {author} {\bibinfo {author} {\bibfnamefont {J.-N.}\ \bibnamefont {Barrot}}\ and\ \bibinfo {author} {\bibfnamefont {J.}~\bibnamefont {Sauvagnat}},\ }\bibfield  {title} {\bibinfo {title} {Input specificity and the propagation of idiosyncratic shocks in production networks},\ }\href@noop {} {\bibfield  {journal} {\bibinfo  {journal} {The Quarterly Journal of Economics}\ }\textbf {\bibinfo {volume} {131}},\ \bibinfo {pages} {1543} (\bibinfo {year} {2016})}\BibitemShut {NoStop}%
\bibitem [{\citenamefont {Inoue}\ and\ \citenamefont {Todo}(2019)}]{inoue2019firm}%
  \BibitemOpen
  \bibfield  {author} {\bibinfo {author} {\bibfnamefont {H.}~\bibnamefont {Inoue}}\ and\ \bibinfo {author} {\bibfnamefont {Y.}~\bibnamefont {Todo}},\ }\bibfield  {title} {\bibinfo {title} {Firm-level propagation of shocks through supply-chain networks},\ }\href@noop {} {\bibfield  {journal} {\bibinfo  {journal} {Nature Sustainability}\ }\textbf {\bibinfo {volume} {2}},\ \bibinfo {pages} {841} (\bibinfo {year} {2019})}\BibitemShut {NoStop}%
\bibitem [{\citenamefont {Carvalho}\ \emph {et~al.}(2021)\citenamefont {Carvalho}, \citenamefont {Nirei}, \citenamefont {Saito},\ and\ \citenamefont {Tahbaz-Salehi}}]{carvalho2021supply}%
  \BibitemOpen
  \bibfield  {author} {\bibinfo {author} {\bibfnamefont {V.~M.}\ \bibnamefont {Carvalho}}, \bibinfo {author} {\bibfnamefont {M.}~\bibnamefont {Nirei}}, \bibinfo {author} {\bibfnamefont {Y.~U.}\ \bibnamefont {Saito}},\ and\ \bibinfo {author} {\bibfnamefont {A.}~\bibnamefont {Tahbaz-Salehi}},\ }\bibfield  {title} {\bibinfo {title} {Supply chain disruptions: Evidence from the great east japan earthquake},\ }\href@noop {} {\bibfield  {journal} {\bibinfo  {journal} {The Quarterly Journal of Economics}\ }\textbf {\bibinfo {volume} {136}},\ \bibinfo {pages} {1255} (\bibinfo {year} {2021})}\BibitemShut {NoStop}%
\bibitem [{\citenamefont {Diem}\ \emph {et~al.}(2022)\citenamefont {Diem}, \citenamefont {Borsos}, \citenamefont {Reisch}, \citenamefont {Kert{\'e}sz},\ and\ \citenamefont {Thurner}}]{diem2022quantifying}%
  \BibitemOpen
  \bibfield  {author} {\bibinfo {author} {\bibfnamefont {C.}~\bibnamefont {Diem}}, \bibinfo {author} {\bibfnamefont {A.}~\bibnamefont {Borsos}}, \bibinfo {author} {\bibfnamefont {T.}~\bibnamefont {Reisch}}, \bibinfo {author} {\bibfnamefont {J.}~\bibnamefont {Kert{\'e}sz}},\ and\ \bibinfo {author} {\bibfnamefont {S.}~\bibnamefont {Thurner}},\ }\bibfield  {title} {\bibinfo {title} {Quantifying firm-level economic systemic risk from nation-wide supply networks},\ }\href {https://doi.org/10.1038/s41598-022-11522-z} {\bibfield  {journal} {\bibinfo  {journal} {Scientific Reports}\ }\textbf {\bibinfo {volume} {12}},\ \bibinfo {pages} {7719} (\bibinfo {year} {2022})}\BibitemShut {NoStop}%
\bibitem [{\citenamefont {Pichler}\ \emph {et~al.}(2023{\natexlab{a}})\citenamefont {Pichler}, \citenamefont {Diem}, \citenamefont {Brintrup}, \citenamefont {Lafond}, \citenamefont {Magerman}, \citenamefont {Buiten}, \citenamefont {Choi}, \citenamefont {Carvalho}, \citenamefont {Farmer},\ and\ \citenamefont {Thurner}}]{pichler2023building}%
  \BibitemOpen
  \bibfield  {author} {\bibinfo {author} {\bibfnamefont {A.}~\bibnamefont {Pichler}}, \bibinfo {author} {\bibfnamefont {C.}~\bibnamefont {Diem}}, \bibinfo {author} {\bibfnamefont {A.}~\bibnamefont {Brintrup}}, \bibinfo {author} {\bibfnamefont {F.}~\bibnamefont {Lafond}}, \bibinfo {author} {\bibfnamefont {G.}~\bibnamefont {Magerman}}, \bibinfo {author} {\bibfnamefont {G.}~\bibnamefont {Buiten}}, \bibinfo {author} {\bibfnamefont {T.~Y.}\ \bibnamefont {Choi}}, \bibinfo {author} {\bibfnamefont {V.~M.}\ \bibnamefont {Carvalho}}, \bibinfo {author} {\bibfnamefont {J.~D.}\ \bibnamefont {Farmer}},\ and\ \bibinfo {author} {\bibfnamefont {S.}~\bibnamefont {Thurner}},\ }\bibfield  {title} {\bibinfo {title} {Building an alliance to map global supply networks},\ }\href@noop {} {\bibfield  {journal} {\bibinfo  {journal} {Science}\ }\textbf {\bibinfo {volume} {382}},\ \bibinfo {pages} {270} (\bibinfo {year} {2023}{\natexlab{a}})}\BibitemShut {NoStop}%
\bibitem [{\citenamefont {Tabachov{\'a}}\ \emph {et~al.}(2024)\citenamefont {Tabachov{\'a}}, \citenamefont {Diem}, \citenamefont {Borsos}, \citenamefont {Burger},\ and\ \citenamefont {Thurner}}]{tabachova2024estimating}%
  \BibitemOpen
  \bibfield  {author} {\bibinfo {author} {\bibfnamefont {Z.}~\bibnamefont {Tabachov{\'a}}}, \bibinfo {author} {\bibfnamefont {C.}~\bibnamefont {Diem}}, \bibinfo {author} {\bibfnamefont {A.}~\bibnamefont {Borsos}}, \bibinfo {author} {\bibfnamefont {C.}~\bibnamefont {Burger}},\ and\ \bibinfo {author} {\bibfnamefont {S.}~\bibnamefont {Thurner}},\ }\bibfield  {title} {\bibinfo {title} {Estimating the impact of supply chain network contagion on financial stability},\ }\href@noop {} {\bibfield  {journal} {\bibinfo  {journal} {Journal of Financial Stability}\ ,\ \bibinfo {pages} {101336}} (\bibinfo {year} {2024})}\BibitemShut {NoStop}%
\bibitem [{\citenamefont {Reisch}\ \emph {et~al.}(2022)\citenamefont {Reisch}, \citenamefont {Heiler}, \citenamefont {Diem}, \citenamefont {Klimek},\ and\ \citenamefont {Thurner}}]{reisch2022monitoring}%
  \BibitemOpen
  \bibfield  {author} {\bibinfo {author} {\bibfnamefont {T.}~\bibnamefont {Reisch}}, \bibinfo {author} {\bibfnamefont {G.}~\bibnamefont {Heiler}}, \bibinfo {author} {\bibfnamefont {C.}~\bibnamefont {Diem}}, \bibinfo {author} {\bibfnamefont {P.}~\bibnamefont {Klimek}},\ and\ \bibinfo {author} {\bibfnamefont {S.}~\bibnamefont {Thurner}},\ }\bibfield  {title} {\bibinfo {title} {Monitoring supply networks from mobile phone data for estimating the systemic risk of an economy},\ }\href@noop {} {\bibfield  {journal} {\bibinfo  {journal} {Scientific reports}\ }\textbf {\bibinfo {volume} {12}},\ \bibinfo {pages} {13347} (\bibinfo {year} {2022})}\BibitemShut {NoStop}%
\bibitem [{\citenamefont {Stangl}\ \emph {et~al.}(2024)\citenamefont {Stangl}, \citenamefont {Borsos}, \citenamefont {Diem}, \citenamefont {Reisch},\ and\ \citenamefont {Thurner}}]{stangl2024firm}%
  \BibitemOpen
  \bibfield  {author} {\bibinfo {author} {\bibfnamefont {J.}~\bibnamefont {Stangl}}, \bibinfo {author} {\bibfnamefont {A.}~\bibnamefont {Borsos}}, \bibinfo {author} {\bibfnamefont {C.}~\bibnamefont {Diem}}, \bibinfo {author} {\bibfnamefont {T.}~\bibnamefont {Reisch}},\ and\ \bibinfo {author} {\bibfnamefont {S.}~\bibnamefont {Thurner}},\ }\bibfield  {title} {\bibinfo {title} {Firm-level supply chains to minimize unemployment and economic losses in rapid decarbonization scenarios},\ }\href@noop {} {\bibfield  {journal} {\bibinfo  {journal} {Nature Sustainability}\ ,\ \bibinfo {pages} {1}} (\bibinfo {year} {2024})}\BibitemShut {NoStop}%
\bibitem [{\citenamefont {Diem}\ \emph {et~al.}(2024{\natexlab{b}})\citenamefont {Diem}, \citenamefont {Schueller}, \citenamefont {Gerschberger}, \citenamefont {Stangl}, \citenamefont {Conrady}, \citenamefont {Gerschberger},\ and\ \citenamefont {Thurner}}]{diem2024supply}%
  \BibitemOpen
  \bibfield  {author} {\bibinfo {author} {\bibfnamefont {C.}~\bibnamefont {Diem}}, \bibinfo {author} {\bibfnamefont {W.}~\bibnamefont {Schueller}}, \bibinfo {author} {\bibfnamefont {M.}~\bibnamefont {Gerschberger}}, \bibinfo {author} {\bibfnamefont {J.}~\bibnamefont {Stangl}}, \bibinfo {author} {\bibfnamefont {B.}~\bibnamefont {Conrady}}, \bibinfo {author} {\bibfnamefont {M.}~\bibnamefont {Gerschberger}},\ and\ \bibinfo {author} {\bibfnamefont {S.}~\bibnamefont {Thurner}},\ }\bibfield  {title} {\bibinfo {title} {Supply network stress-testing of food security on the establishment-level},\ }\href@noop {} {\bibfield  {journal} {\bibinfo  {journal} {International Journal of Production Research}\ ,\ \bibinfo {pages} {1}} (\bibinfo {year} {2024}{\natexlab{b}})}\BibitemShut {NoStop}%
\bibitem [{\citenamefont {EuropeanComission}(2024)}]{ClimateActionEU}%
  \BibitemOpen
  \bibfield  {author} {\bibinfo {author} {\bibnamefont {EuropeanComission}},\ }\href {https://ec.europa.eu/clima/ets/napYearInformation.do?registryCode=HU&periodCode=3&periodYear=2021&languageCode=en&registryCodeLookup=Hungary&installationAllowance=9123288&periodCodeLookup=Phase+4+%282021-2030%29&periodYear=2021&napInstallation.installationIdentifier=&napInstallation.installationName=&currentSortSettings=&backList=%3CBack&resultList.currentPageNumber=2} {\bibinfo {title} {{Climate Action}}} (\bibinfo {year} {2024}),\ \bibinfo {note} {accessed: 2024-03-10}\BibitemShut {NoStop}%
\bibitem [{\citenamefont {Budget}(2023)}]{budget2023global}%
  \BibitemOpen
  \bibfield  {author} {\bibinfo {author} {\bibfnamefont {G.~C.}\ \bibnamefont {Budget}},\ }\bibfield  {title} {\bibinfo {title} {Global carbon budget 2023}\ }\href {https://doi.org/https://doi.org/10.5194/essd-15-5301-2023} {https://doi.org/10.5194/essd-15-5301-2023} (\bibinfo {year} {2023})\BibitemShut {NoStop}%
\bibitem [{\citenamefont {Battiston}\ \emph {et~al.}(2022)\citenamefont {Battiston}, \citenamefont {Monasterolo}, \citenamefont {van Ruijven},\ and\ \citenamefont {Krey}}]{battiston2022nace}%
  \BibitemOpen
  \bibfield  {author} {\bibinfo {author} {\bibfnamefont {S.}~\bibnamefont {Battiston}}, \bibinfo {author} {\bibfnamefont {I.}~\bibnamefont {Monasterolo}}, \bibinfo {author} {\bibfnamefont {B.}~\bibnamefont {van Ruijven}},\ and\ \bibinfo {author} {\bibfnamefont {V.}~\bibnamefont {Krey}},\ }\bibfield  {title} {\bibinfo {title} {The nace--cprs--iam mapping: A tool to support climate risk analysis of financial portfolio using ngfs scenarios.},\ }\href@noop {} {\bibfield  {journal} {\bibinfo  {journal} {Available at SSRN 4223606}\ } (\bibinfo {year} {2022})}\BibitemShut {NoStop}%
\bibitem [{\citenamefont {Hallegatte}(2008)}]{hallegatte2008adaptive}%
  \BibitemOpen
  \bibfield  {author} {\bibinfo {author} {\bibfnamefont {S.}~\bibnamefont {Hallegatte}},\ }\bibfield  {title} {\bibinfo {title} {An adaptive regional input-output model and its application to the assessment of the economic cost of katrina},\ }\href@noop {} {\bibfield  {journal} {\bibinfo  {journal} {Risk Analysis: An International Journal}\ }\textbf {\bibinfo {volume} {28}},\ \bibinfo {pages} {779} (\bibinfo {year} {2008})}\BibitemShut {NoStop}%
\bibitem [{\citenamefont {Boehm}\ \emph {et~al.}(2019)\citenamefont {Boehm}, \citenamefont {Flaaen},\ and\ \citenamefont {Pandalai-Nayar}}]{boehm2019input}%
  \BibitemOpen
  \bibfield  {author} {\bibinfo {author} {\bibfnamefont {C.~E.}\ \bibnamefont {Boehm}}, \bibinfo {author} {\bibfnamefont {A.}~\bibnamefont {Flaaen}},\ and\ \bibinfo {author} {\bibfnamefont {N.}~\bibnamefont {Pandalai-Nayar}},\ }\bibfield  {title} {\bibinfo {title} {Input linkages and the transmission of shocks: Firm-level evidence from the 2011 t{\=o}hoku earthquake},\ }\href@noop {} {\bibfield  {journal} {\bibinfo  {journal} {Review of Economics and Statistics}\ }\textbf {\bibinfo {volume} {101}},\ \bibinfo {pages} {60} (\bibinfo {year} {2019})}\BibitemShut {NoStop}%
\bibitem [{\citenamefont {Pichler}\ \emph {et~al.}(2020)\citenamefont {Pichler}, \citenamefont {Pangallo}, \citenamefont {del Rio-Chanona}, \citenamefont {Lafond},\ and\ \citenamefont {Farmer}}]{pichler2020production}%
  \BibitemOpen
  \bibfield  {author} {\bibinfo {author} {\bibfnamefont {A.}~\bibnamefont {Pichler}}, \bibinfo {author} {\bibfnamefont {M.}~\bibnamefont {Pangallo}}, \bibinfo {author} {\bibfnamefont {R.~M.}\ \bibnamefont {del Rio-Chanona}}, \bibinfo {author} {\bibfnamefont {F.}~\bibnamefont {Lafond}},\ and\ \bibinfo {author} {\bibfnamefont {J.~D.}\ \bibnamefont {Farmer}},\ }\bibfield  {title} {\bibinfo {title} {Production networks and epidemic spreading: How to restart the uk economy?},\ }\href@noop {} {\bibfield  {journal} {\bibinfo  {journal} {arXiv preprint arXiv:2005.10585}\ } (\bibinfo {year} {2020})}\BibitemShut {NoStop}%
\bibitem [{\citenamefont {Battiston}\ and\ \citenamefont {Monasterolo}(2024)}]{battiston2024enhanced}%
  \BibitemOpen
  \bibfield  {author} {\bibinfo {author} {\bibfnamefont {S.}~\bibnamefont {Battiston}}\ and\ \bibinfo {author} {\bibfnamefont {I.}~\bibnamefont {Monasterolo}},\ }\bibfield  {title} {\bibinfo {title} {The {INSPPIRE} {S}ustainable {C}entral {B}anking {T}oolbox - {E}nhanced scenarios for climate stress-tests},\ }\href@noop {} {\bibfield  {journal} {\bibinfo  {journal} {{P}olicy {B}riefing {P}aper 16}\ } (\bibinfo {year} {2024})}\BibitemShut {NoStop}%
\bibitem [{\citenamefont {Fankhauser}\ \emph {et~al.}(2022)\citenamefont {Fankhauser}, \citenamefont {Smith}, \citenamefont {Allen}, \citenamefont {Axelsson}, \citenamefont {Hale}, \citenamefont {Hepburn}, \citenamefont {Kendall}, \citenamefont {Khosla}, \citenamefont {Lezaun}, \citenamefont {Mitchell-Larson} \emph {et~al.}}]{fankhauser2022meaning}%
  \BibitemOpen
  \bibfield  {author} {\bibinfo {author} {\bibfnamefont {S.}~\bibnamefont {Fankhauser}}, \bibinfo {author} {\bibfnamefont {S.~M.}\ \bibnamefont {Smith}}, \bibinfo {author} {\bibfnamefont {M.}~\bibnamefont {Allen}}, \bibinfo {author} {\bibfnamefont {K.}~\bibnamefont {Axelsson}}, \bibinfo {author} {\bibfnamefont {T.}~\bibnamefont {Hale}}, \bibinfo {author} {\bibfnamefont {C.}~\bibnamefont {Hepburn}}, \bibinfo {author} {\bibfnamefont {J.~M.}\ \bibnamefont {Kendall}}, \bibinfo {author} {\bibfnamefont {R.}~\bibnamefont {Khosla}}, \bibinfo {author} {\bibfnamefont {J.}~\bibnamefont {Lezaun}}, \bibinfo {author} {\bibfnamefont {E.}~\bibnamefont {Mitchell-Larson}}, \emph {et~al.},\ }\bibfield  {title} {\bibinfo {title} {The meaning of net zero and how to get it right},\ }\href@noop {} {\bibfield  {journal} {\bibinfo  {journal} {Nature Climate Change}\ }\textbf {\bibinfo {volume} {12}},\ \bibinfo {pages} {15} (\bibinfo {year} {2022})}\BibitemShut {NoStop}%
\bibitem [{\citenamefont {Pichler}\ \emph {et~al.}(2023{\natexlab{b}})\citenamefont {Pichler}, \citenamefont {del Rio-Chanona}, \citenamefont {Farmer}, \citenamefont {Ives},\ and\ \citenamefont {B{\"u}cker}}]{bddotnucker2023employment}%
  \BibitemOpen
  \bibfield  {author} {\bibinfo {author} {\bibfnamefont {A.}~\bibnamefont {Pichler}}, \bibinfo {author} {\bibfnamefont {R.~M.}\ \bibnamefont {del Rio-Chanona}}, \bibinfo {author} {\bibfnamefont {J.~D.}\ \bibnamefont {Farmer}}, \bibinfo {author} {\bibfnamefont {M.}~\bibnamefont {Ives}},\ and\ \bibinfo {author} {\bibfnamefont {J.}~\bibnamefont {B{\"u}cker}},\ }\href@noop {} {\emph {\bibinfo {title} {Employment dynamics in a rapid decarbonization of the power sector}}},\ \bibinfo {type} {Tech. Rep.}\ (\bibinfo  {institution} {Institute for New Economic Thinking at the Oxford Martin School, University~…},\ \bibinfo {year} {2023})\BibitemShut {NoStop}%
\bibitem [{\citenamefont {Way}\ \emph {et~al.}(2022)\citenamefont {Way}, \citenamefont {Ives}, \citenamefont {Mealy},\ and\ \citenamefont {Farmer}}]{way2022empirically}%
  \BibitemOpen
  \bibfield  {author} {\bibinfo {author} {\bibfnamefont {R.}~\bibnamefont {Way}}, \bibinfo {author} {\bibfnamefont {M.~C.}\ \bibnamefont {Ives}}, \bibinfo {author} {\bibfnamefont {P.}~\bibnamefont {Mealy}},\ and\ \bibinfo {author} {\bibfnamefont {J.~D.}\ \bibnamefont {Farmer}},\ }\bibfield  {title} {\bibinfo {title} {Empirically grounded technology forecasts and the energy transition},\ }\href@noop {} {\bibfield  {journal} {\bibinfo  {journal} {Joule}\ }\textbf {\bibinfo {volume} {6}},\ \bibinfo {pages} {2057} (\bibinfo {year} {2022})}\BibitemShut {NoStop}%
\bibitem [{\citenamefont {IEA}(2022)}]{IEA_report}%
  \BibitemOpen
  \bibfield  {author} {\bibinfo {author} {\bibnamefont {IEA}},\ }\href@noop {} {\bibinfo {title} {{E}nergy {P}olicy {R}eview {H}ungary 2022}} (\bibinfo {year} {2022})\BibitemShut {NoStop}%
\bibitem [{\citenamefont {Van Den~Ende}\ \emph {et~al.}(2023)\citenamefont {Van Den~Ende}, \citenamefont {Mandel},\ and\ \citenamefont {Rusinowska}}]{van2023network}%
  \BibitemOpen
  \bibfield  {author} {\bibinfo {author} {\bibfnamefont {R.}~\bibnamefont {Van Den~Ende}}, \bibinfo {author} {\bibfnamefont {A.}~\bibnamefont {Mandel}},\ and\ \bibinfo {author} {\bibfnamefont {A.}~\bibnamefont {Rusinowska}},\ }\href@noop {} {\emph {\bibinfo {title} {Network-Based Allocation of Responsibility for GHG-Emissions}}}\ (\bibinfo  {publisher} {Centre d'{\'e}conomie de la Sorbonne},\ \bibinfo {year} {2023})\BibitemShut {NoStop}%
\bibitem [{\citenamefont {Act}(1991)}]{Act}%
  \BibitemOpen
  \bibfield  {author} {\bibinfo {author} {\bibnamefont {Act}},\ }\bibfield  {title} {\bibinfo {title} {Hungarian {I}nsolvency {C}ode ({A}ct xlix of 1991 on {B}ankruptcy {P}roceedings and {L}iquidation {P}roceedings)},\ }\href@noop {} {\  (\bibinfo {year} {1991})}\BibitemShut {NoStop}%
\bibitem [{\citenamefont {Dhyne}\ \emph {et~al.}(2021)\citenamefont {Dhyne}, \citenamefont {Kikkawa}, \citenamefont {Mogstad},\ and\ \citenamefont {Tintelnot}}]{dhyne2021trade}%
  \BibitemOpen
  \bibfield  {author} {\bibinfo {author} {\bibfnamefont {E.}~\bibnamefont {Dhyne}}, \bibinfo {author} {\bibfnamefont {A.~K.}\ \bibnamefont {Kikkawa}}, \bibinfo {author} {\bibfnamefont {M.}~\bibnamefont {Mogstad}},\ and\ \bibinfo {author} {\bibfnamefont {F.}~\bibnamefont {Tintelnot}},\ }\bibfield  {title} {\bibinfo {title} {Trade and domestic production networks},\ }\href@noop {} {\bibfield  {journal} {\bibinfo  {journal} {The Review of Economic Studies}\ }\textbf {\bibinfo {volume} {88}},\ \bibinfo {pages} {643} (\bibinfo {year} {2021})}\BibitemShut {NoStop}%
\bibitem [{\citenamefont {Glasserman}\ and\ \citenamefont {Young}(2016)}]{glasserman2016contagion}%
  \BibitemOpen
  \bibfield  {author} {\bibinfo {author} {\bibfnamefont {P.}~\bibnamefont {Glasserman}}\ and\ \bibinfo {author} {\bibfnamefont {H.~P.}\ \bibnamefont {Young}},\ }\bibfield  {title} {\bibinfo {title} {Contagion in financial networks},\ }\href@noop {} {\bibfield  {journal} {\bibinfo  {journal} {Journal of Economic Literature}\ }\textbf {\bibinfo {volume} {54}},\ \bibinfo {pages} {779} (\bibinfo {year} {2016})}\BibitemShut {NoStop}%
\bibitem [{\citenamefont {Diem}\ \emph {et~al.}(2020)\citenamefont {Diem}, \citenamefont {Pichler},\ and\ \citenamefont {Thurner}}]{diem2020minimal}%
  \BibitemOpen
  \bibfield  {author} {\bibinfo {author} {\bibfnamefont {C.}~\bibnamefont {Diem}}, \bibinfo {author} {\bibfnamefont {A.}~\bibnamefont {Pichler}},\ and\ \bibinfo {author} {\bibfnamefont {S.}~\bibnamefont {Thurner}},\ }\bibfield  {title} {\bibinfo {title} {What is the minimal systemic risk in financial exposure networks?},\ }\href@noop {} {\bibfield  {journal} {\bibinfo  {journal} {Journal of Economic Dynamics and Control}\ }\textbf {\bibinfo {volume} {116}},\ \bibinfo {pages} {103900} (\bibinfo {year} {2020})}\BibitemShut {NoStop}%
\bibitem [{\citenamefont {Chakraborty}\ \emph {et~al.}(2024)\citenamefont {Chakraborty}, \citenamefont {Reisch}, \citenamefont {Diem}, \citenamefont {Astudillo-Est{\'e}vez},\ and\ \citenamefont {Thurner}}]{chakraborty2024inequality}%
  \BibitemOpen
  \bibfield  {author} {\bibinfo {author} {\bibfnamefont {A.}~\bibnamefont {Chakraborty}}, \bibinfo {author} {\bibfnamefont {T.}~\bibnamefont {Reisch}}, \bibinfo {author} {\bibfnamefont {C.}~\bibnamefont {Diem}}, \bibinfo {author} {\bibfnamefont {P.}~\bibnamefont {Astudillo-Est{\'e}vez}},\ and\ \bibinfo {author} {\bibfnamefont {S.}~\bibnamefont {Thurner}},\ }\bibfield  {title} {\bibinfo {title} {Inequality in economic shock exposures across the global firm-level supply network},\ }\href@noop {} {\bibfield  {journal} {\bibinfo  {journal} {Nature Communications}\ }\textbf {\bibinfo {volume} {15}},\ \bibinfo {pages} {3348} (\bibinfo {year} {2024})}\BibitemShut {NoStop}%
\bibitem [{\citenamefont {Inoue}\ and\ \citenamefont {Todo}(2024)}]{inoue2024disruption}%
  \BibitemOpen
  \bibfield  {author} {\bibinfo {author} {\bibfnamefont {H.}~\bibnamefont {Inoue}}\ and\ \bibinfo {author} {\bibfnamefont {Y.}~\bibnamefont {Todo}},\ }\bibfield  {title} {\bibinfo {title} {Disruption risk evaluation on large-scale production network with establishments and products},\ }\href@noop {} {\bibfield  {journal} {\bibinfo  {journal} {arXiv preprint arXiv:2410.05595}\ } (\bibinfo {year} {2024})}\BibitemShut {NoStop}%
\end{thebibliography}
\end{document}